\documentclass[a4paper,11pt]{article}
\usepackage{a4wide}
\usepackage{amsmath}
\usepackage{subcaption}
\usepackage{cite}
\usepackage[normalem]{ulem}
\usepackage{amssymb}
\usepackage{epsf}
\usepackage{amsfonts}
\usepackage{graphicx}
\usepackage{tikz}
\usepackage{pgfplots}
\usepackage{braket}
\usepackage{ragged2e}
\usepackage{appendix}
\numberwithin{equation}{section}
\usepackage{bbm}
\usepackage{xcolor}
\usepackage[export]{adjustbox}
\usepackage{siunitx,pslatex}
\usepackage{booktabs}
\usepackage{hyperref}
\usepackage{ulem}
\usepackage{caption}
\usetikzlibrary{decorations.pathmorphing, decorations.markings, arrows.meta, calc}
\usepgfplotslibrary{fillbetween}

\hypersetup{
    colorlinks,
    linkcolor={green!50!black},
    citecolor={magenta},
    urlcolor={blue!80!black}
}
\graphicspath{{Figures/}}

\begin{document}
	
	\title{Phase transitions in primary hair planar black holes and solitons}
 
	\author{\textbf{Som Abhisek Mohanty}\thanks{525ph1001@nitrkl.ac.in},~\textbf{Subhash Mahapatra}\thanks{mahapatrasub@nitrkl.ac.in}
		\\\\\textit{{\small Department of Physics and Astronomy, National Institute of Technology Rourkela, Rourkela - 769008, India}}\\
	}
    \date{}
\maketitle

\begin{abstract}
We present a family of planar black hole and soliton solutions with primary scalar hair in asymptotically anti-de Sitter (AdS) space in $D$ dimensions. By solving the coupled Einstein-scalar field equations, we obtain analytic planar hairy black hole and soliton geometries. In these solutions, the scalar field and curvature scalars remain regular everywhere. We also derive analytic expressions for the mass and free energy, which indicate that the hairy soliton represents the ground state of the system. We further analyze the phase transitions between the hairy black hole and the hairy soliton, and find that there exists a first-order phase transition between them, with the transition point controlled by the ratio of the periods of Euclidean time and compact spacelike cycle. The hairy soliton solution obtained here is partly motivated by holographic QCD and may provide a useful gravitational background for modeling the confined phase of QCD from a bottom-up holographic perspective.

\end{abstract}

\section{Introduction}
\label{sec:intro}
Black holes in Anti-de Sitter (AdS) spacetime have emerged as an important arena for investigating the interplay between gravity, thermodynamics, and quantum field theory. The presence of a negative cosmological constant endows AdS spacetime with a natural confining boundary, allowing black holes to exist in thermal equilibrium with their surroundings. This sharply contrasts with asymptotically flat spacetimes, where defining a stable thermodynamic ensemble is more subtle. Consequently, AdS black holes provide a controlled framework in which one can meaningfully study black hole thermodynamics and phase structure \cite{Hawking:1982dh}. These features have also made AdS spacetimes central to the gauge/gravity duality, where gravitational dynamics in the bulk are related to strongly coupled quantum field theories on the boundary \cite{Maldacena:1997re, Gubser:1998bc, Witten:1998qj}.

A distinctive aspect of AdS black holes is the richness of their horizon geometries. While four-dimensional black holes in asymptotically flat spacetimes are restricted to spherical horizon topology, AdS black holes admit a wider class of horizons, including spherical ($\mathbb{S}^2$), planar ($\mathbb{R}^2$), and hyperbolic ($\mathbb{H}^2$) geometries \cite{Lemos:1994xp, Friedman:1993ty, Cai:1996eg, Vanzo:1997gw, Birmingham:1998nr, Mann:1996gj}. This topological freedom has major implications for thermodynamic behavior, leading to a wide variety of phase structures with no analog in non-AdS settings, as the dependence of thermodynamic quantities on horizon topology greatly enriches the landscape of possible black hole phases \cite{Brill:1997mf, Priyadarshinee:2021rch, Sheykhi:2007wg}.

A particularly important development in this context is the discovery that AdS black holes exhibit rich phase structure, closely resembling that of ordinary thermodynamic systems. The seminal work of Hawking and Page showed that, in AdS space, there exists a first-order phase transition between spherical AdS black hole and global AdS spacetime \cite{Hawking:1982dh}. Above the critical
temperature $T_c$, the black hole becomes thermodynamically favored, while below $T_c$ global AdS becomes thermodynamically favored. This Hawking-Page transition is now understood, through the gauge/gravity correspondence, as a confinement-deconfinement transition in the dual gauge theory \cite{Witten:1998zw}. Consequently, black hole phase transitions in AdS have acquired direct relevance for the study of strongly interacting matter, such as the quark-gluon plasma. Beyond the Hawking-Page transition, more intricate phase behavior arises when additional parameters such as charge, rotation, scalar fields, or external fields are introduced. Charged and rotating AdS black holes can display phase transitions analogous to those of Van der Waals fluids, including critical points and second-order phase transitions \cite{Chamblin:1999tk, Chamblin:1999hg, Cvetic:1999ne, Kubiznak:2012wp, Caldarelli:1999xj}. For reviews on this subject, see \cite{Kubiznak:2016qmn}. The inclusion of scalar fields or nonlinear matter sectors further enriches the phase diagram and allows one to model phenomena such as chiral symmetry breaking, anisotropy, and magnetic-field effects in the dual field theory \cite{Gunasekaran:2012dq, Hendi:2012um, Zou:2013owa, Bohra:2020qom, Jena:2024cqs, Jena:2025xcf}.
 
The occurrence of phase transitions depends crucially on the spacetime topology, both in the gravitational description and in the dual gauge theory. For Schwarzschild-AdS black holes with planar horizon geometry, for example, there is no Hawking-Page transition relative to thermal AdS: the planar black hole solution remains thermodynamically favored at all nonzero temperatures. As a result, the corresponding dual thermal gauge theory, defined on a planar spatial manifold, does not undergo a phase transition and stays in the deconfined phase for all temperatures \cite{Vanzo:1997gw, Birmingham:1998nr}. Interestingly, by compactifying one of the horizon coordinates, the planar AdS black hole can undergo an analogous Hawking-Page type phase transition \cite{Surya:2001vj}. In this case, the lowest-energy configuration is the AdS soliton, which can be obtained via a double analytic continuation of the planar black hole geometry \cite{Horowitz:1998ha}. The AdS soliton thus emerges as a competing thermodynamic phase, and the phase transition between the planar black hole and the AdS soliton is governed by the dimensionless ratio $\beta_b/L_b$, where 
$\beta_b$ denotes the period of the Euclidean time (thermal) cycle and 
$L_b$ corresponds to the period of the compact spacelike cycle \cite{Surya:2001vj}. The AdS soliton constitutes the ground state of the gravitational theory, whose uniqueness was outlined in \cite{Horowitz:1998ha} and further explored in \cite{Galloway:2001uv, Galloway:2002ai, Woolgar:2016axs}.

Given the importance of the AdS soliton in holographic applications, see for instance \cite{Witten:1998zw, Nishioka:2006gr, Nishioka:2009zj}, the work of \cite{Surya:2001vj}  triggered interest in many directions. This includes supersymmetric soliton solutions \cite{Anabalon:2021tua}, and adding a linear and nonlinear gauge field in the Einstein action, leading to a charged AdS soliton \cite{Banerjee:2007by, Anabalon:2022ksf, Quijada:2023fkc}. With a $U(1)$ charge, the thermodynamic phase structure between the charged black hole and charged soliton becomes even more interesting, with magnetic flux (due to the magnetic charge of the AdS soliton) entering as a control variable in a new ensemble. This leads to the possibility of phase transitions even at zero temperature. Another interesting
extension of this work was in the context of Gauss-Bonnet-dilaton gravity \cite{Cai:2007wz}. More recently, the phase transition between neutral hairy planar black hole and solitons was explored in \cite{Anabalon:2019tcy}. In the present work, our main objective is to construct a class of black hole and soliton solutions with regular primary scalar hair in asymptotically AdS spacetime, and to investigate the impact of the scalar hair on the corresponding phase structure.

The interplay between the scalar field and AdS geometries has attracted much attention of late. For instance, hairy black holes in AdS space have 
appeared abundantly in applied holographic theories, with the most noticeable application occurring in the context of holographic superconductors (and other condensed matter phenomena) \cite{Gubser:2008px, Hartnoll:2008vx}. Similarly, much of the area of holographic QCD is based on a combined system of Einstein-scalar gravity theory. In holographic QCD, the scalar field plays the role of the running coupling constant in the dual field theory and is essential for realistic QCD model building \cite{Gubser:2008ny, DeWolfe:2010he, Gursoy:2008za, Jarvinen:2011qe}. See \cite{Rougemont:2023gfz} for a recent review on holographic QCD. In addition, it is well known that the addition of the scalar field can modify the thermodynamic properties of the AdS black holes in a nontrivial way \cite{Astefanesei:2023sep, Guo:2021ere, Giribet:2014fla, Hennigar:2015wxa}. 

The investigation of the Einstein-scalar gravity system is also important from a purely gravitational point of view. Particularly, black holes are conjectured to follow the famous no-hair theorem \cite{Ruffini:1971bza}. The black hole no-hair conjecture simply asserts that, in asymptotically flat spacetime, a stationary black hole with a spherical horizon is fully specified by only three global charges: its mass, angular momentum, and electric charge. In other words, such black holes are not expected to sustain additional matter degrees of freedom, such as the scalar fields, outside the event horizon. Although this idea has been supported by a number of early studies \cite{Bekenstein:1971hc, Israel:1967wq, Wald:1971iw, Carter:1971zc, Robinson:1975bv, Mazur:1982db, Teitelboim:1972qx}, it does not constitute a rigorous theorem in the strict mathematical sense. In fact, various counterexamples have since been constructed in different theoretical settings \cite{Volkov:1990sva, Bizon:1990sr, Kuenzle:1990is, Garfinkle:1990qj, Greene:1992fw, Torii:1993vm, Herdeiro:2014goa, Berti:2013gfa, Ovalle:2020kpd, Mahapatra:2022xea, Navarro:2026lrf, Guimaraes:2025jsh, Meert:2021khi, daRocha:2026kko}. For a comprehensive review on the existence of scalar
hair, see \cite{Herdeiro:2015waa}. Over the past several years, extensive investigations of hairy black hole solutions in diverse asymptotic geometries have been carried out; see \cite{Torii:1998ir, Torii:2001pg, Martinez:2004nb, Winstanley:2002jt, Kolyvaris:2010yyf, Dias:2011tj, Bhattacharyya:2010yg, Kleihaus:2013tba, Kolyvaris:2011fk, Anabalon:2013qua, Mahapatra:2020wym, Priyadarshinee:2023cmi, Daripa:2024ksg} for a representative (though not exhaustive) list of references.

In this work, we present a family of analytic stable primary-hair black hole and soliton solutions in the Einstein-scalar gravity system in general dimensions. In particular, the Einstein-scalar system is considered, and the coupled Einstein-scalar field equations are solved simultaneously in terms of a scale function $A(z)$ (see the next section for details) using the potential reconstruction technique \cite{Dudal:2017max, Bohra:2019ebj, Mahapatra:2018gig, He:2013qq, Arefeva:2018hyo, Arefeva:2018cli, Dudal:2021jav, Alanen:2009xs, Cai:2012xh, Toniato:2025gts}.  The different choices of $A(z)$ allow us to construct a different family of hairy black hole and soliton solutions. To present a more systematic and comprehensive analysis, we choose two particular forms of $A(z)=-a z$
and $A(z)=-a z^2$, with the parameter $a$ controlling the strength of the scalar hair. The prime motivation for considering such $A(z)$ forms is that they have been extensively studied within the framework of gauge/gravity duality, particularly in the construction of holographic models of QCD. In these approaches, suitable choices of the form factor $A(z)$ are known to effectively capture several qualitative features of strongly coupled gauge theories, such as confinement, linear Regge trajectories, and other nonperturbative aspects of QCD-like dynamics \cite{Karch:2006pv, Gursoy:2007cb, Dudal:2018ztm, Mahapatra:2019uql, Jena:2022nzw, Herzog:2006ra}. 
Motivated by these developments, the class of solutions constructed here provides a natural gravitational setup to further explore the properties of strongly coupled gauge theories from a holographic perspective.  In particular, the novel hairy soliton solutions constructed here are expected to play a significant role in probing various aspects of confined QCD phases from the bottom-up perspective.

The constructed hairy solutions exhibit several appealing features. First, the scalar field remains regular and finite throughout the region outside the horizon (or outside the tip of the soliton) and falls off at the
asymptotic boundary. Second, the curvature invariants, including the Kretschmann and Ricci scalars, stay finite outside the horizon, indicating the absence of additional singularities in the spacetime. Third, in the limit $a\rightarrow 0$, the solutions smoothly reduce to the standard planar black hole and solitonic geometries. Finally, the scalar potential remains bounded above by its UV boundary value, thereby satisfying the Gubser criterion for a well-defined boundary theory \cite{Gubser:2000nd}, and approaches the $D$-dimensional negative cosmological constant value near the asymptotic boundary. We then investigate the thermodynamics of the constructed hairy solutions. We obtain analytic expressions of the Gibbs free energy and mass of hairy black holes and solitons using the holographic renormalization procedure, and find that they satisfy the standard thermodynamic relations. Interestingly, we find that there is a phase transition between the hairy black hole and hairy soliton phase. The transition appears when the periods of Euclidean time and compact spacelike cycle become equal, i.e., $\beta_b=L_b$. At this point, the free energy of the hairy black hole and soliton exchange dominance, with soliton phase dominating the phase structure when $L_b<\beta_b$, while the black hole phase dominates when $L_b>\beta_b$. We further examine the effect of the scalar hair parameter $a$ on the transition temperature and find that it increases with $a$ when the soliton tip position $z_0$ is held fixed. This increase reflects the corresponding change in the boundary circumference, rather than an independent hair-induced shift at fixed boundary geometry. Consequently, the temperature range over which the soliton phase is preferred expands as $a$ increases under this choice of fixed $z_0$. This result remains true irrespective of the form of $A(z)$ considered in this work, as well as in different dimensions. 

It is important to stress that the hairy black hole and solitonic solutions constructed in this work correspond to primary scalar hair. In this context, it is useful to distinguish between primary and secondary hair \cite{Coleman:1991ku}. Secondary hair typically arises as a consequence of already existing primary hair, such as gauge charges, and therefore does not introduce genuinely new independent characteristics to the black hole. In other words, primary hair provides an additional independent parameter (or quantum number) describing the black hole, whereas secondary hair does not \cite{Greene:1992fw}. In our setup, the scalar hair can be smoothly turned off, reducing the solutions to the planar black hole and soliton geometries in the limit  $a\rightarrow 0$. It is worth noting that most studies of hairy black holes in the literature focus on secondary hair, while examples involving primary hair are comparatively rare. A few such cases can be found in \cite{Gonzalez:2013aca, Anabalon:2012sn, Charmousis:2014zaa, Coleman:1991ku, Kitagawa:2026tcl}. 

Let us also mention that a scalar-hair AdS soliton has previously been constructed in \cite{Anabalon:2019tcy, Anabalon:2016izw}. Here, the scalar field has a logarithmic profile but remains regular over the physical radial interval, including the soliton tip and the UV boundary. Our solutions provide another example of hairy AdS solitons with a regular scalar-field profile. We believe that this feature is of independent geometrical interest, irrespective of the potential applications of these solutions to holographic QCD.

The paper is organized as follows. In the next section, we introduce the Einstein-scalar gravity framework and discuss the analytic solutions describing hairy black hole and soliton geometries in general spacetime dimensions. In Section~\ref{hairysolD5}, we study the geometrical and thermodynamical
properties of hairy black hole and soliton solutions in five dimensions for different forms of $A(z)$. The corresponding analysis in four dimensions is carried out in Section~\ref{hairysolD4}. Finally, in Section~\ref{conclusions}, we summarize our main findings and discuss possible directions for future research.

\section{Hairy AdS black hole and AdS soliton solutions}
\label{hairysol}
To construct and study primary hair black hole and soliton solutions in general spacetime dimensions, we begin with the following Einstein-scalar action,
\begin{eqnarray}
S_{ES} = \frac{1}{16 \pi G_D} \int \mathrm{d^D}x \ \sqrt{-g}  \ \left[R  -\frac{1}{2}\partial_{M}\phi \partial^{M}\phi -V(\phi)\right]\,,
\label{actionES}
\end{eqnarray}
where $G_D$ is the $D$-dimensional Newton constant, $\phi$ is the scalar field, and  $V(\phi)$ is the potential of the field $\phi$. The variation of the action (\ref{actionES}) leads to the following Einstein and scalar field equations:
\begin{eqnarray}
& & R_{MN}- \frac{1}{2}g_{MN} R + \frac{1}{2} \biggl(\frac{g_{MN}}{2} \partial_{P}\phi \partial^{P}\phi -\partial_{M}\phi \partial_{N}\phi  + g_{MN} V(\phi)  \biggr)  =0 \,, 
\label{EinsteinEE}
\end{eqnarray}
\begin{eqnarray}
& & \frac{1}{\sqrt{-g}}\partial_{M} \biggl[ \sqrt{-g}  \partial^{M}\phi \biggr] - \frac{\partial V(\phi)}{\partial \phi} = 0 \,.
\label{dilatonEE}
\end{eqnarray}

\subsection{Planar hairy AdS black holes}

Let us first construct a planar hairy black hole solution. For this purpose, we consider the following \textit{Ans\"atze} for the metric $g_{MN}$ and scalar field $\phi$:
\begin{eqnarray}
& & ds_{b}^2=\frac{\ell^2 e^{2A(z)}}{z^2}\biggl[-g_b(z)dt_b^2 + \frac{dz^2}{g_b(z)} + dx_{b}^2 + \sum\limits_{i=2}^{D-2} dx_i^2 \biggr] \,, \nonumber \\
& & \phi=\phi(z)\,,
\label{bhansatz}
\end{eqnarray}
where $\ell$ is the AdS length scale, which will be set to one from now on, $A(z)$ is the scale factor, and $g_b(z)$ is the blackening function. We take the spatial coordinate $x_b$ to be compact, i.e., $0 \leq x_b \leq L_b$, while the remaining spatial coordinates $x_i$ are taken to be noncompact (extended). As usual, $z$ is the radial coordinate, and it runs from $z=0$ (asymptotic boundary) to $z=z_h$ (horizon radius), or to $z=\infty$ for thermal-AdS (without horizon). A primary motivation for considering the above metric \textit{Ans\"atze} is their usefulness in holographic QCD model building. In particular, with an appropriate choice of the form factor $A(z)$, one can realize both confined and deconfined phases in the dual boundary field theory, while simultaneously incorporating essential features of QCD, such as the Regge trajectories of heavy meson spectra \cite{Dudal:2017max}. Consequently, the resulting hairy black hole and soliton configurations provide an important and versatile framework for holographic QCD studies.

Substituting the above \textit{Ans\"atze} into Eq.~(\ref{EinsteinEE}), we get three Einstein equations of motion,
\begin{eqnarray}
t_bt_b: && \ A''(z) + A'(z) \left(\frac{3-D}{z}+\frac{g_{b}'(z)}{2g_b(z)} +\frac{(D-3)}{2}A'(z) \right) 
-\frac{g_{b}'(z)}{2zg_b(z)}   \nonumber \\
& & + \frac{(D-1)}{2 z^2}+\frac{\phi'(z)^2}{4(D-2)} +\frac{e^{2A(z)}\ell^2V(z)}{2(D-2)z^2g_b(z)} =0\,,
\label{Einsteintt}
\end{eqnarray}
\begin{eqnarray}
zz: & & \ g_b(z)\left(\frac{D-1}{z^2}-\frac{2(D-1)A'(z)}{z}+(D-1)A'(z)^2-\frac{\phi'(z)^2}{2(D-2)}  \right)\nonumber \\
& & + g_{b}'(z) \left(A'(z)-\frac{1}{z}\right)+ \frac{e^{2A(z)}\ell^2V(z)}{(D-2)z^2}  =0 \,,
\label{Einsteinzz}
\end{eqnarray}
\begin{eqnarray}
 x_{i}x_{i}: & & \ g_{b}''(z) + 2(D-2) g_{b}'(z) \left(A'(z)-\frac{1}{z}\right)  +\frac{e^{2A(z)}\ell^2V(z)}{z^2} + \frac{\phi'(z)^2 g_b(z)}{2} \nonumber \\
& &  + (D-2) g_b(z)\left(\frac{D-1}{z^2} -2(D-3)\frac{A'(z)}{z} +(D-3)A'(z)^2  + 2 A''(z) \right)  =0 \,.
\label{Einsteinxixi}
\end{eqnarray}
While the above expressions may seem complicated, they can be reorganized into the following simpler forms, which are then much easier to handle:
\begin{eqnarray}
& & g_{b}''(z) + (D-2)g_{b}'(z) \left(A'(z)-\frac{1}{z}\right)  = 0\,,
\label{EOM11}
\end{eqnarray}
\begin{eqnarray}
& & A''(z) - A'(z) \left(A'(z)-\frac{2}{z}\right)+\frac{\phi'(z)^2}{2(D-2)} = 0 \,,
\label{EOM22}
\end{eqnarray}
\begin{eqnarray}
& & \frac{g_{b}''(z)}{4g_{b}(z)}+\frac{(D-2)}{2}A''(z) + (D-2)^2 A'(z)\left(-\frac{1}{z}+\frac{A'(z)}{2}+\frac{3}{4(D-2)}\frac{g_{b}'(z)}{g_{b}(z)}  \right) \nonumber \\
& &  -\frac{3(D-2)}{4}\frac{g_{b}'(z)}{zg_{b}(z)} +\frac{e^{2A(z)}\ell^2V(z)}{2z^2g_b(z)} +\frac{(D-1)(D-2)}{2z^2}   = 0 \,.
\label{EOM33}
\end{eqnarray}
Similarly, we get the following equation of motion for the scalar field:
\begin{eqnarray}
 \phi ''(z) +\phi '(z) \left(\frac{g_{b}'(z)}{g_b(z)}+(D-2)A'(z)-\frac{D-2}{z}\right) 
    -\frac{\ell^2 e^{2A(z)}}{z^2 g_b(z)} \frac{\partial V(\phi)}{\partial \phi} = 0 \,.
\label{dilatonEOM}
\end{eqnarray}
Hence, we obtain a total of four equations of motion. It can be straightforwardly shown, however, that only three of these are independent. In our analysis, we treat the scalar equation (\ref{dilatonEOM}) as a constraint and regard the remaining equations as independent. To solve this system, we impose the following boundary conditions:
\begin{eqnarray}
&& g_b(0)=1 \,, ~~~ \ \ g_b(z_h)=0, \nonumber \\
&& A(0) = 0 \,, ~~~ \phi(0)=0\,,
\label{boundaryconditions}
\end{eqnarray}
which are required to ensure correct AdS asymptotics and have a black hole at $z=z_h$. Apart from these boundary conditions, we additionally require that the scalar field $\phi$ remains real everywhere in the bulk.\\

We adopt the following strategy to solve the Einstein-scalar equations of motion simultaneously:

\begin{itemize}
\item We first solve Eq.~(\ref{EOM11}) and find a solution for $g_b(z)$ in terms of $A(z)$.

\item Subsequently, we solve Eq.~(\ref{EOM22}) and find $\phi(z)$ in terms of $A(z)$.

\item Last, we solve Eq.~(\ref{EOM33}) and obtain $V(z)$ in terms of $A(z)$.
\end{itemize}

Adopting the above-mentioned strategy and solving Eq.~(\ref{EOM11}), we obtain the following solution for $g_b(z)$:
\begin{eqnarray}
& & g_b(z) =  C_1 - C_{b} \int_0^z \, d\xi \ e^{-(D-2)A(\xi)} \xi^{(D-2)} \,,
\label{gsol}
\end{eqnarray}
where the constants $C_1$ and $C_b$ can be fixed from Eq.~(\ref{boundaryconditions}) and we get
\begin{eqnarray}
C_1 = 1,  \ \ \ \ \ C_b = \frac{1}{ \int_0^{z_h} \, d\xi e^{-(D-2)A(\xi)} \xi^{(D-2)} }  \,.
\label{intconstexp}
\end{eqnarray}
Note that the integration constant $C_b$ is related to the mass of the black hole. In the next section, we demonstrate that the constructed hairy solutions are of a primary nature by explicitly computing the conserved charges and showing that they depend only on the corresponding independent integration constants.  Similarly, the scalar field $\phi$ can be solved in terms of $A(z)$ from Eq.~(\ref{EOM22}),
\begin{eqnarray}
\phi(z) = \int dz \sqrt{2(D-2)\biggl[-A''(z) + A'(z) \left(A'(z)-\frac{2}{z}\right) \biggr]} + C_{2} \,,
\label{phisol}
\end{eqnarray}
where the constant $C_{2}$ is fixed by requiring that the scalar field $\phi$ vanishes near the asymptotic boundary, i.e., $\phi \to 0$ as $z \to 0$. Finally, $V$ can be determined from Eq.~(\ref{EOM33}),
\begin{eqnarray}
V(z) & = & -\frac{2z^2g_b(z)e^{-2A(z)}}{\ell^2} \biggl[\frac{(D-1)(D-2)}{2z^2}-\frac{3(D-2)}{4}\frac{g_{b}'(z)}{zg_b(z)}+\frac{g_{b}''(z)}{4g_b(z)}+\frac{(D-2)}{2}A''(z) \nonumber\\
& &  + (D-2)^2 \left(-\frac{1}{z}+\frac{A'(z)}{2}+\frac{3}{4(D-2)}\frac{g_{b}'(z)}{g_{b}(z)}  \right)A'(z)   \biggr]   \,.
\label{Vsol}
\end{eqnarray}
The equations derived above show that the Einstein-scalar system defined in Eq.~(\ref{actionES}) admits analytic solutions once a specific choice for the form factor $A(z)$ is made. This implies that for any admissible form of $A(z)$, one can explicitly construct a planar black hole geometry endowed with scalar hair in arbitrary spacetime dimensions. Since the scalar potential is determined by the choice of $A(z)$, different form factors naturally lead to physically distinct black hole solutions. Consequently, the Einstein-scalar theory described by Eq.~(\ref{actionES}) supports a broad class of analytic hairy black hole geometries.

From the perspective of gauge/gravity duality, however, this freedom is typically restricted by physical considerations stemming from the dual boundary theory. For instance, in holographic QCD model building, the form factor is often chosen so that the boundary theory reproduces essential features of QCD, such as confinement and the confinement/deconfinement phase transition. A particularly well-studied choice is
\[
A(z) = -a~z^{n},
\]
with nonnegative $n$, which has been shown to capture several qualitative properties observed in lattice QCD simulations \cite{Dudal:2017max,Dudal:2018ztm,Gursoy:2008za}. Additional phenomenological constraints are commonly imposed: the parameter $a$ is typically taken to be positive so that the Hawking-Page transition temperature (dual to the confined/deconfined transition temperature) matches lattice expectations, while values of $n>1$ are favored to ensure confinement at low temperatures \cite{Dudal:2017max,Dudal:2018ztm}. In the present work, we adopt a more flexible approach and explore a wider range of values for $a$ and $n$, corresponding to different scalar potentials, in order to systematically investigate how scalar hair influences black hole physics and phase transitions between different geometric solutions across various spacetime dimensions.

From a geometric point of view, the choice $A(z) = -a z^{n}$ with $n \geq 1$ is also well motivated, as it guarantees that the spacetime approaches AdS near the asymptotic boundary $z \to 0$. In this limit, the scalar potential admits the expansion
\begin{eqnarray}
V(z)\big|_{z \to 0} &=& -\frac{(D-1)(D-2)}{\ell^{2}} + \frac{m^{2}\phi^{2}}{2} + \cdots \nonumber \\
&=& 2\Lambda + \frac{m^{2}\phi^{2}}{2} + \cdots ,
\label{Vsolexp}
\end{eqnarray}
where $\Lambda = -(D-1)(D-2)/2\ell^{2}$ denotes the negative cosmological constant. Together with the condition $g_b(z)\big|_{z \to 0} = 1$, this ensures that the geometry is asymptotically AdS. The scalar field mass $m^{2}$ satisfies the Breitenlohner-Freedman (BF) bound, $m^{2} \geq -(D-1)^{2}/4$, ensuring the stability of the AdS background \cite{Breitenlohner:1982jf}. Moreover, the scalar potential obeys the Gubser criterion and remains bounded from above by its UV value, i.e., $V(0) \geq V(z)$, ensuring a well-defined dual boundary theory \cite{Gubser:2000nd}.       

To compute the temperature of the constructed hairy black hole, we Wick rotate the metric ($t_b \rightarrow -i\tau_b$) to obtain the corresponding Euclidean black hole geometry
\begin{eqnarray}
& & ds_{b}^2=\frac{\ell^2 e^{2A(z)}}{z^2}\biggl[g_b(z)d\tau_{b}^2 + \frac{dz^2}{g_b(z)} + dx_{b}^2 + \sum\limits_{i=2}^{D-2} dx_i^2 \biggr]\,.
\label{bhansatz}
\end{eqnarray}
Requiring regularity of the Euclidean metric at the horizon $z = z_h$ imposes a periodic identification of the imaginary time coordinate with period $\beta_b = 1/T_b$. This leads to
\begin{eqnarray}
 T_{b} & = & -\frac{g_{b}'(z_h)}{4 \pi} = \frac{C_b z_h^{D-2} e^{-(D-2)A(z_h)}}{4 \pi }  \\ \nonumber & = & \frac{z_{h}^{D-2} e^{-(D-2)A(z_h)}}{ 4 \pi \int_0^{z_h} \, d\xi e^{-(D-2)A(\xi)} \xi^{(D-2)} } \,,
\label{BHTexp}
\end{eqnarray}
where $T_b$ is the temperature of the black hole. Let us also now record the expression of the black hole entropy, which will be useful for the subsequent discussion of black hole thermodynamics,
\begin{eqnarray}
& & S_{b}=\frac{\ell^{D-2}L_b V_{(D-3)} e^{(D-2)A(z_h)}}{4 G_D z_{h}^{D-2}} \,,
\label{BHentexp}
\end{eqnarray}
where $V_{(D-3)}$ is the volume of the $(D-3)$-dimensional plane.

It is important to emphasize again that Eqs.~(\ref{gsol})-(\ref{Vsol}) constitute solutions of the Einstein-scalar action (\ref{actionES}) for any $A(z)$. Consequently, we have an infinite family of analytic hairy black hole solutions. Different choices of $A(z)$, however, lead to different scalar potentials $V$, and hence correspond to different Einstein-scalar theories. Nonetheless, once we have fixed the forms of $A(z)$, then the form of $V(z)$ is also fixed, and in return, Eqs.~(\ref{gsol})-(\ref{Vsol}) provide a self-consistent solution to a particular Einstein-scalar action with the associated scalar potential.

\subsection{Hairy thermal-AdS}
It is important to recognize that the Einstein-scalar equations also admit an alternative solution, one without a horizon, corresponding to thermal-AdS. This configuration emerges in the limit $z_h \to \infty$ of the black hole solution described above, which effectively sets $g_b(z) = 1$. The metric for thermal-AdS then takes the form
\begin{equation}
ds^2_{AdS} = \frac{\ell^2 e^{2A(z)}}{z^2} \Big[-dt^2 + dz^2 + dx_{1}^2 + \sum_{i=2}^{D-2} dx_i^2 \Big] \,.
\label{thermalAds}
\end{equation}
Although the spacetime remains asymptotically AdS, the bulk geometry can exhibit nontrivial structure depending on the functional form of $A(z)$. In particular, for the forms of $A(z)$ considered in this work, the hairy thermal-AdS geometry develops a curvature singularity in the infrared limit $z\to\infty$, as curvature invariants diverge there. Nevertheless, the presence of this singularity does not necessarily render the background physically inadmissible. In particular, the scalar potential remains bounded from above by its ultraviolet value along the solution, satisfying the Gubser criterion for an admissible infrared singularity \cite{Gubser:2000nd}. Thus, although the thermal-AdS geometry is not regular in the deep infrared, its singularity is of the acceptable type according to this criterion. We therefore regard the deformed thermal-AdS solution as an admissible infrared background for the thermodynamic comparison considered here. Note that this type of deformed thermal-AdS geometries, especially with $A(z)=-a z^2$, are of great importance to construct confined structure in the dual theory and have been widely used in the literature \cite{Karch:2006pv, Gursoy:2007cb}.

\subsection{Hairy AdS solitons}
Following \cite{Horowitz:1998ha}, the $D$-dimensional hairy AdS soliton can be obtained by a double
analytic continuation $(t_b\rightarrow i x_s,~x_b\rightarrow i t_s)$ of the hairy black hole metric (\ref{bhansatz}). The metric for the hairy AdS soliton is written as
\begin{eqnarray}
& & ds_{s}^2=\frac{\ell^2 e^{2A(z)}}{z^2}\biggl[-dt_{s}^2 + \frac{dz^2}{g_s(z)} + g_s(z) dx_{s}^2 + \sum\limits_{i=2}^{D-2} dx_i^2 \biggr] \,,
\label{sansatz}
\end{eqnarray}
where the coordinate $x_s$ is periodic and has an arbitrary period $L_s$, i.e., $0\leq x_s \leq L_s$. The function $g_s(z)$ has the same expression as in Eq.~(\ref{gsol}), with $z_h$ replaced by $z_0$
\begin{eqnarray}
& & g_s(z)  = 1 -  C_s \int_0^z \, d\xi \ e^{-(D-2)A(\xi)} \xi^{(D-2)} = 1 -  \frac{\int_0^z \, d\xi \ e^{-(D-2)A(\xi)} \xi^{(D-2)}}{\int_{0}^{z_0} \, d\xi \ e^{-(D-2)A(\xi)} \xi^{(D-2)}} \,.
\label{gssol}
\end{eqnarray}
The regular center point $z=z_0$ corresponds 
to the largest root of $g(z)=0$. At $z=z_0$, the geometry is smooth, and the 
radial coordinate now runs over the interval $0 \leq z \leq z_0$. The regularity 
of the solitonic spacetime at $z=z_0$ further fixes the period of the coordinate 
$x_s$ to be
\begin{eqnarray}
\label{solperiod1}
L_{s} & = & -\frac{4 \pi}{g_{s}'(z_0)} \\ \nonumber
& = & \frac{4 \pi \int_0^{z_0} \, d\xi e^{-(D-2)A(\xi)} \xi^{(D-2)}}{z_{0}^{D-2} e^{-(D-2)A(z_0)}}\,.
\end{eqnarray}
The Euclidean planar solitonic geometry is ($t_s\rightarrow -i\tau_s$)
\begin{eqnarray}
& & ds_{s}^2=\frac{\ell^2 e^{2A(z)}}{z^2}\biggl[d\tau_{s}^2 + \frac{dz^2}{g_s(z)} + g_s(z) dx_{s}^2 + \sum\limits_{i=2}^{D-2} dx_i^2 \biggr] \,,
\label{sansatz}
\end{eqnarray}
where the coordinate $\tau_s$ have the period $\beta_s$, which is not restricted by any new regularity condition. Note that, in discussing the phase transition between the black hole and soliton solutions, we must match their asymptotic geometries. This corresponds to
\begin{eqnarray}
& & \sqrt{g_{\tau_b\tau_b}}\beta_b|_{z \rightarrow 0} = \sqrt{g_{\tau_s\tau_s}}\beta_s|_{z \rightarrow 0} \,, \nonumber \\
& & \sqrt{g_{x_b x_b}}L_b|_{z \rightarrow 0} = \sqrt{g_{x_s x_s}}L_s|_{z \rightarrow 0} \,.
\label{sansatz}
\end{eqnarray}
Since the on-shell Euclidean actions, after regularization, remain finite for each configuration (see the next section), it is sufficient to keep only the leading contribution in the cutoff expansion. This requires the Euclidean time periods to be equal, i.e., $\beta_b=\beta_s$, and also the periods of the compact spatial coordinate to match, i.e., $L_b=L_s$.

A word about the potential reconstruction methods and associated limitations with it are in order. One limitation with this method is that the potential seems to depend on the parameters of the theory, for instance, the temperature (or $z_h$) in our case. One might also worry that the dependence of $V$ on $z_h$ is troublesome as it indicates that different values
of it correspond to a different action, and therefore, to a different gravity model altogether. Note that the off-shell action in Eq.~(\ref{actionES}) does not contain $z_h$ explicitly. Only when the equations of motion and, more crucially, the boundary
conditions are imposed, these parameters appear in the theory (just like in Einstein
gravity without matter fields to the best of our understanding). The input $A(z)$ is
independent of these parameters. It is true that, in order to get an explicit expression
for the potential, we use the equations of motion. However, the potential will depend
on $z_h$ only when the boundary conditions are imposed. At the level of equations of
motion alone, there is no explicit dependence on $z_h$. Indeed, notice that there is a second independent solution
to equations of motion, which corresponds to thermal-AdS. For the thermal-AdS, we have $g(z) = 1$. Then it is easy to
infer from Eq.~(\ref{gsol}) that $V$ is independent of $z_h$.
If the potential were dependent on $z_h$ from the beginning in the action itself, then the potential in the thermal-AdS background
would have depended on $ z_h$ as well, which is certainly not the case as we have alluded to above. Therefore, one
should interpret the dependence of the potential $V$ on $z_h$ as
an on-shell dependence, and not as an off-shell one.     

Nevertheless, it is desirable that $V$ should depend only mildly on $z_h$. If this dependence is sufficiently weak, the reconstructed background can still be regarded as a reliable approximate solution. In particular, demonstrating that the potential $V(\phi)$, evaluated along the flow, varies only mildly with the parameters of the model provides evidence for the internal consistency of the approximation. In the next section, we will show explicitly that this is the case.

We would also like to emphasize another important motivation for using the potential reconstruction approach. A key advantage of this framework is that it often provides analytic or semi-analytic control over the background solutions, allowing for a clearer understanding of the underlying physics and parameter dependence. In contrast, many alternative holographic QCD models rely heavily on numerical solutions \cite{DeWolfe:2011ts}, which can sometimes obscure the physical origin of various phenomena. Moreover, widely used soft-wall models \cite{Karch:2006pv} (which are ad hoc in construction and are not consistent gravitational solutions) can often be viewed as specific realizations of backgrounds generated within the potential reconstruction framework. In this way, using the reconstruction method, one can partially associate a consistent framework with the soft-wall model. Therefore, besides its phenomenological relevance and despite its limitations, the potential reconstruction method offers a useful balance between analytical tractability and model-building flexibility, making it an attractive framework for holographic QCD studies.

The solutions developed above are completely general and apply to any spacetime dimension. Below, we will thoroughly discuss the gravitational and thermodynamical aspects of the constructed hairy black hole and soliton solutions in $D=4$ and $D=5$ dimensions. The corresponding analysis for $D=6$, which is more directly relevant to holographic QCD, is presented in the Appendix. The $D=4$ and $D=5$ AdS-soliton backgrounds are dual to confined phases of field theories in $1+1$ and $1+2$ dimensions, respectively (with an additional compact spatial direction). Although these theories are not directly dual to $1+3$-dimensional QCD, lower-dimensional QCD-like theories have long served as important theoretical laboratories for exploring confinement, mass generation, and other nonperturbative aspects of strongly coupled gauge theories \cite{tHooft:1974pnl, Polyakov:1976fu, Gross:1992tu}. Consequently, the $D=4$ and $D=5$ analyses presented below are of independent interest, while the $D=6$ case provides the most direct connection to holographic QCD. From a purely gravitational perspective, these solutions are also of independent interest, as they provide a class of stable hairy AdS solitons with a regular scalar field profile.

It is worth emphasizing that the conserved mass, including the contribution from the scalar field, is determined only after a specific form of $A(z)$ is chosen. Since different choices of $A(z)$ lead to different scalar potentials and different near-boundary expansions of the scalar field, the corresponding holographic renormalization counterterms and the explicit form of the renormalized mass are, in general, $A(z)$-dependent. Thus, although the scalar contribution is consistently included in the holographically renormalized boundary stress tensor, there is no single universal explicit expression for this contribution applicable to arbitrary $A(z)$. In the following sections, we therefore evaluate the renormalized mass explicitly for the particular choices of $A(z)$ considered in our analysis.

\section{Stability and thermodynamic phase transitions in $D=5$}
\label{hairysolD5}
In this section, we discuss the thermodynamic stability and phase transitions of the hairy black holes and solitons constructed in the previous section. Here, we concentrate on $D=5$. In the next section, we shall analyze the results with $D=4$.

\subsection{Case: $n=1$}
Let us first discuss the geometric and thermodynamic properties of the gravity system with $n=1$, corresponding to $A(z)=-a z$. With $A(z)=-a z$, the expressions for the blackening function $g_b(z)$ and
the scalar field $\phi(z)$ in five dimensions reduce to
\begin{eqnarray}
 g_b(z) & = & 1 - \frac{C_b \left(e^{3 a z} (3 a z (3 a z (a z-1)+2)-2)+2\right)}{27a^4},  \nonumber \\
 \phi(z)  & = & \sqrt{6} \left(\sqrt{a z (a z+2)}-\log \left(a z-\sqrt{a z (a
   z+2)}+1\right)\right)\,,
\label{gbsoln1D5}
\end{eqnarray}
where the integration constant $C_b$ is
\begin{eqnarray}
C_b  & = & \frac{27 a^4}{e^{3 a z_h} \left(3 a z_h \left(3 a z_h \left(a
   z_h-1\right)+2\right)-2\right)+2}\,.
   \label{inteconstD5n1}
\label{Cbsoln1D5}
\end{eqnarray}
Note that the nonzero values of the scalar field appear only when $a\neq 0$, and it vanishes when
$a=0$. This behavior is consistent with the expectation that, in the limit $a \to 0$, the solution should reduce to the standard planar Schwarzschild black hole. Indeed, this can be explicitly verified by taking the limit $a \to 0$ in the blackening function $g_b(z)$, which then smoothly reduces to the corresponding planar Schwarzschild black hole form.

\begin{figure}[ht]
\centering
\includegraphics[width=0.45\textwidth]{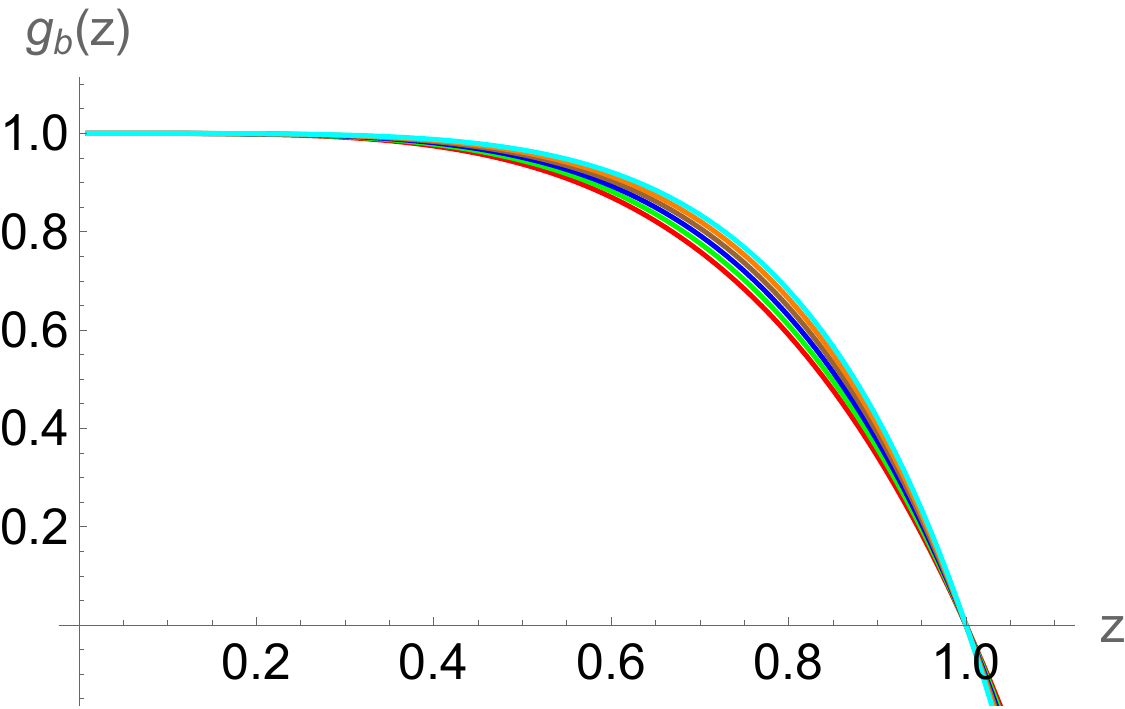}
\includegraphics[width=0.45\textwidth]{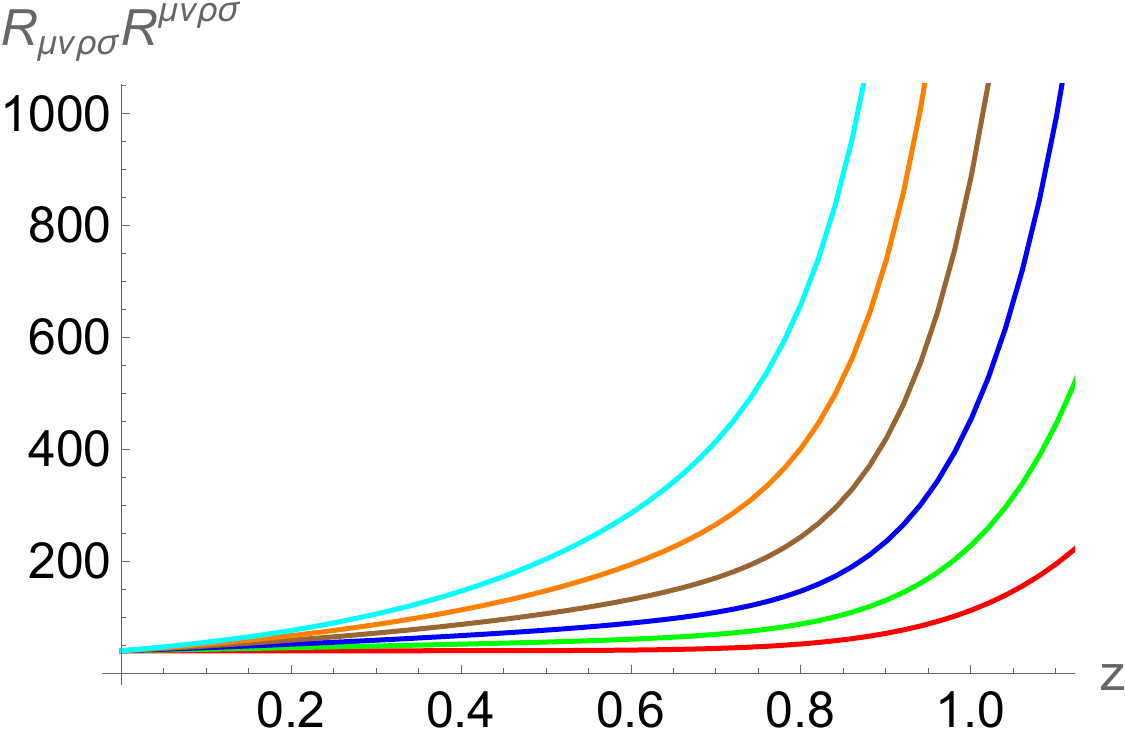}

\vspace{0.3cm}

\includegraphics[width=0.45\textwidth]{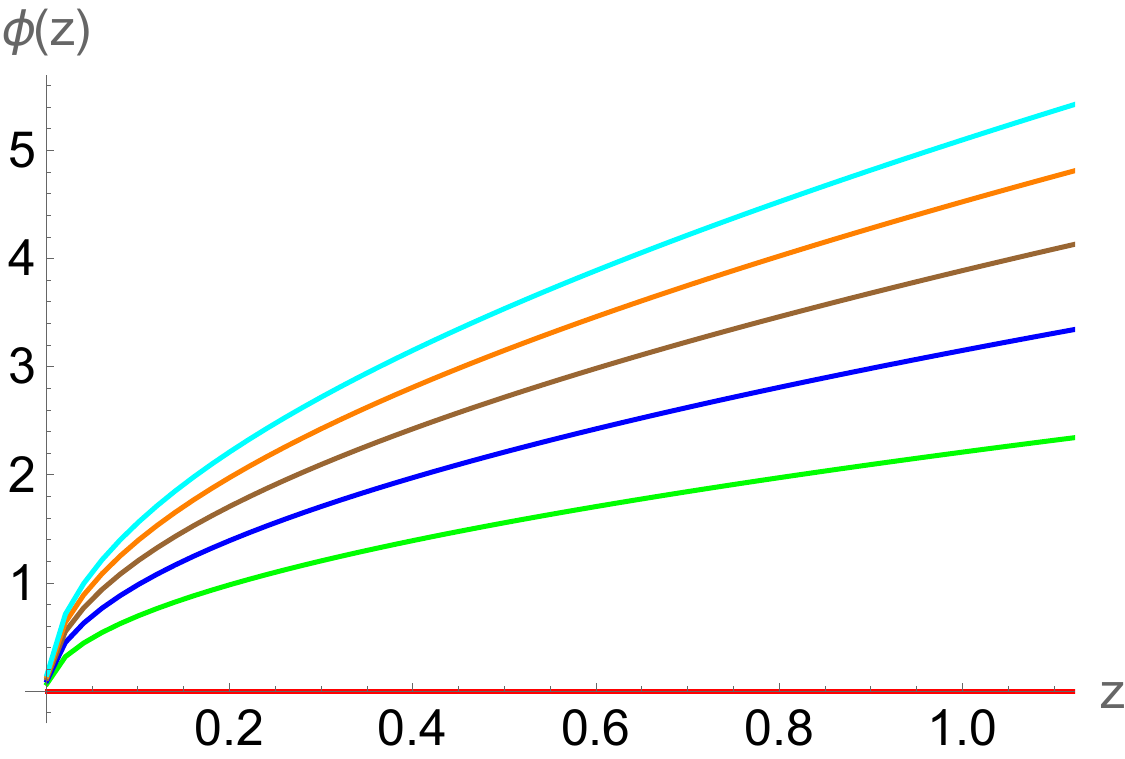}
\includegraphics[width=0.45\textwidth]{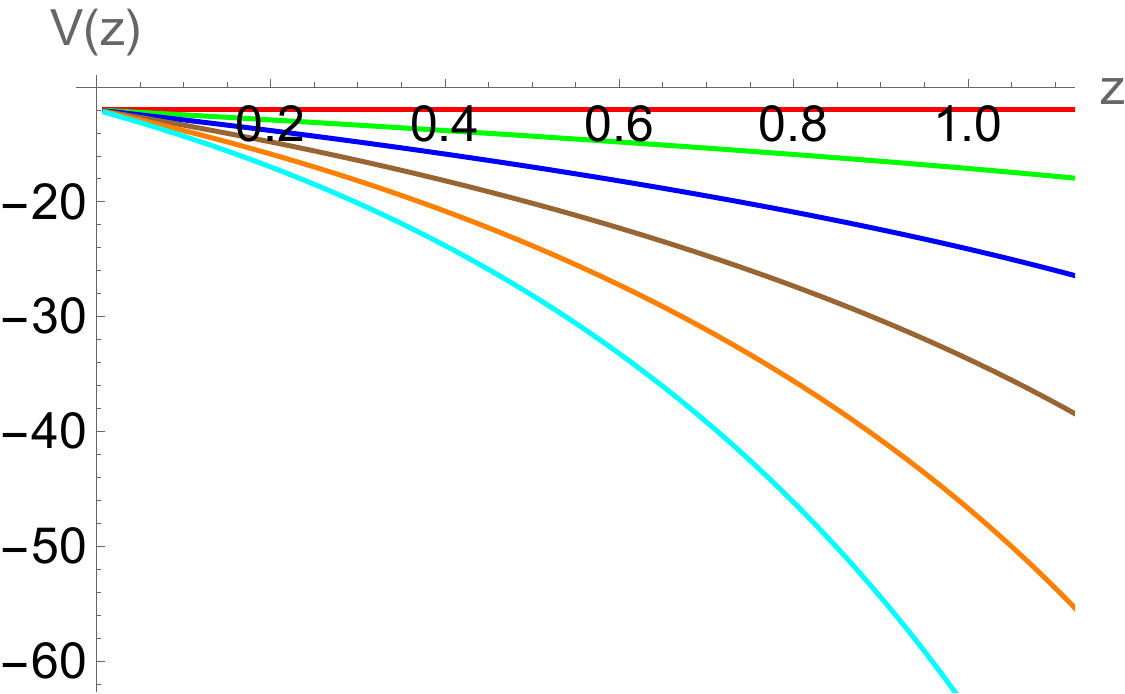}
\caption{\small The behavior of $g_b(z)$, $R_{\mu\nu\rho\sigma}R^{\mu\nu\rho\sigma}$, $\phi(z)$, and $V(z)$ for different values of the hair parameter $a$. Here $z_h=1$ is used. Red, green, blue, brown, orange, and cyan curves correspond to $a=0$, $0.1$, $0.2$, $0.3$, $0.4$, and $0.5$, respectively.}
\label{geometryBHn1D5}
\end{figure}

In Fig.~\ref{geometryBHn1D5}, we illustrate the behavior of $g_b(z)$, Kretschmann
scalar $R_{\mu\nu\rho\lambda}R^{\mu\nu\rho\lambda}$, $\phi(z)$, and $V(z)$ for
different values of hairy parameters $a$. Here,
we have shown results for a particular value of horizon radius $z_h=1$; however,
analogous results hold for other values of $z_h$ as well. For all values of $a$, the blackening function $g_b(z)$ changes 
sign at $z=z_h$, indicating the presence of an event horizon. Moreover, the 
Kretschmann scalar remains finite everywhere outside the horizon, confirming the absence of curvature singularities in the bulk spacetime. This feature of well-behaved geometry remains true for the hairy AdS soliton as well. The metric function of the AdS soliton $g_s(z)$ can be obtained from the blackening function $g_b(z)$ by replacing $z_h$ with $z_0$. Since the Kretschmann scalar remains finite everywhere outside $z_0$, the geometry of the hairy soliton also remains well-behaved everywhere. With the scalar hair, the magnitude of the Kretschmann scalar increases, indicating that the geometry becomes more curved as the scalar hair is switched on.

Furthermore, we have examined the behavior of the scalar field and found it to be real and regular everywhere outside the horizon. As follows from Eq.~(\ref{gbsoln1D5}), the scalar field vanishes only at the asymptotic AdS boundary. Its finiteness both at and outside the horizon indicates the existence of a well-behaved planar hairy black hole solution. Similarly, the scalar potential remains regular throughout the spacetime and asymptotically approaches $V=-12/\ell^{2}$ at the boundary for all values of the parameter $a$. For $a=0$, the potential is constant, while for finite $a$ it monotonically decreases with increasing $z$. Importantly, for all values of $a$, the potential is bounded from above by its ultraviolet boundary value.

\begin{figure}[h!]
\begin{minipage}[b]{0.5\linewidth}
\centering
\includegraphics[width=2.8in,height=2.3in]{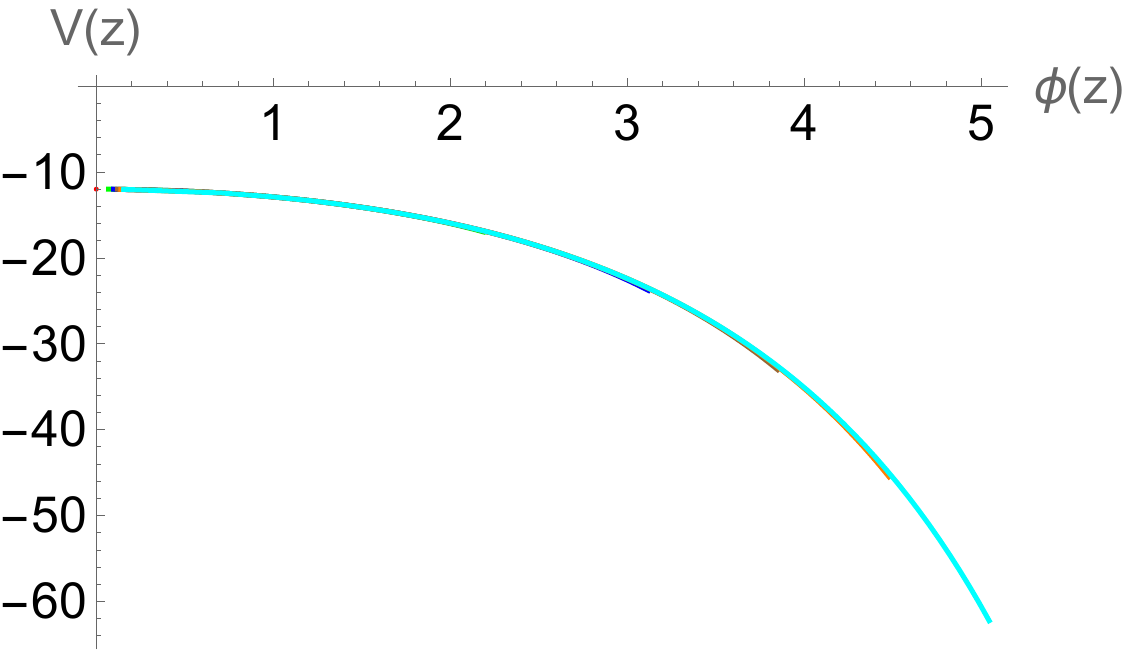}
\caption{ \small Scalar potential as a function of scalar field for various values of $a$. Here $G_5=1$, $V_2=1$, and $z_h=1$ are used. Red, green, blue, brown, orange, and cyan curves correspond to $a=0$, $0.1$, $0.2$, $0.3$, $0.4$, and $0.5$, respectively.}
\label{phivsVvsazh1n1D5}
\end{minipage}
\hspace{0.4cm}
\begin{minipage}[b]{0.5\linewidth}
\centering
\includegraphics[width=2.8in,height=2.3in]{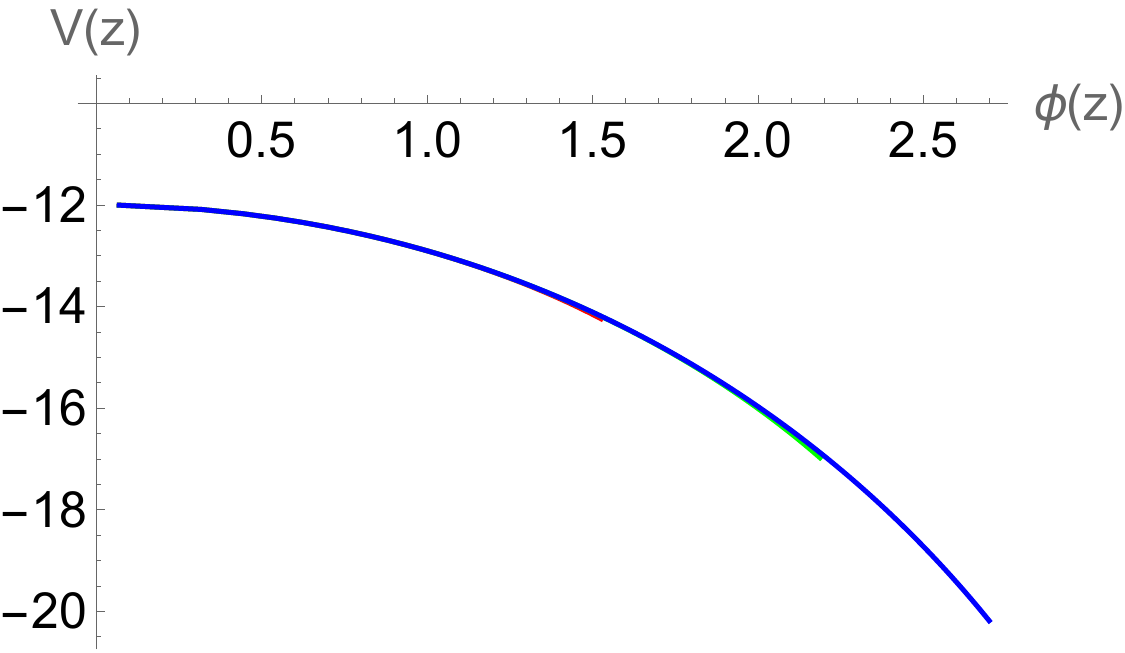}
\caption{\small Scalar potential as a function of scalar field for various values of $z_h$. Here $G_5=1$, $V_2=1$, and $a=0.1$ are used. Red, green, and blue curves correspond to $z_h=0.5$, $1.0$, and $1.5$, respectively.}
\label{phivsVvszh1aPt1n1D5}
\end{minipage}
\end{figure}

It is also instructive to see the behavior of the potential with
respect to the scalar field. This is shown in Figs.~\ref{phivsVvsazh1n1D5} and \ref{phivsVvszh1aPt1n1D5}. We observe that the potential profiles corresponding to different values of $a$ and $z_h$ almost completely overlap, demonstrating that the potential exhibits only a very mild dependence on these parameters. In particular, the profiles are nearly indistinguishable over most of the bulk geometry and become only slightly separable in the immediate vicinity of the horizon. This strongly suggests that, to a good approximation, the potential is effectively a function of the scalar field $\phi$ alone, as required for a consistent Einstein-scalar theory. Consequently, the reconstructed backgrounds should be regarded as good approximate solutions of the underlying Einstein-scalar system. 

\begin{figure}[h!]
\begin{minipage}[b]{0.5\linewidth}
\centering
\includegraphics[width=2.8in,height=2.3in]{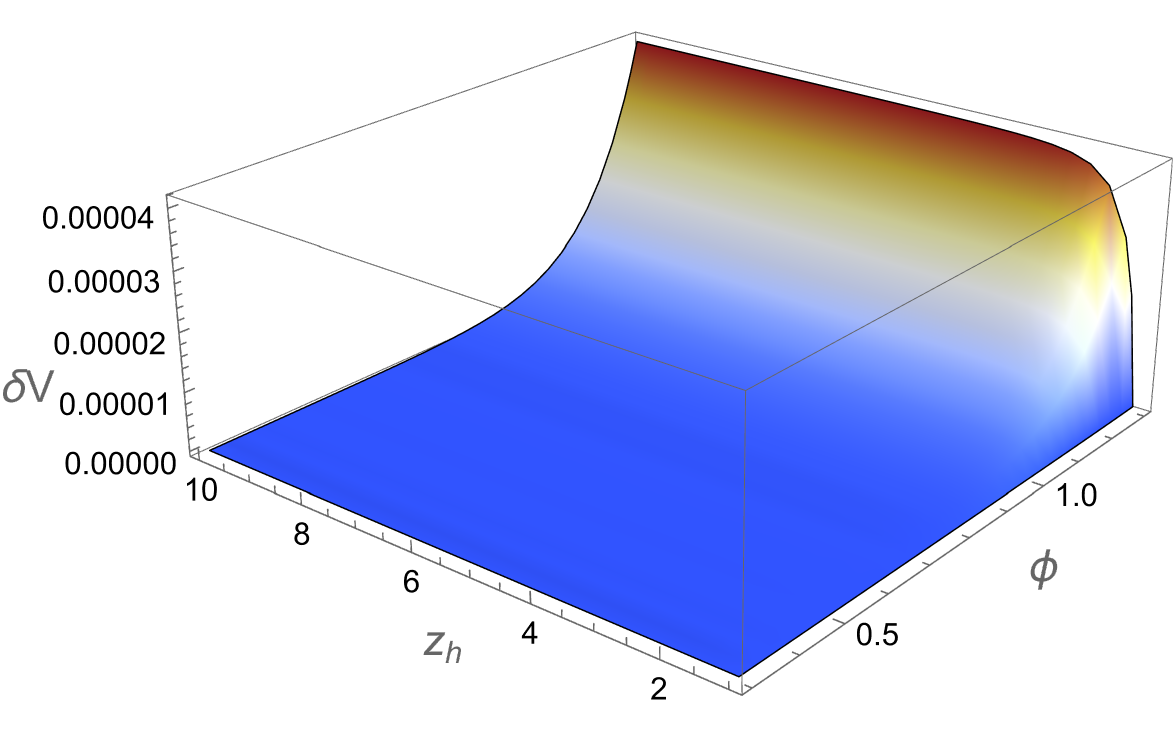}
\caption{ \small The dependence of $\delta V$ on $z_h$ and $\phi$. Here $a=0.1$ is used.}
\label{zhvsphivsdeltaVaPt1D5n1}
\end{minipage}
\hspace{0.4cm}
\begin{minipage}[b]{0.5\linewidth}
\centering
\includegraphics[width=2.8in,height=2.3in]{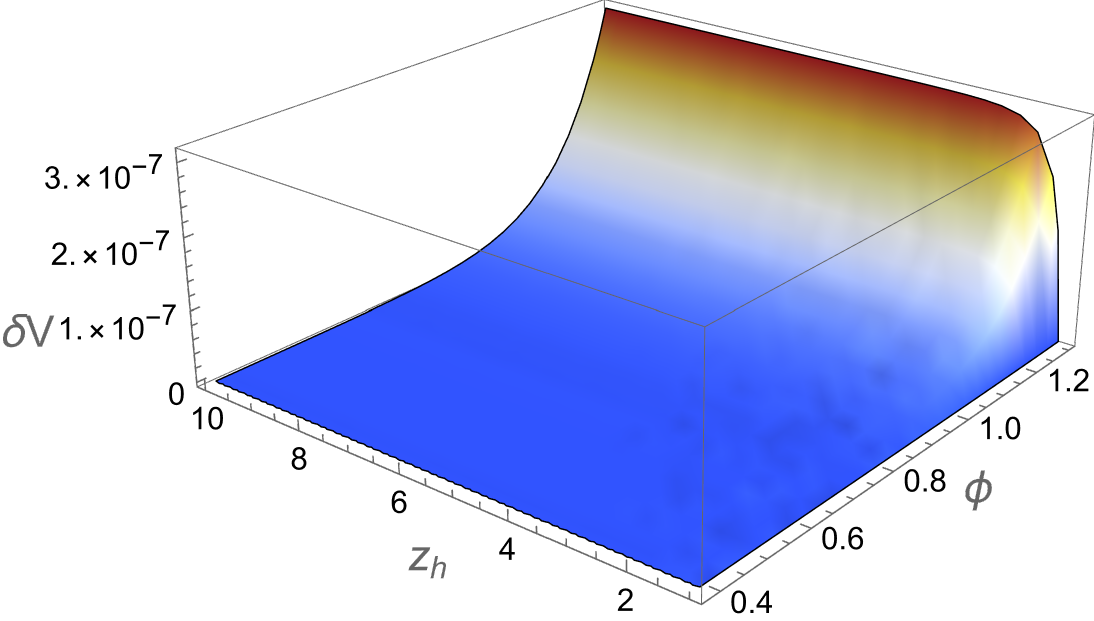}
\caption{\small The dependence of $\delta V$ on $z_h$ and $\phi$. Here $a=0.3$ is used.}
\label{zhvsphivsdeltaVaPt3D5n1}
\end{minipage}
\end{figure}

To further assess the approximate nature of the solutions, it is useful to examine how the reconstructed potential varies with $z_h$ at fixed $\phi$. This can be quantified via
\begin{equation}
\delta V(\phi,z_h)=
\frac{V\bigl(z(\phi),z_h\bigr)
- V\bigl(z(\phi), z_{\rm ref}\bigr)}
{\left|V\bigl(z(\phi);z_{\rm ref}\bigr)\right|}\,,
\end{equation}
where $z_{ref}$ is an arbitrary reference value, chosen here to be $z_{ref}=1$ for simplicity. The dependence of $\delta V$ on $z_h$ and $\phi$ is shown in Figs.~\ref{zhvsphivsdeltaVaPt1D5n1} and \ref{zhvsphivsdeltaVaPt3D5n1}. We observe that, for fixed $\phi$, the relative variation of the potential remains small over the parameter range considered, indicating that its dependence on $z_h$ is mild. This suggests that the hairy black hole solutions can be regarded as good approximate solutions of an Einstein-scalar theory with a fixed potential, with the residual $z_h$-dependence treated as a small correction. The resulting solutions may therefore provide a useful approximate framework for investigating the phase structure of the system. We emphasize, however, that this mild variation does not establish that the entire family of solutions is supported by exactly the same scalar potential; the interpretation of the solutions and their thermodynamic properties must be understood within this approximation.

\subsubsection{Thermodynamics of the hairy black hole for $A(z)=-a z$}
The global thermodynamic stability of the hairy black hole phase against the hairy soliton will be discussed shortly when we 
analyze its free energy. To analyze the thermodynamic stability of the hairy black hole solution, it is essential to examine its local stability. Local stability is determined by the response of the equilibrium system to small fluctuations in thermodynamic variables and is characterized by the positivity of the specific heat $\mathcal{C}=T_b\partial S_b/\partial T_b$. The temperature and entropy of the hairy black hole are given by
\begin{eqnarray}
T_b & = & \frac{27 a^4 z_h^3 e^{3 a z_h}}{4 \pi  \left(e^{3 a z_h} \left(9 a^3 z_h^3-9
   a^2 z_h^2+6 a z_h-2\right)+2\right)},  \nonumber \\
S_b  & = & \frac{L_b V_2 e^{-3 a z_h}}{4 G_5 z_h^3}\,.
\label{Tempentn1D5}
\end{eqnarray}
The condition $\mathcal{C}>0$ ensures the local stability of the thermodynamic system. We find that the slope of the $S_b-T_b$ curve is always positive, which implies that the specific heat remains positive. Accordingly, the constructed hairy black hole solution is thermodynamically stable.

We now present a detailed analysis of the gravitational action and the quasilocal stress tensor for the hairy black hole solution. This analysis is carried out using the holographic renormalization procedure~\cite{Balasubramanian:1999re}, which provides a systematic framework for computing the thermodynamic properties of the system. Within this approach, thermodynamic quantities are obtained from the regularized on-shell action by supplementing the bulk action with appropriate boundary counterterms. To implement this procedure it is desirable to have asymptotic expansions of various fields:
\begin{eqnarray}
& & g(z) = 1-\frac{27 a^4 z^4}{4 \left(e^{3 a z_h} (3 a z_h (3 a z_h (a
   z_h-1)+2)-2)+2\right)}+O\left(z^5\right)\,, \nonumber \\
& & \phi(z) = 4 \sqrt{3} \sqrt{a} \sqrt{z}+\frac{a^{3/2}
   z^{3/2}}{\sqrt{3}}+O\left(z^{5/2}\right) \,, \nonumber \\
& & V(z) = -12-42 a z-69 a^2 z^2+O\left(z^3\right) = 2\Lambda + \frac{m^{2}\phi^{2}}{2} + \cdots
\label{boundaryexpn1D5}
\end{eqnarray}
with $m^2=-7/4$. Since $m^2>-4$, it satisfies the BF bound. It further satisfies the standard mass-operator relation $m^2=\Delta (\Delta-4)$ of the scalar field. We see that the asymptotic expansion of the scalar field is of the standard AdS form, i.e., $\phi(z)=\phi_s z^{\Delta_-} + \cdots$,  with the leading coefficient fixed by the deformation parameter $a$. Below, we impose standard Dirichlet boundary conditions, treating the leading mode as the source for the dual operator. 

For the Einstein-scalar theory described by Eq.~(\ref{actionES}), the renormalized action is constructed by subtracting the divergent boundary contributions from the bulk action. In particular, we have
\begin{eqnarray}
S^{b}_{ren} = S_{ES}^{b} + S_{GH}^{b} + S_{BK}^{b} + S_{ct}^{b} \,,
\label{actionregGibbs}
\end{eqnarray}
where
\begin{eqnarray}
& & S_{ES}^{b}  =  \frac{1}{16 \pi G_5} \int_{\mathcal{M}} \mathrm{d^5}x \sqrt{-g}\left( \frac{2 V(z)}{3}  \right)\,, ~~~
S_{GH}^{b}  =  \frac{1}{8 \pi G_5} \int_{\partial \mathcal{M}} \mathrm{d^4}x \ \sqrt{-\gamma} \ \Theta\,,  \nonumber \\
& & S_{BK}^{b} =  -\frac{1}{16 \pi G_5} \int_{\partial \mathcal{M}} \mathrm{d^4}x \ \sqrt{-\gamma} \left(6 - R^{(4)}\right)\,, ~~~
S_{ct}^{b} =
 \frac{2}{16 \pi G_5} \int_{\partial \mathcal{M}} \mathrm{d^4}x \ \sqrt{-\gamma} \ \sum_{i=1}^{i=4} b_i \phi^{2i} \,.
\label{countertermn1D5}
\end{eqnarray}
Here, the first term ($S_{ES}^{b}$) represents the on-shell bulk action, the second term ($S_{GH}^{b}$) corresponds to the standard Gibbons-Hawking (GH) surface term, the third term ($S_{BK}^{b}$) denotes the Balasubramanian-Kraus (BK) counterterms, and the fourth term ($S_{ct}^{b}$) accounts for the scalar field counterterms. The quantity $\gamma$ denotes the induced metric on the boundary $\partial\mathcal{M}$, while $R^{(4)}$ is the Ricci scalar constructed from the boundary metric $\gamma$, which vanishes identically for the planar geometries under consideration. The symbol $\Theta$ denotes the trace of the extrinsic curvature $\Theta_{\mu\nu}$.  In the presence of scalar hair, the on-shell action acquires additional divergences, and the coefficients $b_i$ are determined by requiring the complete cancellation of divergences originating from the scalar sector of the action.\footnote{We only need scalar counterterms up to $\phi^6$ term to cancel all divergences. The $\phi^8$ counterterm contributes only a finite term. The coefficient $b_4$ therefore cannot be fixed by divergence cancellation alone. We fix $b_4$ by requiring that the conserved mass obtained from the renormalized boundary stress tensor agrees with the expression determined by the coefficient of the $z^4$ term in the near-boundary expansion of $g_b(z)$. We adopt the same prescription for the corresponding hairy soliton solution. Since the $b_4$ term contributes equally to the black hole and soliton actions for matched boundary geometries and scalar sources, it cancels in the difference of their renormalized actions and therefore does not affect the phase-transition criterion. } Explicitly, we have the following:
\begin{eqnarray}
S_{ES}^{b} & = & \frac{\beta_b L_b V_2}{16 \pi G_5} \left(-\frac{2 \left(-e^{-3 a z} (a z+1) \left(27 a^4-2 C_b\right)-3 a^2 z^2 C_b + 4
   a z C_b-2 C_b\right)}{27 a^4 z^4} \right)\bigg\rvert_{z=\epsilon}^{z=z_h} \,, \\
S_{GH}^{b} & = & \frac{\beta_b L_b V_2}{16 \pi G_5}  \left(-9 a^4+\frac{12 a^2}{\epsilon^2}-\frac{16 a}{\epsilon^3}-C_b+\frac{8}{\epsilon^4}\right)\,, \\
S_{BK}^{b} & = &  \frac{\beta_b L_b V_2}{16 \pi G_5} \left(-64 a^4+\frac{64 a^3}{\epsilon}-\frac{48 a^2}{\epsilon^2}+\frac{24 a}{\epsilon^3}+\frac{3
   C_b}{4}-\frac{6}{\epsilon^4}  \right)\,,\\
S_{ct}^{b} & = &  \frac{\beta_b L_b V_2}{16 \pi G_5} \left( \frac{283 a^4}{4}-\frac{64 a^3}{\epsilon}+\frac{39 a^2}{\epsilon^2}-\frac{12 a}{\epsilon^3} \right) \,,
\label{countertermexpn1D5}
\end{eqnarray}
where $V_2$ is the volume of the two-dimensional plane and $\epsilon$ is a UV cutoff. 
From the above expressions, we get the renormalized free energy $\mathcal{F}_b=-S^{b}_{ren}/\beta_b$ of the hairy black hole as
\begin{eqnarray}
\mathcal{F}_b & = & - \frac{L_b V_2 C_b}{64 \pi G_5} \,, \nonumber \\
 & = & -\frac{27 a^4 L_b V_2}{64 \pi  G_5 \left(e^{3 a z_h} \left(9 a^3 z_h^3-9 a^2 z_h^2+6 a
   z_h-2\right)+2\right)}\,,
\label{countertermexpn1D5}
\end{eqnarray}
where we have explicitly used the expression of the integration constant $C_b$ from Eq.~(\ref{inteconstD5n1}). 

From the renormalized action, and using the Arnowitt-Deser-Misner (ADM) decomposition, we can further compute the corresponding stress-energy tensor
\begin{eqnarray}
 T^{\mu\nu} =\frac{1}{8 \pi G_5} \left[ \Theta \gamma^{\mu\nu} - \Theta^{\mu\nu} + \frac{2}{\sqrt{-\gamma}}\frac{\delta\mathcal{L}_{ct}}{\delta \gamma_{\mu\nu}} \right] \,,
\label{stresstensordefplanarn1D5}
\end{eqnarray}
where $\mathcal{L}_{ct}$ is the Lagrangian of the counterterms only. Explicitly, we have
\begin{eqnarray}
T_{\mu\nu} =\frac{1}{8 \pi G_5} \left[ \Theta \gamma_{\mu\nu} - \Theta_{\mu\nu}- 3 \gamma_{\mu\nu} + \gamma_{\mu\nu} \sum_{i=1}^{i=4} b_i \phi^{2i} \right]  \,.
\label{stresstensorn1d5}
\end{eqnarray}
The mass of the black hole is then obtained from the $tt$ component of the stress-energy tensor $T_{\mu\nu}$. More generally, if $K^\mu$ denotes a Killing vector generating an isometry of the boundary spacetime, the corresponding conserved charge is given by
\begin{eqnarray}
 M_b & = & \int_\Sigma \ d^3 x \sqrt{\sigma} u^{\mu} T_{\mu\nu} K^{\nu}  \,,
\label{massn1D5}
\end{eqnarray}
where $\Sigma$ is a spacelike surface in $\partial\mathcal{M}$, with induced metric $\sigma$, and $u_{\mu}$ is the timelike unit normal to $\Sigma$. For the hairy black hole solution, we get
\begin{eqnarray}
 M_b & = &  \frac{3 L_b V_2 C_b}{64 \pi G_5} \,, \nonumber \\
 & = & \frac{81 a^4 L_b V_2}{64 \pi  G_5 \left(e^{3 a z_h} \left(3 a z_h \left(3 a z_h
   \left(a z_h-1\right)+2\right)-2\right)+2\right)} \,.
\label{massexpn1D5}
\end{eqnarray}
Notice that the mass is proportional to the integration constant $C_b$, suggestive of the primary nature of the hair. At this point, some clarification is required. The parameter $a$ controls the scalar profile $\phi(z)$ independently of the black hole integration constant $C_b$, which determines the mass of the black hole. In particular, at fixed $a$, varying the horizon position $z_h$ changes $C_b$ and hence the black hole mass, without changing the scalar profile. Conversely, varying $a$ modifies the scalar profile independently of $C_b$. This indicates that the scalar configuration is not determined by the black hole mass alone. Moreover, the scalar hair can be continuously switched off in the limit $a\to 0$, recovering the corresponding hairless solution. These features motivate us to regard $a$ as a continuously tunable scalar hair parameter and to use the terminology primary scalar hair within our potential-reconstruction framework. A related use of this terminology in another variant of the potential-reconstruction approach can be found in \cite{Gonzalez:2013aca}.

We emphasize, however, that the term primary hair is used here to characterize the role of $a$ as a parameter controlling the scalar profile, rather than to assert that the strict criterion for primary scalar hair has been established. In particular, such a classification would require demonstrating the existence of an independent scalar hair parameter for solutions of a common, fixed scalar potential under specified scalar boundary conditions. In our construction, the reconstructed potential exhibits a mild residual dependence on the horizon parameter, while varying $a$ also changes the scalar boundary data. Accordingly, our solutions do not correspond to the primary hair in a strict mathematical sense.

Furthermore, this expression is consistent with the coefficient of the $z^{4}$ term in the near-boundary expansion of the blackening function $g_b(z)$, as required by our renormalization prescription. In particular,
\begin{eqnarray}
 M_b & = & - \frac{3 L_b V_2}{16 \pi G_5} \times [z^4~ \text{coefficient of}~g_b(z)] \,.
\end{eqnarray}
Importantly, the expressions obtained for the free energy and mass satisfy the expected thermodynamic relation
$\mathcal{F}_b = M_b - T_b S_b$, providing a nontrivial consistency check of our thermodynamic analysis for the hairy black hole solutions. Furthermore, the pressure $P_b$, extracted from the spatial components of the stress-energy tensor $T_{\mu\nu}$, obeys the standard thermodynamic relation $\mathcal{F}_b = -P_b$. It is noteworthy that, despite the complexity of the hairy black hole solutions, analytic expressions for the relevant thermodynamic observables can be derived. Moreover, all these thermodynamic quantities smoothly reduce to their standard five-dimensional planar black hole counterparts in the limit $a\rightarrow 0$, further underscoring the internal consistency of the solution obtained.

Unfortunately, the free energy does not satisfy the differential first law like relation $dF=-SdT$. Usually, in the presence of a scalar field, the standard first law may acquire additional contributions, as discussed in several works; see, for instance, \cite{Liu:2013gja, Lu:2014maa, Li:2020spf}. In the present model, however, the discrepancy is most likely related to the mild dependence of the reconstructed scalar potential on the horizon parameter, as discussed earlier. Since the potential is not strictly independent of the parameters characterizing the solutions, the origin of this discrepancy remains unclear, and a complete understanding requires further investigation. This issue of non-satisfaction of the standard differential first law-like relation in the potential reconstruction method-inspired models has been pointed out earlier as well, see for instance \cite{Priyadarshinee:2021rch}. Nevertheless, the free energy and mass obtained through holographic renormalization satisfy the thermodynamic identity $\mathcal{F}_b=M_b-T_bS_b$, while the pressure extracted from the boundary stress-energy tensor obeys $\mathcal{F}_b=-P_b$. These relations provide important consistency checks of the thermodynamic quantities derived in this work.

\begin{figure}[h!]
\begin{minipage}[b]{0.5\linewidth}
\centering
\includegraphics[width=2.8in,height=2.3in]{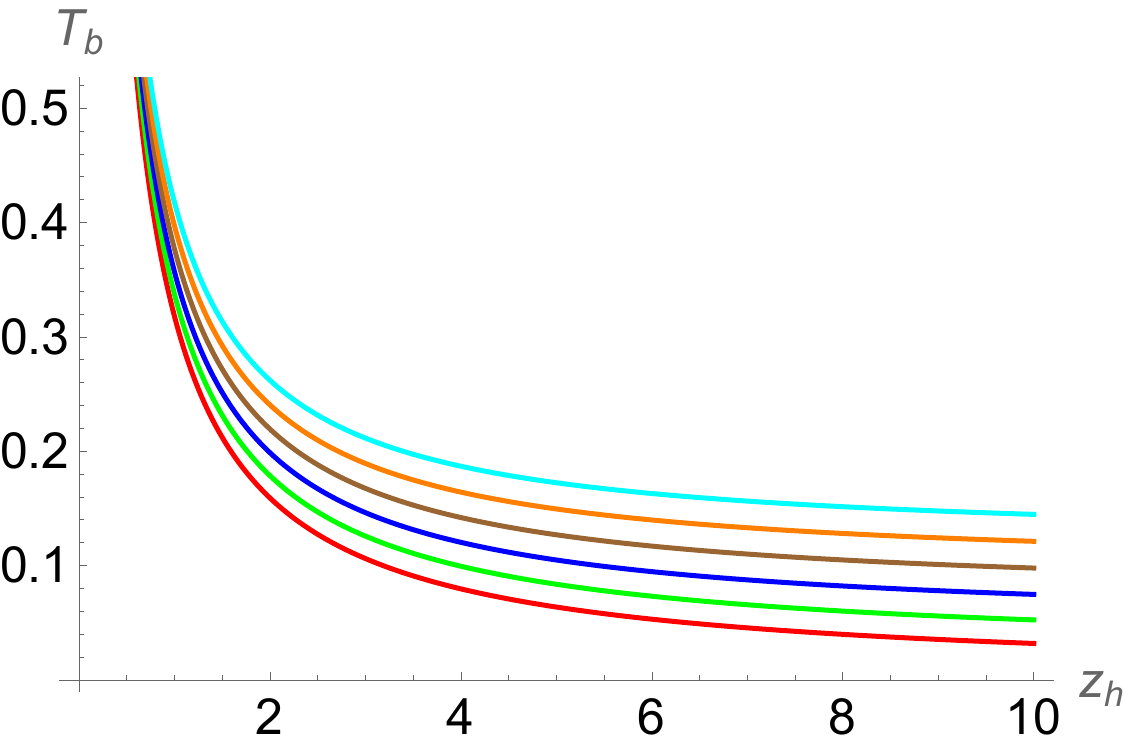}
\caption{ \small Hawking temperature $T_b$ as a function of horizon radius $z_h$ for various values of $a$. Here $G_5=1$ and $V_2=1$ are used. Red, green, blue, brown, orange, and cyan curves correspond to $a=0$, $0.1$, $0.2$, $0.3$, $0.4$, and $0.5$, respectively.}
\label{zhvsTempBHn1D5}
\end{minipage}
\hspace{0.4cm}
\begin{minipage}[b]{0.5\linewidth}
\centering
\includegraphics[width=2.8in,height=2.3in]{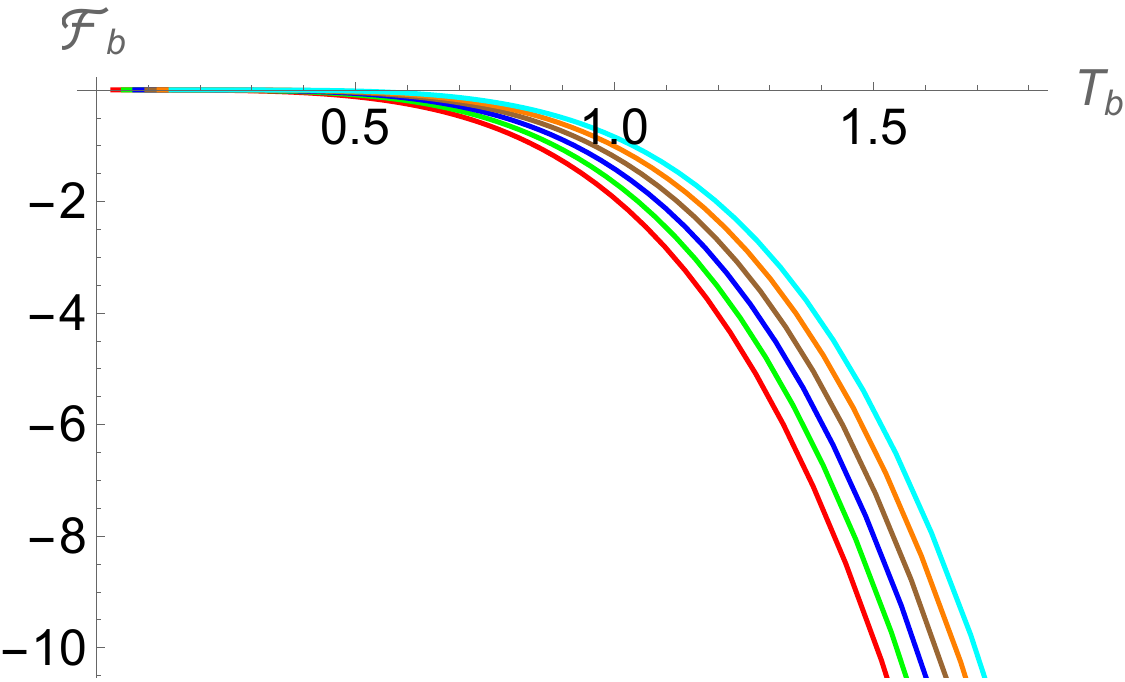}
\caption{\small Free energy $\mathcal{F}_b$ of the hairy black hole as a function of temperature $T_b$ for various values of $a$. Here $G_5=1$ and $V_2=1$ are used. Red, green, blue, brown, orange, and cyan curves correspond to $a=0$, $0.1$, $0.2$, $0.3$, $0.4$, and $0.5$, respectively.}
\label{TempvsFreeBHn1D5}
\end{minipage}
\end{figure}

The thermodynamics of the constructed hairy black hole is shown in Figs.~\ref{zhvsTempBHn1D5} and \ref{TempvsFreeBHn1D5}. We observe that there is a single black hole branch, and its temperature decreases monotonically with the inverse horizon radius $z_h$. The black hole entropy also exhibits a decreasing profile with $z_h$. This indicates that the slope of the $S_b-T_b$ curve is always positive. Accordingly, since the specific heat remains positive, it implies that the constructed hairy black hole solutions are locally stable against small thermal fluctuations. For global stability, we further analyze its free energy structure. The results are shown in Fig.~\ref{TempvsFreeBHn1D5}, where the thermal profile of the free energy of the hairy black hole is shown. Here, the free energy is normalized with respect to the thermal-AdS.\footnote{We normalize the free energy of the black hole with respect to the thermal-AdS by obtaining their free energy (on-shell action) difference and matching their asymptotic boundaries via
\begin{equation*}
\beta_{th}\sqrt{g_{\tau\tau}^{th}}=\beta_{b}\sqrt{g_{\tau\tau}^{b}}\,,~~~L_{th}\sqrt{g_{x_1x_1}^{th}}=L_{b}\sqrt{g_{x_1x_1}^{b}}
\end{equation*}
with $\beta_{th}=\beta_b$ and $L_{th}=L_b$. Here, $\beta_{th}$ and $L_b$ are the periods of Euclidean time and the compact spacelike cycle of thermal-AdS. Note that since the asymptotic structure and divergent behavior are the same in both black hole and thermal-AdS solutions, both the background subtraction method and holographic renormalization methods give the same result for the free energy difference.} We observe that the free energy is always negative, indicating that the hairy black hole phase always has a lower free energy than the thermal-AdS, and is thermodynamically favored at all temperatures. This is true for all values of the scalar hair strength $a$. 

\subsubsection{Thermodynamics of the hairy AdS soliton for $A(z)=-a z$} 
We now proceed to obtain all the necessary thermodynamical quantities for the hairy AdS soliton case. Similarly to the black hole case, the regularized action and the thermodynamic quantities can be obtained via the holographic renormalization procedure. The near boundary expansion and scalar mass expression are identical to those given in Eq.~(\ref{boundaryexpn1D5}), with the replacement $z_h \rightarrow z_0$. The various terms in the regularized action are now given by
\begin{eqnarray}
S_{ES}^{s} & = & \frac{\beta_s L_s V_2}{16 \pi G_5} \left(-\frac{2 \left(-e^{-3 a z} (a z+1) \left(27 a^4-2 C_s\right)-3 a^2 z^2 C_s+4
   a z C_s-2 C_s\right)}{27 a^4 z^4} \right)\bigg\rvert_{z=\epsilon}^{z=z_0} \,, \\
S_{GH}^{s} & = & \frac{\beta_s L_s V_2}{16 \pi G_5}  \left(-9 a^4+\frac{12 a^2}{\epsilon^2}-\frac{16 a}{\epsilon^3}-C_s+\frac{8}{\epsilon^4}\right)\,, \\
S_{BK}^{s} & = &  \frac{\beta_s L_s V_2}{16 \pi G_5} \left(-64 a^4+\frac{64 a^3}{\epsilon}-\frac{48 a^2}{\epsilon^2}+\frac{24 a}{\epsilon^3}+\frac{3
   C_s}{4}-\frac{6}{\epsilon^4}  \right)\,,\\
S_{ct}^{s} & = &  \frac{\beta_s L_s V_2}{16 \pi G_5} \left( \frac{283 a^4}{4}-\frac{64 a^3}{\epsilon}+\frac{39 a^2}{\epsilon^2}-\frac{12 a}{\epsilon^3} \right) \,,
\label{countertermexpsoln1D5}
\end{eqnarray}
with $C_s$ given by
\begin{eqnarray}
C_s = \frac{1}{ \int_0^{z_0} \, d\xi ~ e^{3 a \xi}~ \xi^{3} } = \frac{27 a^4}{e^{3 a z_0} \left(3 a z_0 \left(3 a z_0 \left(a
   z_0-1\right)+2\right)-2\right)+2}  \,.
\label{csexpn1D5}
\end{eqnarray}
Note that since the asymptotic structure of the hairy soliton and hairy black hole is similar, the corresponding boundary counterterms take the same form. The renormalized action for the hairy soliton is therefore obtained by combining all of the above contributions, and, as expected, all divergences cancel out, giving
\begin{eqnarray}
S_{ren}^s &  = &  \frac{\beta_s L_s V_2 C_s}{64 \pi G_5} \,, \nonumber \\
&  = & \frac{27 a^4 \beta_s L_s V_2}{64 \pi  G_5 \left(9 a^3 z_0^3 e^{3 a z_0}-9 a^2 z_0^2 e^{3 a
   z_0}+6 a z_0 e^{3 a z_0}-2 e^{3 a z_0}+2\right)}\,.
\label{countertermexpsoln1D5}
\end{eqnarray}
This leads to the free energy of the hairy AdS soliton $\mathcal{F}_s=-S_{ren}^{s}/\beta_s = M_s$ as
\begin{eqnarray}
\mathcal{F}_s &  = & -\frac{27 a^4 L_s V_2}{64 \pi  G_5 \left(9 a^3 z_0^3 e^{3 a z_0}-9 a^2 z_0^2 e^{3 a
   z_0}+6 a z_0 e^{3 a z_0}-2 e^{3 a z_0}+2\right)}\,.
\label{fsmexpsoln1D5}
\end{eqnarray}
As expected, the mass or the free energy of the hairy AdS soliton is negative. In particular, substituting the value of $L_s$ from Eq.~(\ref{solperiod1}) and simplifying, we get
\begin{eqnarray}
\mathcal{F}_s &  = & -\frac{V_2 e^{-3 a z_0}}{16  G_5  z_0^3 }\,,
\end{eqnarray}
which is always negative for all values of the hairy parameter. 

As in the black hole case, one can compute the stress-energy tensor for the hairy soliton analogously. The energy of the hairy AdS soliton is then obtained from the $tt$ component of the stress-energy tensor and is computed as
\begin{eqnarray}
M_s &=& -\frac{C_s L_s V_2}{64 \pi G_5}\,.
\end{eqnarray}
This expression is in agreement with Eq.~(\ref{fsmexpsoln1D5}), providing a further consistency check of the analysis. 

It is also instructive to compute the energy of the hairy black hole with respect to the hairy soliton. The regulated energy is given by
\begin{eqnarray}
\Delta M &=& M_b - M_s = \frac{L_b V_2}{64 \pi G_5} \left(3 C_b + C_s   \right)\,.
\end{eqnarray}
Since both $C_b$ and $C_s$ are positive, the energy of the hairy soliton is always lower than that of the hairy black hole. This energy ordering is consistent with the expectation of the Horowitz-Myers conjecture for the black hole and soliton configurations considered \cite{Horowitz:1998ha}. We should emphasize that this comparison, however, does not establish the conjecture in its full generality, since the energy of the soliton has not been compared with that of all admissible configurations satisfying the same boundary conditions.

\subsubsection{Phase transition between hairy black hole and soliton for $A(z)=-a z$}
Let us now study the free energy and the phase transitions between the hairy AdS black hole and the hairy AdS soliton. As mentioned earlier,  to match the boundary asymptotics of both solutions at a radial cutoff $z=\epsilon$, we require temporal and spatial periodicities to be identical, i.e., $\beta_b=\beta_s$, and $L_b=L_s$. The free energy of the hairy black hole with respect to that of the hairy soliton is given by 
\begin{eqnarray}
\Delta \mathcal{F} &=& \frac{L_b V_2}{64 \pi G_5} \left(C_s - C_b \right)\,.
\end{eqnarray}
It is clear that, depending on the integration constants $C_b$ and $C_s$ that characterize the black hole and soliton solutions, respectively, the free energy difference would change sign, signaling a transition between these two solutions. In particular, if  $\Delta \mathcal{F}>0$, the hairy soliton dominates the phase space, while for $\Delta \mathcal{F}<0$, the hairy black hole dominates the phase space.  To make this transition more explicit, we need to study in an ensemble with a common temperature and $L_b$.

\begin{figure}[h!]
\begin{minipage}[b]{0.5\linewidth}
\centering
\includegraphics[width=2.8in,height=2.3in]{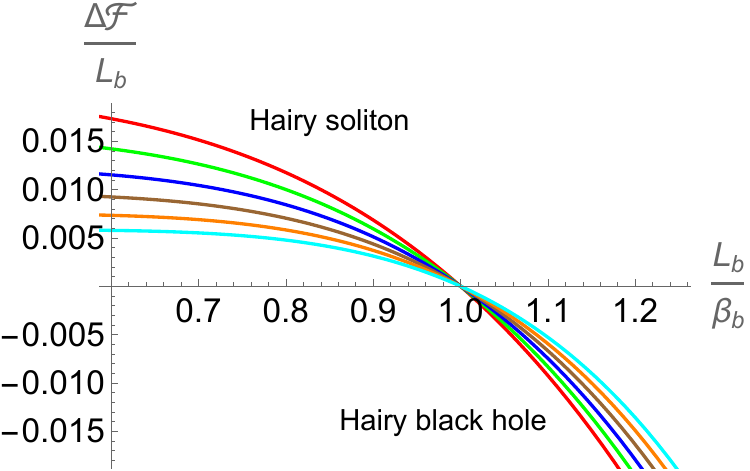}
\caption{ \small Free energy difference $\Delta\mathcal{F}$ as a function of periodicity ratio $L_b/\beta_b$ for various values of $a$. Here $G_5=1$, $V_2=1$, and $z_0=1$ are used. Red, green, blue, brown, orange, and cyan curves correspond to $a=0$, $0.1$, $0.2$, $0.3$, $0.4$, and $0.5$, respectively.}
\label{Lbbybetabvsfreediffvsan1D5}
\end{minipage}
\hspace{0.4cm}
\begin{minipage}[b]{0.5\linewidth}
\centering
\includegraphics[width=2.8in,height=2.3in]{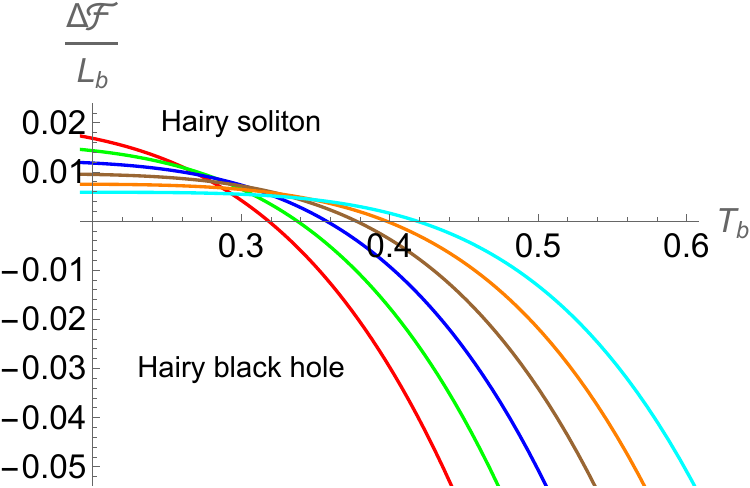}
\caption{\small Free energy difference $\Delta\mathcal{F}$ as a function of black hole temperature $T_b$ for various values of $a$. Here $G_5=1$, $V_2=1$, and $z_0=1$ are used. Red, green, blue, brown, orange, and cyan curves correspond to $a=0$, $0.1$, $0.2$, $0.3$, $0.4$, and $0.5$, respectively.}
\label{Tempvsfreediffvsan1D5}
\end{minipage}
\end{figure}

\begin{figure}[h!]
\centering
\includegraphics[width=3.8in,height=2.8in]{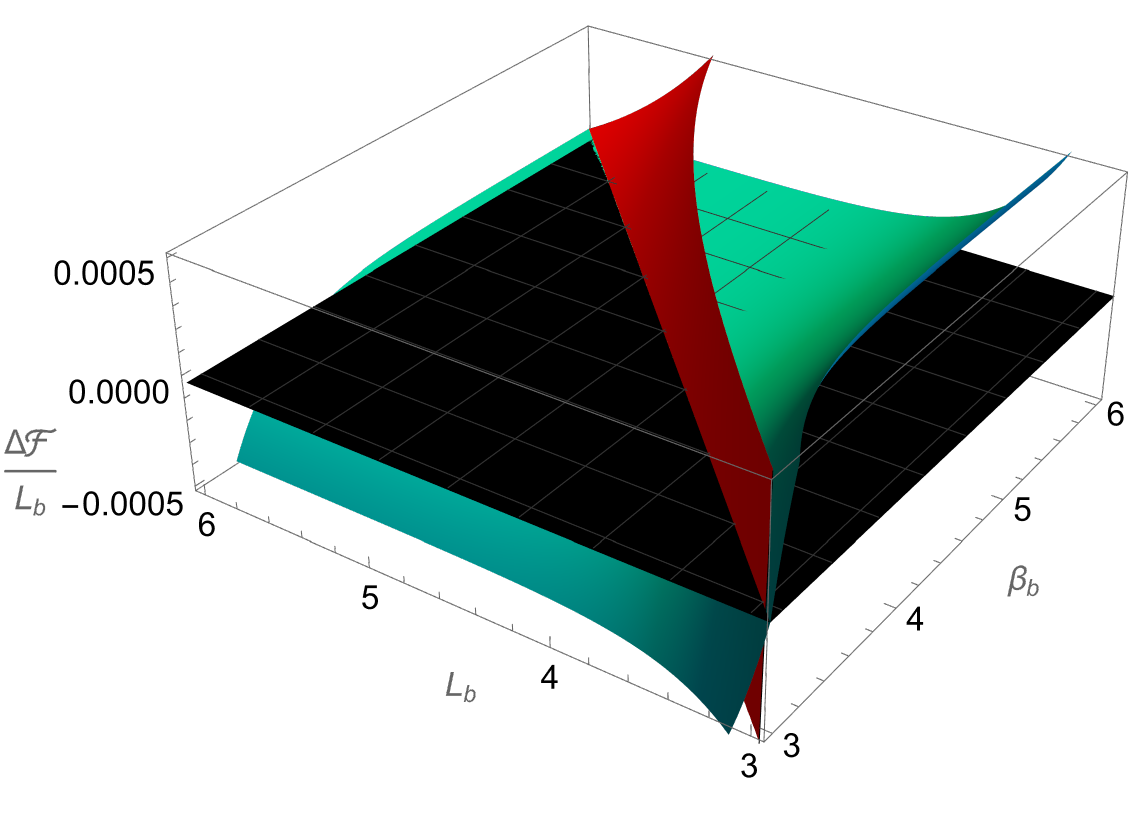}
\caption{ \small Free energy difference $\Delta\mathcal{F}$ as a function of $L_b$ and $\beta_b$ for $a=0.1$ (red surface) and $a=0.5$ (cyan surface). The black plane indicates $\Delta\mathcal{F}=0$ surface.  Here $G_5=1$ and $V_2=1$ are used.}
\label{Lbbybetabvsfreediffvs3Dn1D5}
\end{figure}

In Fig.~\ref{Lbbybetabvsfreediffvsan1D5}, the free energy difference as a function of periodicity ratio $L_b/\beta_b$ is shown for various values of $a$. Here we use $z_0=1$ for illustration; however, our results remain the same for other values of $z_0$. We observe that the transition between the hairy black hole and hairy soliton is controlled by the ratio $L_b/\beta_b$. In particular, the hairy black hole dominates the phase space when $L_b > \beta_b$, whereas the hairy soliton dominates the phase space when $L_b < \beta_b$. The point $L_b = \beta_b$ corresponds to a metastable state where both these hairy phases coexist. Interestingly, this point remains the same for all $a$ values, as can be explicitly observed from Fig.~\ref{Lbbybetabvsfreediffvsan1D5}.

In Fig.~\ref{Tempvsfreediffvsan1D5}, the free energy difference as a function of black hole temperature is shown. Naively, it suggests that the critical temperature $T_{crit}$ at which the free energies exchange dominance depends nontrivially on the hair parameter $a$. However, it is important to clarify that the dependence of the critical temperature on the hair parameter $a$ must be interpreted in conjunction with the boundary geometry. The phase transition occurs when $\beta_b=L_b$, yielding $T_{crit}=1/L_b$. The soliton regularity condition (\ref{solperiod1}) gives $L_s=L_s(a,z_0)$, which depends on $a$ even when the tip position $z_0$ is held fixed. Since matching the asymptotic geometries requires $L_b=L_s$, the increase in $T_{crit}=1/L_b$ observed when varying $a$ at fixed $z_0$ reflects the accompanying change in the boundary circumference, rather than an independent increase of the critical temperature at fixed $L_b$. Indeed, at fixed boundary circumference, the crossing condition gives $T_{\mathrm{crit}}=1/L_b$, independent of $a$. Thus, the results shown here demonstrate how varying the hair parameter changes the critical temperature when the soliton tip position is held fixed; they do not imply a hair-induced shift of the critical temperature at fixed boundary geometry.

To provide a more comprehensive picture, we have shown a three-dimensional profile of free energy difference as a function of $\beta_b$ and $L_b$ for two different $a$ values in Fig.~\ref{Lbbybetabvsfreediffvs3Dn1D5}. It is evident that the free energy surface intersects the $\Delta\mathcal{F}=0$ plane along a straight line defined by $\beta_b = L_b$. Consequently, in the region above this plane, the hairy soliton phase is thermodynamically preferred, whereas below it the hairy black hole phase becomes favored.

\subsection{Case: $n=2$}
\begin{figure}[ht]
\centering
\includegraphics[width=0.45\textwidth]{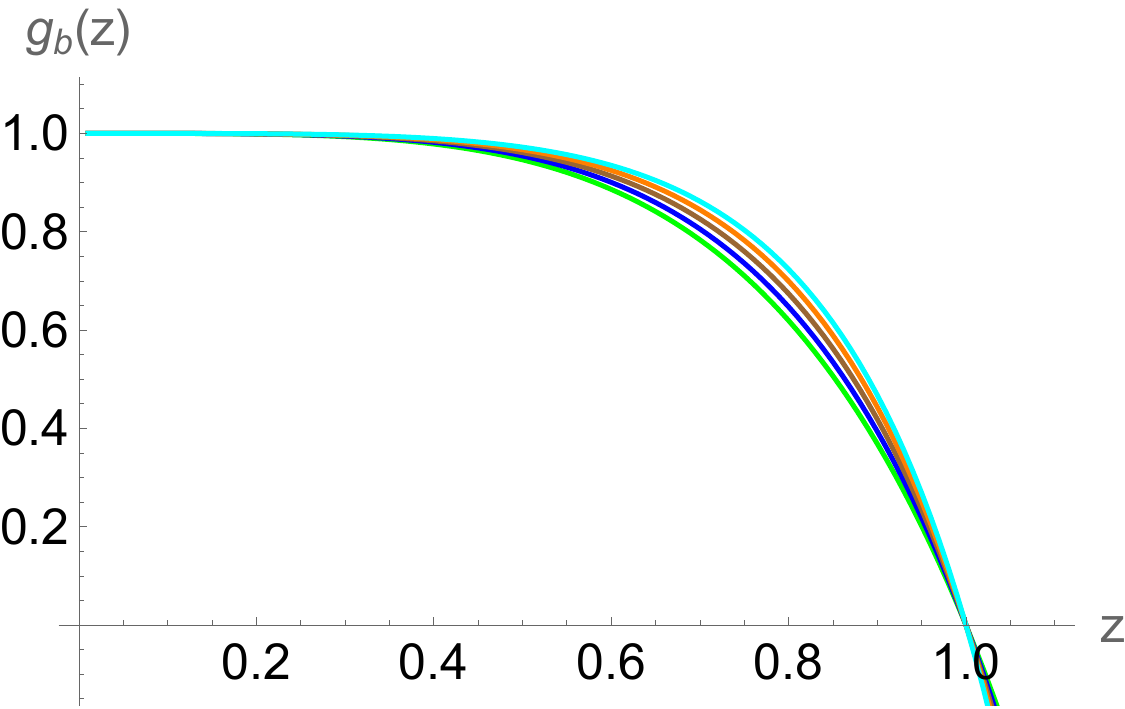}
\includegraphics[width=0.45\textwidth]{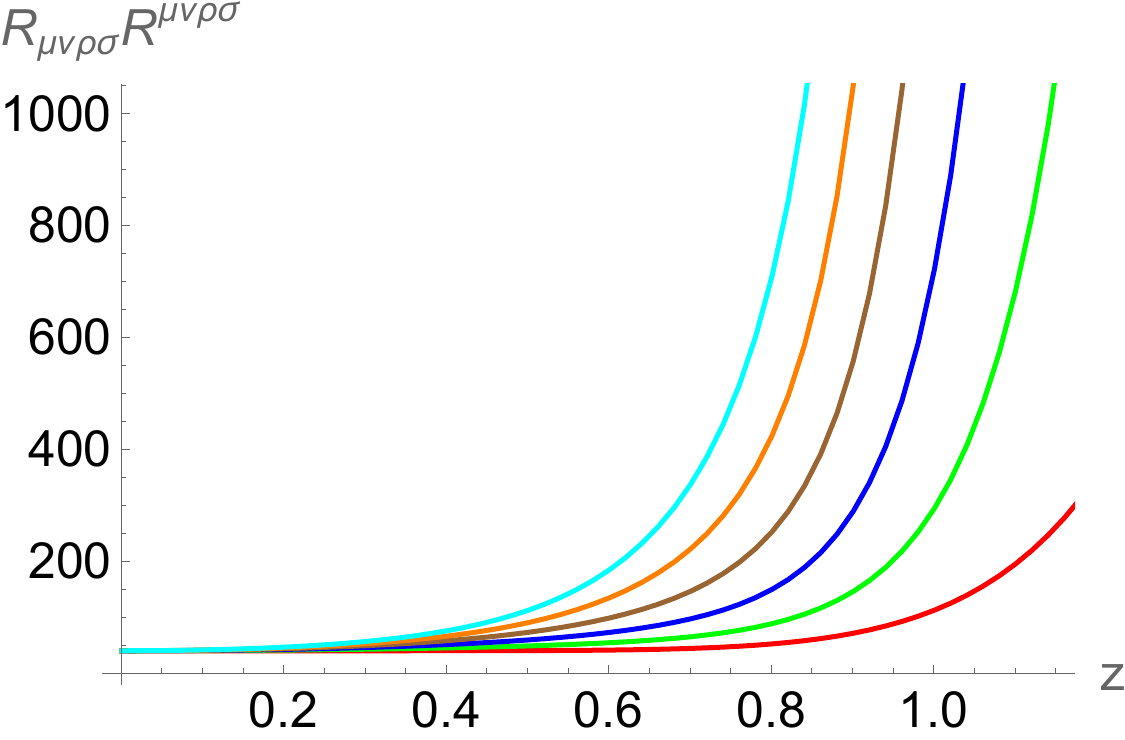}

\vspace{0.6cm}

\includegraphics[width=0.45\textwidth]{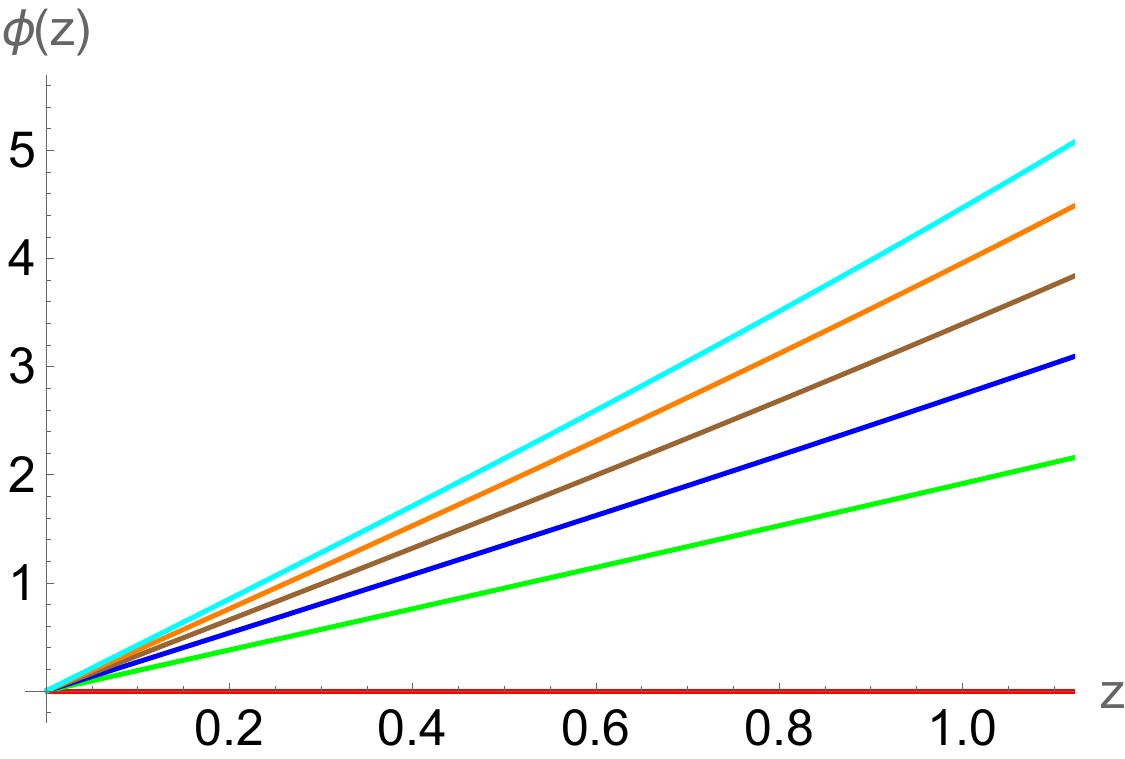}
\includegraphics[width=0.45\textwidth]{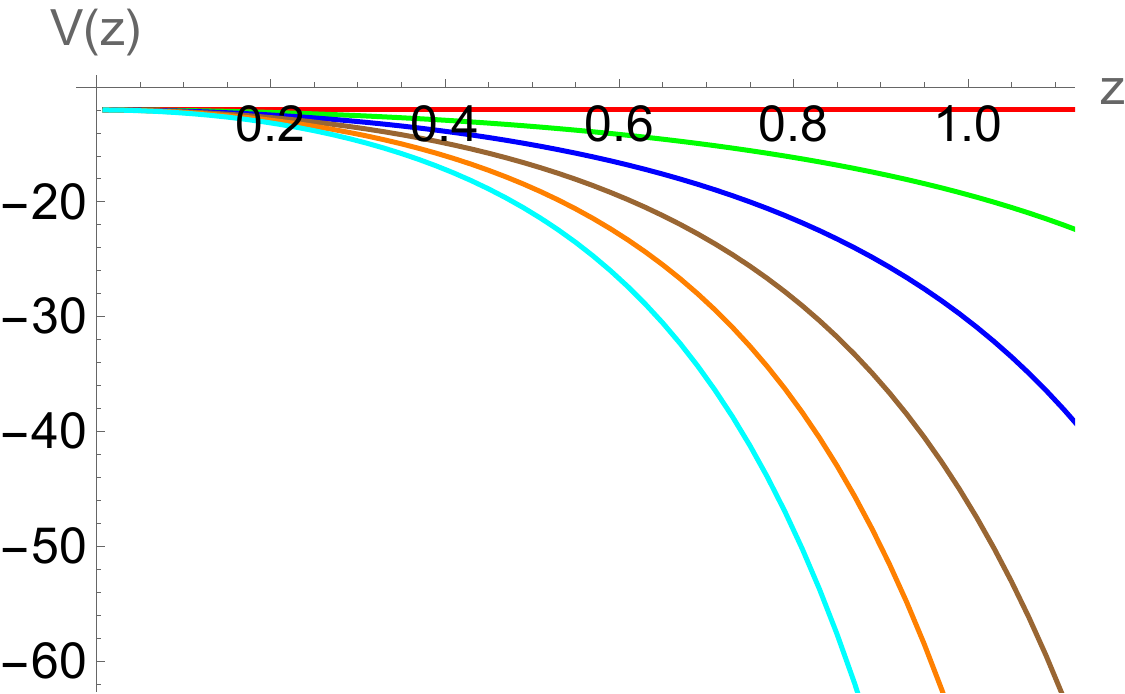}
\caption{\small The behavior of $g_b(z)$, $R_{\mu\nu\rho\sigma}R^{\mu\nu\rho\sigma}$, $\phi(z)$, and $V(z)$ for different values of the hair parameter $a$. Here $z_h=1$ is used. Red, green, blue, brown, orange, and cyan curves correspond to $a=0$, $0.1$, $0.2$, $0.3$, $0.4$, and $0.5$, respectively.}
\label{geometryBHn2D5}
\end{figure}

In this section, we proceed to analyze the geometric and thermodynamic features of the hairy solutions for $n=2$, corresponding to the choice $A(z)=-a z^2$. This case is particularly important in the context of holographic QCD. In particular, the scale factor $e^{-a z^2}$ is crucial for realizing confinement and reproducing the linear Regge trajectories of heavy meson spectra within holographic models \cite{Karch:2006pv}. For this reason, we provide a detailed discussion of this setup. As in the $n=1$ case, one can again obtain analytic expressions for the blackening function $g_b(z)$ and the scalar field $\phi(z)$, which are given by
\begin{eqnarray}
 g_b(z) & = & 1 - C_b \frac{\left(e^{3 a z^2} \left(3 a z^2-1\right)+1\right)}{18 a^2},  \nonumber \\
 \phi(z)  & = & \sqrt{3} z \sqrt{a \left(2 a z^2+3\right)}+\frac{3}{2} \sqrt{\frac{3}{2}}
   \left(\log (3)-2 \log \left(\sqrt{2 a z^2+3}-\sqrt{2} \sqrt{a}
   z\right)\right)\,,
\label{gbsoln2D5}
\end{eqnarray}
where the integration constant $C_b$ is
\begin{eqnarray}
C_b  & = & \frac{18 a^2}{e^{3 a z_h^2} \left(3 a z_h^2-1\right)+1} \,. 
\label{Cbsoln2D5}
\end{eqnarray}
Once again, the scalar field becomes nontrivial only for $a \neq 0$ and vanishes identically when $a=0$. Thus, as desired, in the limit $a \to 0$, the hairy black hole solution smoothly reduces to the standard planar Schwarzschild black hole. The scalar field remains regular and finite everywhere outside the horizon, and vanishes only at the asymptotic AdS boundary. Similarly, the blackening function $g_b(z)$ changes sign at $z=z_h$, signaling the presence of an event horizon. This is true for all values of $a$. Furthermore, the Kretschmann scalar remains finite everywhere outside the horizon, confirming the absence of curvature singularities in the bulk spacetime. The finiteness of the scalar field and Kretschmann scalar both at and outside the horizon again indicates the existence of a well-behaved planar hairy black hole solution. This regularity property also extends to the corresponding hairy AdS soliton solution. In particular, the soliton metric function $g_s(z)$ can again be obtained from $g_b(z)$ by replacing $z_h$ with $z_0$. Likewise, the scalar potential is smooth everywhere in the bulk and approaches the AdS value $V=-12/\ell^{2}$ near the boundary for all choices of the parameter $a$. When $a=0$, the potential remains constant, whereas for nonzero $a$ it decreases monotonically as $z$ increases. Importantly, in all cases, the potential stays bounded from above by its ultraviolet boundary value. The overall radial dependence of these functions is shown in Fig.~\ref{geometryBHn2D5}. In this case as well, the potential profile almost overlaps for different $z_h$ and $a$ values, again indicating mild dependence on these parameters.

\subsubsection{Thermodynamics of the hairy black hole for $A(z)=-a z^2$}
As in the $n=1$ case, we can similarly compute various thermodynamic observables for the $n=2$ case. The holographic renormalization procedure is completely analogous. The only difference will arise in the nature of the scalar counterterms. In particular, compared to the case $A(z)=-az$,  the on-shell action now contains fewer scalar-field-induced divergences. Consequently, the scalar counterterm can be taken to be\footnote{Here, the $\phi^2$ counterterm suffices to cancel all divergences, while the $\phi^4$ term contributes only a finite part. We fix its coefficient by matching the renormalized mass to the $z^4$ coefficient in the near-boundary expansion of $g_b(z)$, adopting the same prescription for the hairy soliton. Since this finite contribution is identical for both backgrounds with matched boundary data, it cancels in the difference of their renormalized actions and does not affect the phase-transition structure. }
\begin{eqnarray}
S_{ct}^{b} & = &
 \frac{2}{16 \pi G_5} \int_{\partial \mathcal{M}} \mathrm{d^4}x \ \sqrt{-\gamma} \ \sum_{i=1}^{i=2} b_i \phi^{2i} \,.
 \label{countertermn2D5}
\end{eqnarray}
Using the boundary expansion
\begin{eqnarray}
& & g(z) = 1-\frac{9 a^2 z^4}{2 \left(e^{3 a z_h^2} \left(3 a
   z_h^2-1\right)+1\right)}+O\left(z^6\right) \,, \nonumber \\
& & \phi(z) = 6 \sqrt{a} z+\frac{2}{3} a^{3/2} z^3+O\left(z^4\right)  \,, \nonumber \\
& & V(z) = -12-54 a z^2-120 a^2 z^4+O\left(z^5\right) = 2\Lambda + \frac{m^{2}\phi^{2}}{2} + \cdots
\label{boundaryexpn2D5}
\end{eqnarray}
with $m^2=-3$, thereby again satisfying the BF bound, we get the following expressions of various terms in the renormalized action:
\begin{eqnarray}
S_{ES}^{b} & = & \frac{\beta_b L_b V_2}{16 \pi G_5}  \left(\frac{e^{-3 a z^2}}{9 a^2 z^4} \left(36 a^3 z^2+18 a^2-a z^2 \left(e^{3 a z^2}+2\right)
   C_b+e^{3 a z^2} C_b-C_b\right) \right)\bigg\rvert_{z=\epsilon}^{z=z_h} \,, \\
S_{GH}^{b} & = & \frac{\beta_b L_b V_2}{16 \pi G_5}  \left( -12 a^2-\frac{8 a}{\epsilon^2}-C_b+\frac{8}{\epsilon^4}  \right)\,, \\
S_{BK}^{b} & = &  \frac{\beta_b L_b V_2}{16 \pi G_5} \left(-48 a^2+\frac{24 a}{\epsilon^2}+\frac{3 C_b}{4}-\frac{6}{\epsilon^4}  \right)\,, \\
S_{ct}^{b} & = &  \frac{\beta_b L_b V_2}{16 \pi G_5} \left( 57 a^2-\frac{18 a}{\epsilon^2}  \right) \,.
\label{countertermexpn2D5}
\end{eqnarray}
From the above expressions, the renormalized free energy $\mathcal{F}_b=-S^{b}_{ren}/\beta_b$ of the hairy black hole is obtained as
\begin{eqnarray}
\mathcal{F}_b & = & - \frac{L_b V_2 C_b}{64 \pi G_5} \,, \nonumber \\
 & = & - \frac{9 a^2 V_2 L_b}{32 \pi  G_5\left(e^{3 a z_h^2} \left(3 a
   z_h^2-1\right)+1\right)}\,,
\label{freeenergyCexpn2D5}
\end{eqnarray}
where we have explicitly used the expression of the integration constant $C_b$ from Eq.~(\ref{Cbsoln2D5}). 
The stress-energy tensor now takes form
\begin{eqnarray}
T_{\mu\nu} =\frac{1}{8 \pi G_5} \left[ \Theta \gamma_{\mu\nu} - \Theta_{\mu\nu}- 3 \gamma_{\mu\nu} + \gamma_{\mu\nu} \left(  b_1 \phi^2 + b_2 \phi^4 \right) \right]  \,.
\label{stresstensorn2d5}
\end{eqnarray}
Using $T_{\mu\nu}$ and Eq.~(\ref{massn1D5}), the conserved mass is now given by
\begin{eqnarray}
 M_b & = &  \frac{3 L_b V_2 C_b}{64 \pi G_5} \,, \nonumber \\
 & = & \frac{27 a^2 V_2 L_b}{32 \pi  G_5 \left(e^{3 a z_h^2} \left(3 a
   z_h^2-1\right)+1\right)} \,.
\label{massexpn2D5}
\end{eqnarray}
Again, the mass is proportional to the integration constant $C_b$, indicating the primary nature of the black hole hair. This mass expression also matches with the coefficient of the $z^{4}$ term in the near-boundary expansion of the blackening function $g_b(z)$. Similarly, the temperature and entropy of the black hole are given by
\begin{eqnarray}
T_b & = & \frac{9 a^2 z_h^3 e^{3 a z_h^2}}{2 \pi \left(  e^{3 a z_h^2} \left(3 a
   z_h^2-1\right)+1 \right)}\,,  \nonumber \\
S_b  & = & \frac{L_b V_2 e^{-3 a z_{h}^{2}}}{4 G_5 z_h^3}\,.
\label{tempentn2D5}
\end{eqnarray}
From Eqs.~(\ref{freeenergyCexpn2D5}), (\ref{massexpn2D5}), and (\ref{tempentn2D5}), it is straightforward to verify that the thermodynamic observables for the $A(z)=-a z^2$ case also satisfy the expected relation
$\mathcal{F}_b = M_b - T_b S_b$. In addition, we have obtained an analytic expression for the pressure and find that it again obeys the standard identity $\mathcal{F}_b = -P_b$, providing a nontrivial consistency check of our thermodynamic analysis for the hairy black hole solutions. Remarkably, despite the intricate structure of these solutions, we are again able to obtain closed-form expressions for the relevant thermodynamic quantities. Furthermore, in the limit $a \to 0$, all results smoothly reduce to those of the usual five-dimensional planar Schwarzschild black hole, further confirming the consistency of the expressions.

\begin{figure}[h!]
\begin{minipage}[b]{0.5\linewidth}
\centering
\includegraphics[width=2.8in,height=2.3in]{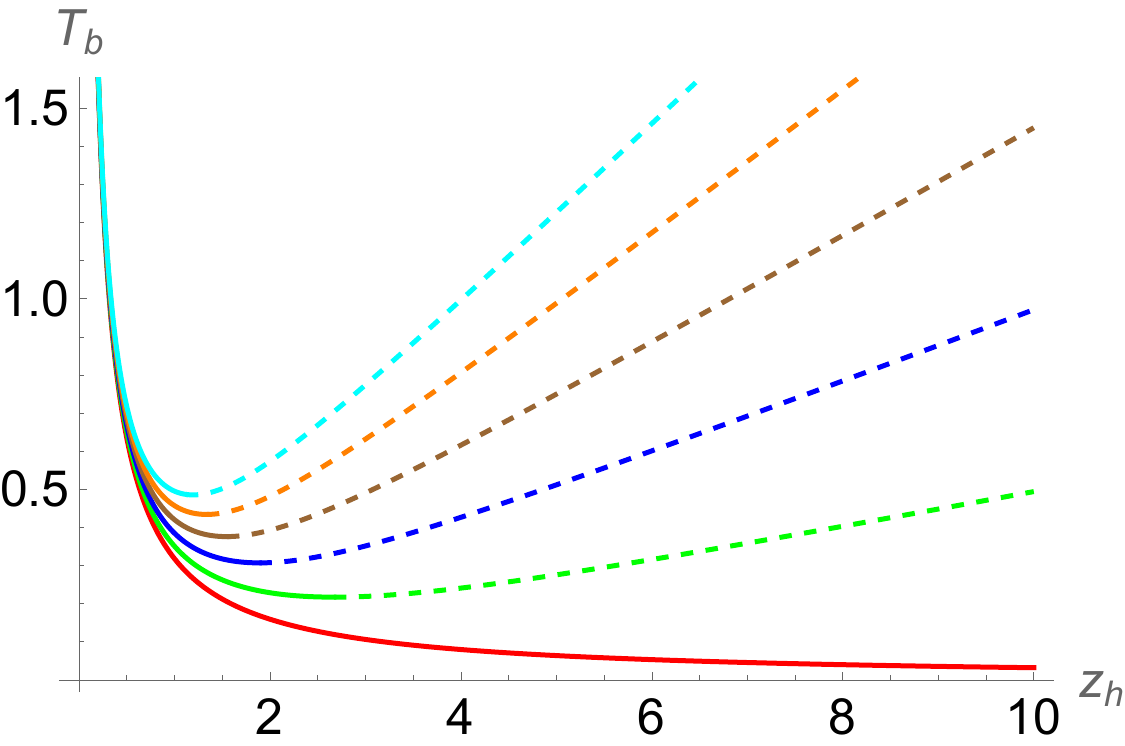}
\caption{ \small Hawking temperature $T_b$ as a function of horizon radius $z_h$ for various values of $a$. Here $G_5=1$ is used. Red, green, blue, brown, orange, and cyan curves correspond to $a=0$, $0.1$, $0.2$, $0.3$, $0.4$, and $0.5$, respectively.}
\label{zhvsTempBHn2D5}
\end{minipage}
\hspace{0.4cm}
\begin{minipage}[b]{0.5\linewidth}
\centering
\includegraphics[width=2.8in,height=2.3in]{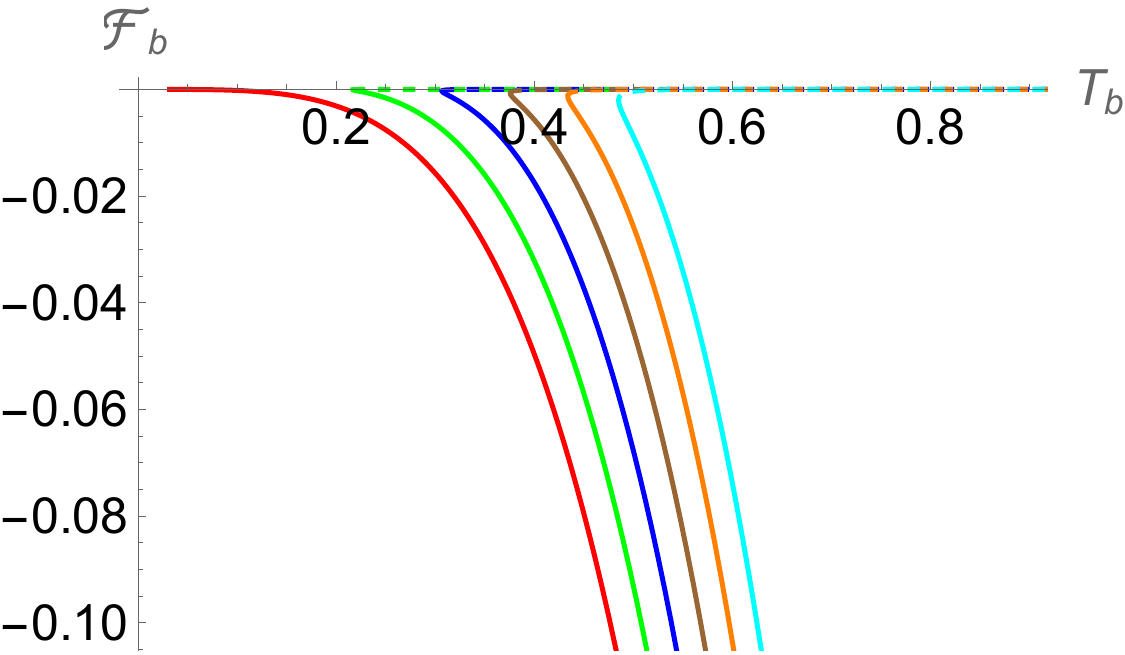}
\caption{\small Free energy $\mathcal{F}_b$ of the hairy black hole as a function of temperature $T_b$ for various values of $a$. Here $G_5=1$ and $V_2=1$ are used. Red, green, blue, brown, orange, and cyan curves correspond to $a=0$, $0.1$, $0.2$, $0.3$, $0.4$, and $0.5$, respectively.}
\label{TempvsFreeBHn2D5}
\end{minipage}
\end{figure}

The thermodynamic phase structure of the constructed hairy black hole for $A(z)=-a z^2$ is shown in Figs.~\ref{zhvsTempBHn2D5} and \ref{TempvsFreeBHn2D5}. We find that, as opposed to the $A(z)=-a z$ case, switching on the hair parameter drastically changes the phase structure of the black hole. In particular, for any finite $a$, there are now two hairy black hole branches -- one large and one small -- for each value of the temperature. The large black hole branch (shown by solid lines), for which the temperature decreases with $z_h$, has positive specific heat and is thermodynamically stable. While the small black hole branch (shown by dashed lines), for which the temperature increases with $z_h$, has negative specific heat and is thermodynamically unstable. These large and small black hole branches appear for any finite value of the hair parameter $a$, while for $a=0$ we get only a single large black hole branch, which is the case for the planar Schwarzschild black hole. Interestingly, the small and large black hole branches exist only above a certain minimum temperature $T_{min}$, and they cease to exist below it. Therefore, below this minimum temperature, only the thermal-AdS or AdS-soliton solution persists (see below). 

The corresponding free energy behavior is shown in Fig.~\ref{TempvsFreeBHn2D5}. It is observed that the free energy of the large black hole branch is always smaller than the small black hole branch. This indicates that the large black hole branch is thermodynamically favored over the small black hole branch at all temperatures. This is true for all values of $a$. However, interestingly, since the large black hole branch exists only above $T_{min}$, a phase transition is expected to occur between the large black hole and AdS-soliton as the temperature is lowered. To explicitly see this, we first need to compute the free energy of the hairy AdS-soliton.\footnote{In \cite{Dudal:2017max}, a phase transition between the large black hole and thermal-AdS backgrounds was also discussed as the temperature is lowered. This transition is interpreted as the confinement/deconfinement phase transition in the dual field theory, with thermal-AdS and the black hole background corresponding to the confined and deconfined phases, respectively. We emphasize that the phase transition investigated in the present work is different: it concerns the competition between the black-hole and AdS-soliton phases. }

\subsubsection{Thermodynamics of the hairy AdS soliton for $A(z)=-a z^2$}
Since the hairy soliton shares the same asymptotic structure as the hairy black hole, the scalar counterterms required to regularize the action take the same form as in Eq.~(\ref{countertermn2D5}). The asymptotic boundary expansion and scalar mass expression are the same as those presented in Eq.~(\ref{boundaryexpn2D5}), after replacing $z_h \rightarrow z_0$. The individual contributions to the regularized Euclidean action for the hairy AdS soliton are then given by
\begin{eqnarray}
S_{ES}^{s} & = & \frac{\beta_s L_s V_2}{16 \pi G_5} \left(\frac{e^{-3 a z^2}}{9 a^2 z^4} \left(36 a^3 z^2+18 a^2-a z^2 \left(e^{3 a z^2}+2\right)
   C_s+e^{3 a z^2} C_s-C_s\right) \right)\bigg\rvert_{z=\epsilon}^{z=z_0} \,, \\
S_{GH}^{s} & = & \frac{\beta_s L_s V_2}{16 \pi G_5}  \left( -12 a^2-\frac{8 a}{\epsilon^2}-C_s+\frac{8}{\epsilon^4}  \right)\,, \\
S_{BK}^{s} & = &  \frac{\beta_s L_s V_2}{16 \pi G_5} \left(-48 a^2+\frac{24 a}{\epsilon^2}+\frac{3 C_s}{4}-\frac{6}{\epsilon^4}  \right)\,,\\
S_{ct}^{s} & = &  \frac{\beta_s L_s V_2}{16 \pi G_5} \left(  57 a^2-\frac{18 a}{\epsilon^2} \right) \,,
\label{countertermexpsoln2D5}
\end{eqnarray}
with $C_s$ now given by
\begin{eqnarray}
C_s = \frac{1}{ \int_0^{z_0} \, d\xi ~ e^{3 a \xi^2}~ \xi^{3} } = \frac{18 a^2}{e^{3 a z_0^2} \left(3 a z_0^2-1\right)+1}  \,.
\label{csexpn2D5}
\end{eqnarray}
The renormalized action for the hairy soliton then reduces to a simpler expression
\begin{eqnarray}
S_{ren}^s &  = &  \frac{\beta_s L_s V_2 C_s}{64 \pi G_5} \,, \nonumber \\
&  = & \frac{9 a^2 \beta_s L_s V_2 }{32 \pi  G_5 \left(e^{3 a z_0^2} \left(3 a
   z_0^2-1\right)+1\right)}\,.
\label{renomactionsoln2D5}
\end{eqnarray}
This leads to the free energy of the hairy AdS soliton $\mathcal{F}_s=-S_{ren}^{s}/\beta_s = M_s$ as
\begin{eqnarray}
& & \mathcal{F}_s   =  -\frac{9 a^2 L_s V_2 }{32 \pi  G_5 \left(e^{3 a z_0^2} \left(3 a
   z_0^2-1\right)+1\right)} = -\frac{V_2 e^{-3 a z_0^2}}{16 G_5 z_0^3} \,.
\label{fsmexpsoln2D5}
\end{eqnarray}
As expected, the free energy of the hairy AdS soliton is again negative.  

\begin{figure}[h!]
\begin{minipage}[b]{0.5\linewidth}
\centering
\includegraphics[width=2.8in,height=2.3in]{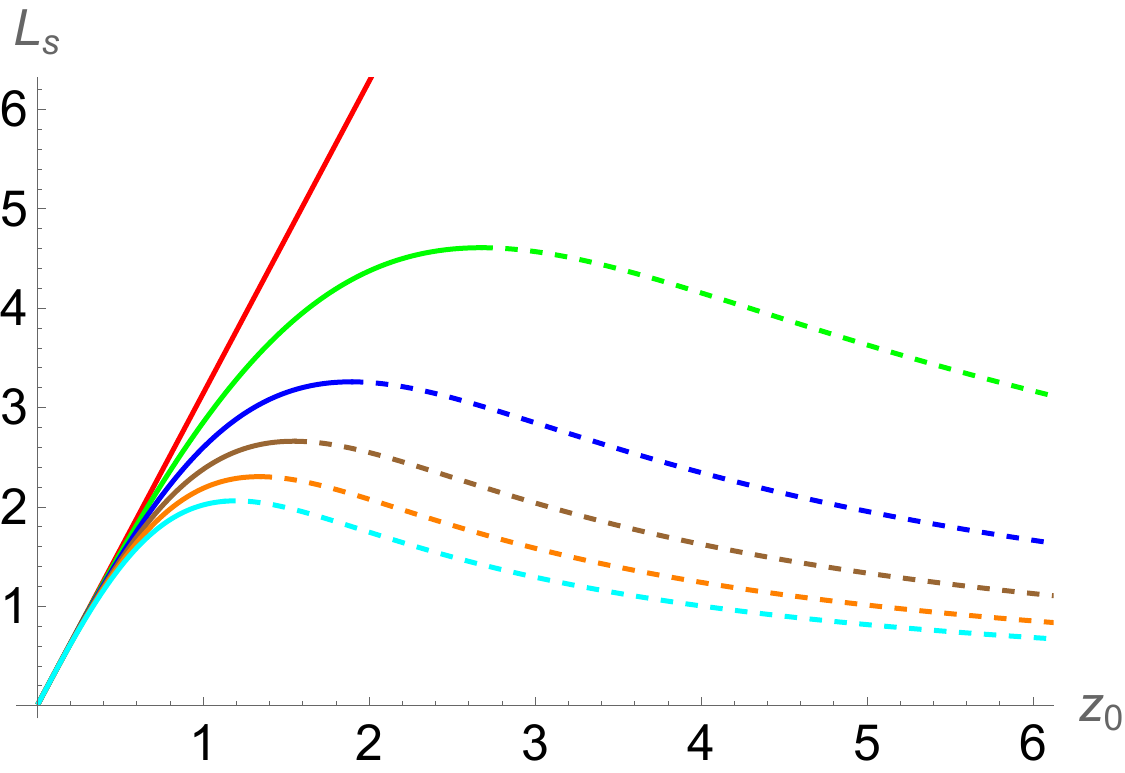}
\caption{ \small Spatial period $L_s$ as a function of $z_0$ for various values of $a$. Here $G_5=1$ and $V_2=1$ are used. Red, green, blue, brown, orange, and cyan curves correspond to $a=0$, $0.1$, $0.2$, $0.3$, $0.4$, and $0.5$, respectively.}
\label{z0vsLsvsan2D5}
\end{minipage}
\hspace{0.4cm}
\begin{minipage}[b]{0.5\linewidth}
\centering
\includegraphics[width=2.8in,height=2.3in]{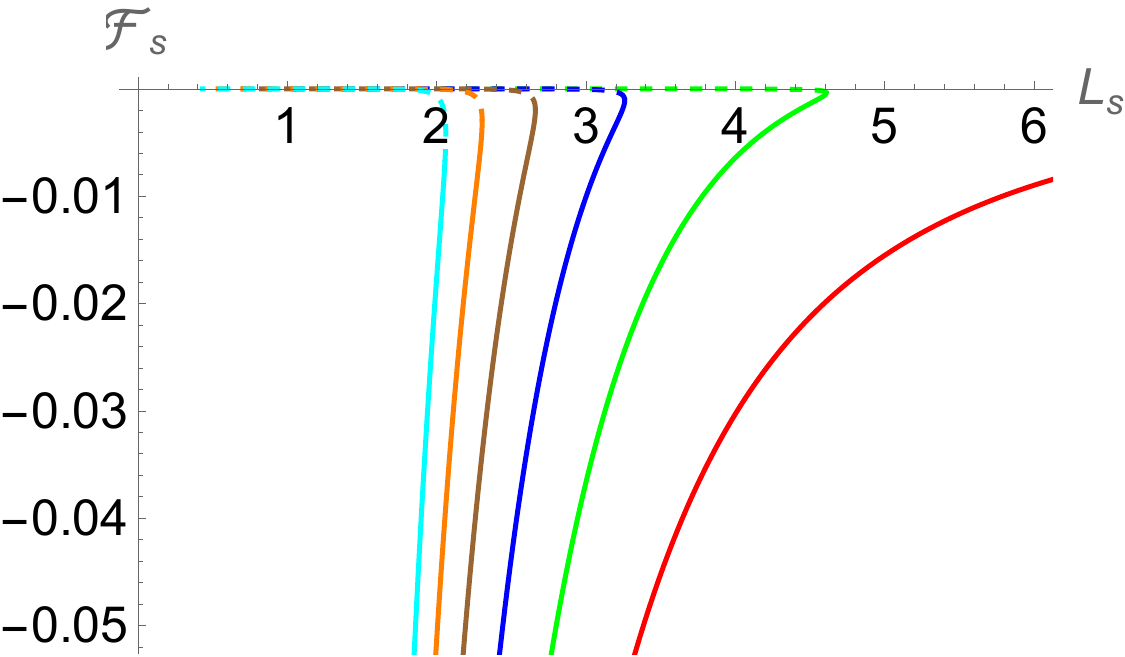}
\caption{\small Free energy $\mathcal{F}_s$ of the hairy soliton as a function of $L_s$ for various values of $a$. Here $G_5=1$ and $V_2=1$ are used. Red, green, blue, brown, orange, and cyan curves correspond to $a=0$, $0.1$, $0.2$, $0.3$, $0.4$, and $0.5$, respectively.}
\label{LsvsFreesolitonn2D5}
\end{minipage}
\end{figure}

We see from the above equation that the free energy of the hairy soliton is not only negative but also exhibits one to one relation with $z_0$. However, it is also instructive to analyze the profile of $L_s$ with $z_0$ and the profile of free energy with $L_s$. Then it is not hard to see that, just like in the black hole case, there are actually two $z_0$ solutions for each value of $L_s$. In particular, for finite $a$, there exists a maximum  $L_s$ beyond which there is no hairy soliton solution. Below this maximum  $L_s$, there are two solitonic solutions: one for small $z_0$ (shown by solid lines) and one for large $z_0$ (shown by dashed lines). The free energy of the small $z_0$ solution is always smaller than the free energy of the large $z_0$ solution. Consequently, the small $z_0$ soliton phase always corresponds to the local minima of the solution. This issue is clearly illustrated in Fig.~\ref{z0vsLsvsan2D5} and \ref{LsvsFreesolitonn2D5}. This behavior is completely analogous to the black hole case discussed in the previous subsection, where two black hole solutions appear for a given value of the period $\beta_b$. In the subsequent section, we will consider only the thermodynamically favored small $z_0$ hairy soliton phase to study the phase structure between the hairy black hole and hairy soliton solutions.

We can similarly compute the conserved energy of the hairy soliton from the stress-energy tensor (\ref{stresstensorn2d5}). We get
\begin{eqnarray}
M_s &=& - \frac{C_s L_s V_2}{64 \pi G_5}\,.
\end{eqnarray}
This expression is again in agreement with Eq.~(\ref{fsmexpsoln2D5}), providing a further consistency check of the analytic results. Furthermore, the energy difference between the hairy black hole and the hairy soliton is given by
\begin{eqnarray}
\Delta M =  M_b - M_s = \frac{L_b V_2}{64 \pi G_5} \left(3 C_b + C_s \right)\,.
\end{eqnarray}
This quantity is always positive, indicating that the hairy black hole has higher energy than the corresponding hairy soliton. Thus, the Horowitz-Myers conjecture continues to hold in the present black hole-soliton case.

\subsubsection{Phase transition between hairy black hole and soliton for $A(z)=-a z^2$}
\begin{figure}[h!]
\centering
\includegraphics[width=3.8in,height=2.6in]{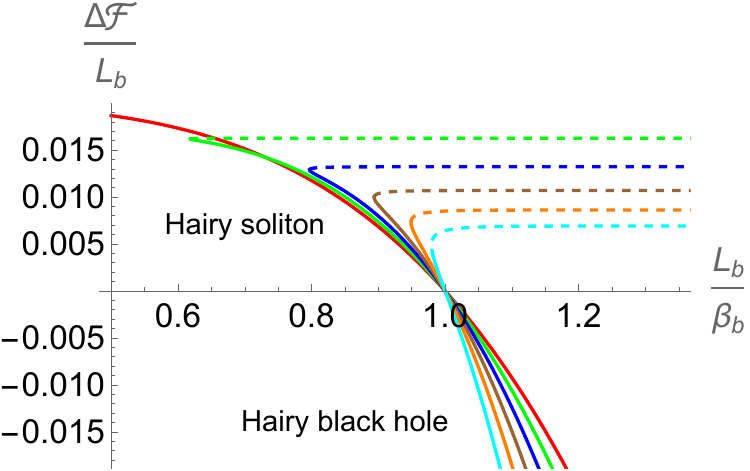}
\caption{ \small Free energy difference $\Delta\mathcal{F}$ as a function of periodicity ratio $L_b/\beta_b$ for various values of $a$. Here $G_5=1$, $V_2=1$, and $z_0=1$ are used. Red, green, blue, brown, orange, and cyan curves correspond to $a=0$, $0.1$, $0.2$, $0.3$, $0.4$, and $0.5$, respectively.}
\label{Lbbybetabvsfreediffvsan2D5}
\end{figure}

Now we analyze the thermodynamic phase structure of the hairy black hole and hairy soliton solutions. As mentioned earlier, for $A(z)=-a z^2$, there are two black hole and two solitonic phases, with thermodynamically favored phases occurring for small $z_h$ and $z_0$ values. Accordingly, the phase structure is expected to be more interesting compared to the $A(z)=-az$ case. In Fig.~\ref{Lbbybetabvsfreediffvsan2D5}, the free energy difference of the hairy black hole and soliton solutions is shown. Here we have used $z_0=1$. The value $z_0=1$ ensures that we are considering the thermodynamically favored hairy soliton branch. We again notice that $\Delta\mathcal{F}$ changes sign and hence a phase transition between the black hole and soliton occurs as the ratio $L_b/\beta_b$ is varied. In particular, the large black hole solution (that appears for small $z_h$) is thermodynamically favored in the $L_b>\beta_b$ region, whereas the soliton phase is favored in the $L_b<\beta_b$ region. The transition point $L_b=\beta_b$ again remains the same for all $a$ values.  This structure is clearly illustrated in Fig.~\ref{Lbbybetabvsfreediffvsan2D5}, and is completely analogous to the $A(z)=-a z$ case. Interestingly, the small black hole phase (that appears for large $z_h$) always has a higher free energy than the stable solitonic phase. The free energy behavior of the small black hole phase is shown by horizontal dashed lines in Fig.~\ref{Lbbybetabvsfreediffvsan2D5}. Therefore, the unstable small black hole phase is not only thermodynamically disfavored with respect to the large black hole phase but also with respect to the solitonic phase. 

\begin{figure}[h!]
\centering
\includegraphics[width=3.8in,height=2.8in]{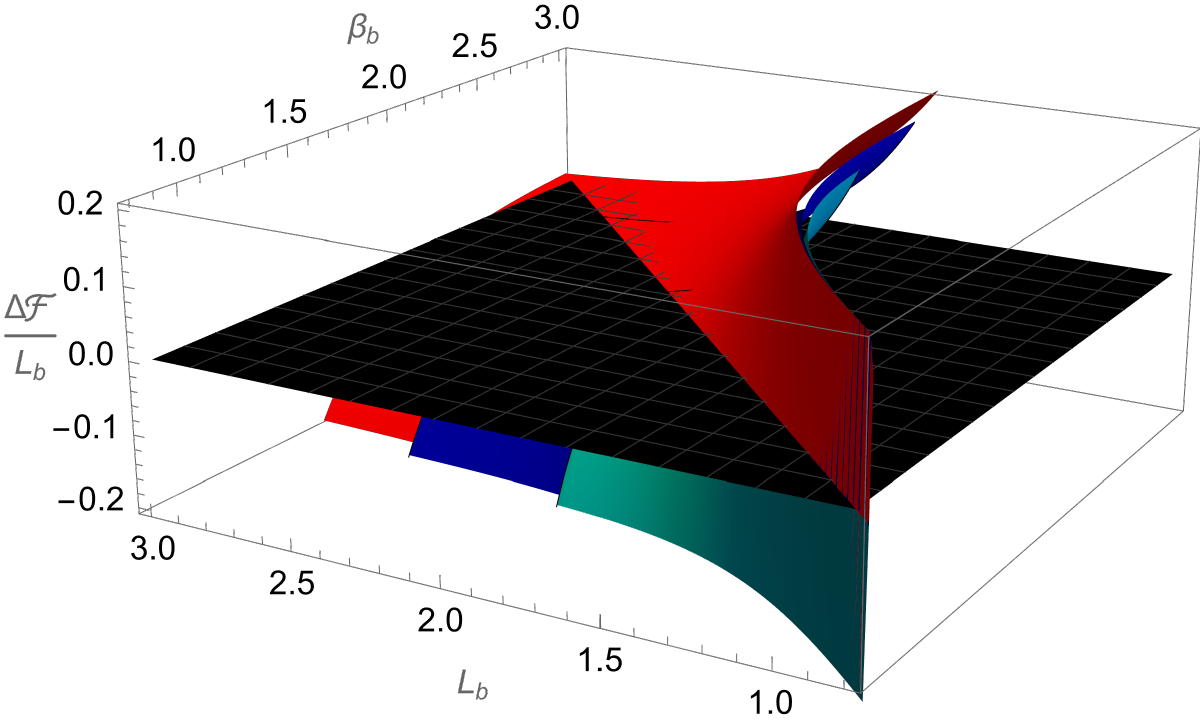}
\caption{ \small Free energy difference $\Delta\mathcal{F}$ as a function of $L_b$ and $\beta_b$ for $a=0.1$ (red surface),  $a=0.3$ (blue surface), and $a=0.5$ (cyan surface). The black plane indicates $\Delta\mathcal{F}=0$ surface.  Here $G_5=1$ and $V_2=1$ are used.}
\label{Lbbybetabvsfreediffvs3Dn2D5}
\end{figure}

The three-dimensional profile of the free energy difference as a function of $\beta_b$ and $L_b$ for three different $a$ values is shown in Fig.~\ref{Lbbybetabvsfreediffvs3Dn2D5}. Since the maximum value of $\beta_b$ and $L_b$ depends on $a$, the allowed range of $\beta_b$ and $L_b$ also depends on $a$. In particular, the allowed maximum value of $\beta_b$ and $L_b$ decreases with $a$. Accordingly, the parameter region of the cyan surface (for $a=0.5$) is smaller than the red surface (for $a=0.1$). It is again evident that the free energy surface intersects the $\Delta\mathcal{F}=0$ plane along a straight line defined by $\beta_b = L_b$. Consequently, in the region above the black plane, the hairy soliton phase is thermodynamically preferred, whereas below it, the hairy black hole phase becomes favored.

\section{Stability and thermodynamic phase transitions in $D=4$}
\label{hairysolD4}
Having thoroughly examined the geometric and thermodynamic properties of hairy black holes and solitons in five dimensions, we now extend our analysis to four dimensions in order to make the discussion more complete. Since the computations closely follow those of the five-dimensional case, we present the results only briefly. We focus mainly on the case $A(z)=-az$, as analogous conclusions can be readily drawn for other form factors.

\begin{figure}[ht]
\centering
\includegraphics[width=0.45\textwidth]{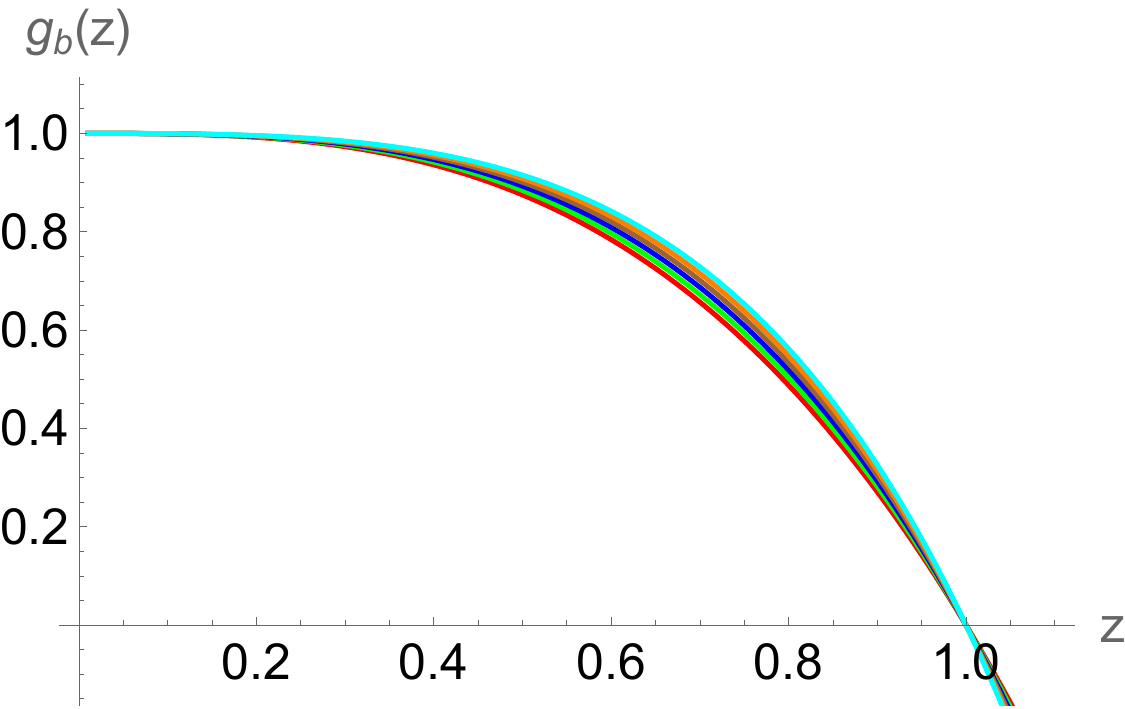}
\includegraphics[width=0.45\textwidth]{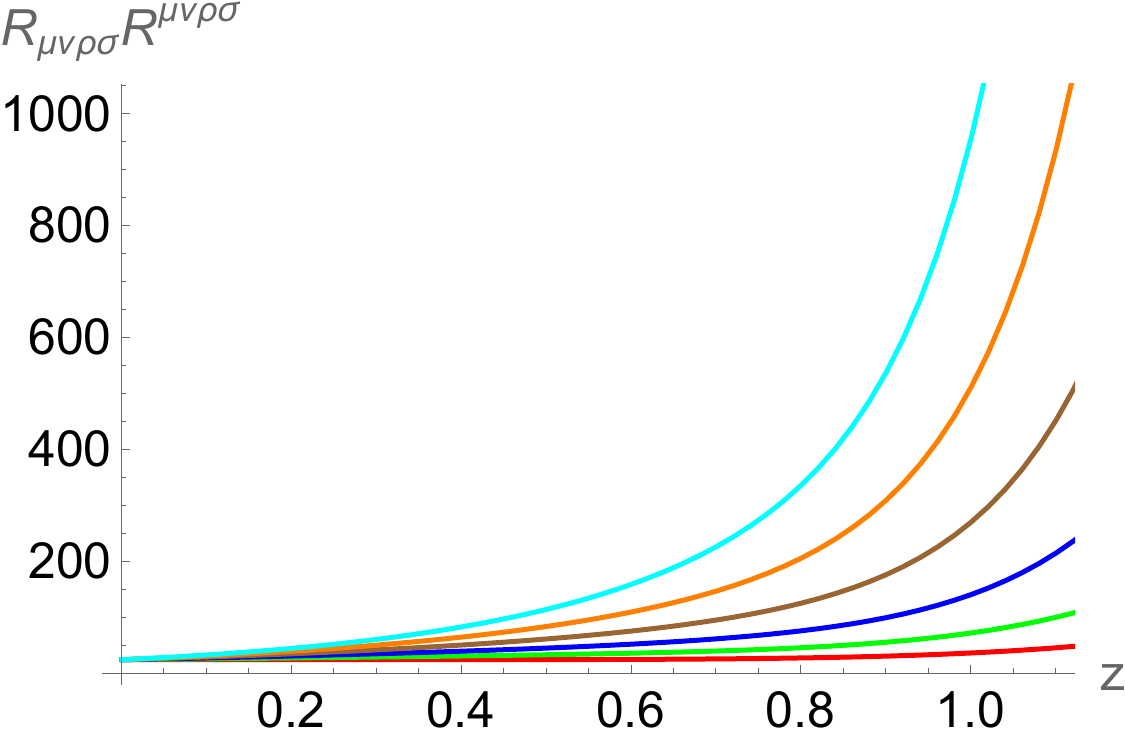}

\vspace{0.3cm}

\includegraphics[width=0.45\textwidth]{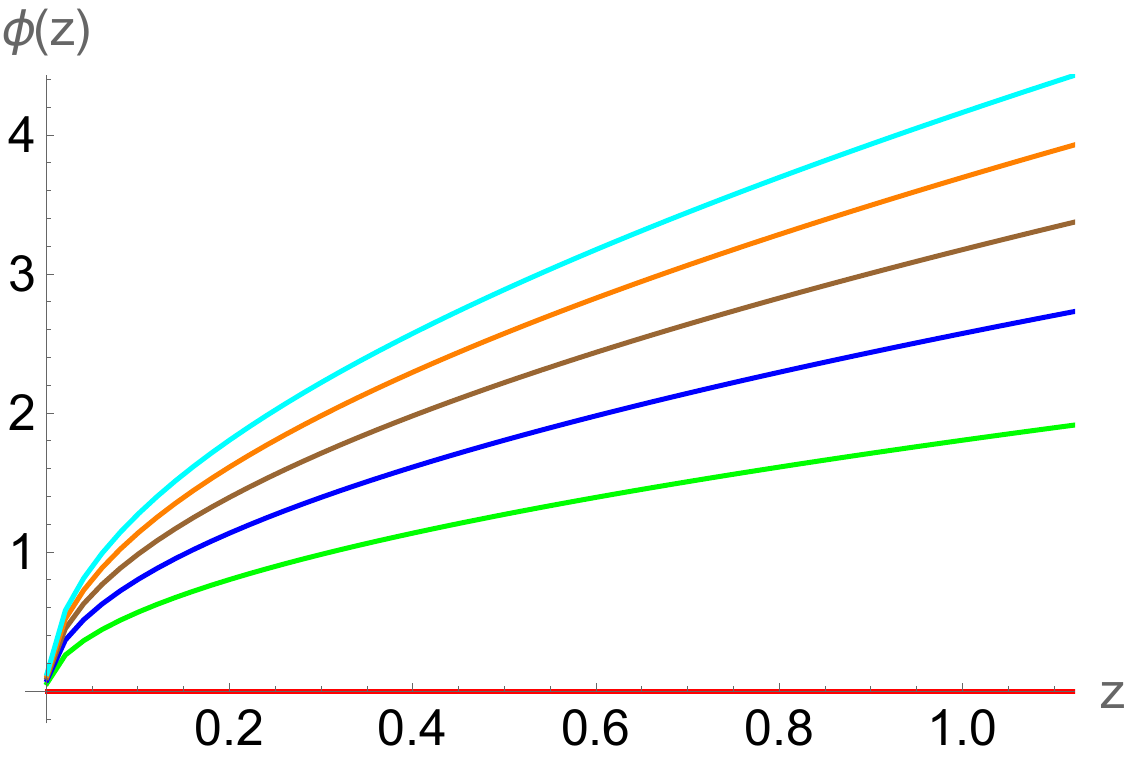}
\includegraphics[width=0.45\textwidth]{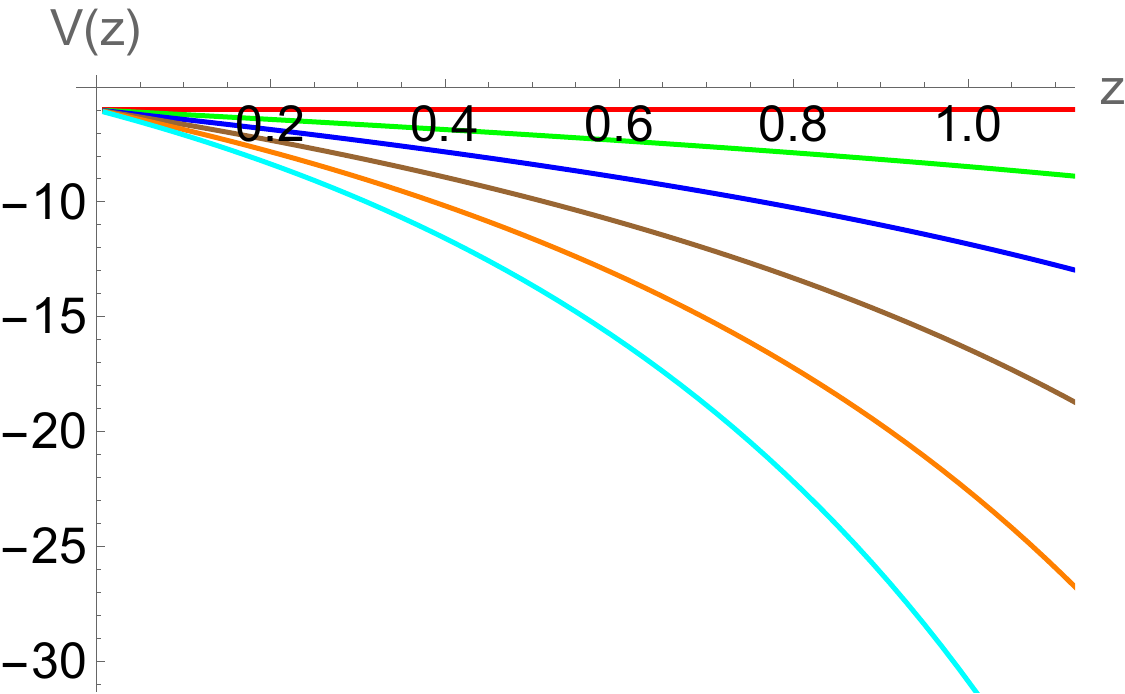}
\caption{\small The behavior of $g_b(z)$, $R_{\mu\nu\rho\sigma}R^{\mu\nu\rho\sigma}$, $\phi(z)$, and $V(z)$ for different values of the hair parameter $a$. Here $z_h=1$ is used. Red, green, blue, brown, orange, and cyan curves correspond to $a=0$, $0.1$, $0.2$, $0.3$, $0.4$, and $0.5$, respectively.}
\label{geometryBHn1D4}
\end{figure}

With $A(z)=-a z$, the expressions for the blackening function $g_b(z)$ and the scalar field $\phi(z)$ in four dimensions reduce to
\begin{eqnarray}
 g_b(z) & = & 1 - C_b \frac{\left(e^{2 a z} (2 a z (a z-1)+1)-1\right)}{4 a^3}   \nonumber \\
 \phi(z)  & = & 2 \sqrt{a z (a z+2)}-4 \log \left(\frac{\sqrt{a z+2}-\sqrt{a z}}{\sqrt{2}}\right) \,,
\label{gbsoln1D4}
\end{eqnarray}
where the integration constant $C_b$ is
\begin{eqnarray}
C_b  & = & \frac{4 a^3}{e^{2 a z_h} \left(2 a z_h \left(a z_h-1\right)+1\right)-1}\,. 
\label{Cbsoln1D4}
\end{eqnarray}

The blackening function $g_b(z)$ again smoothly reduces to the corresponding four-dimensional planar Schwarzschild black hole form in the limit $a \to 0$. The radial profile of $g_b(z)$ for different values of $a$ is shown in Fig.~\ref{geometryBHn1D4}. It is finite and goes to zero on the horizon. The Kretschmann scalar similarly remains finite everywhere outside the horizon, thereby indicating the well-behaved nature of the hairy black hole geometry. As in the five-dimensional case, the Kretschmann scalar grows in magnitude as the scalar hair parameter increases, indicating that the spacetime becomes more strongly curved for larger $a$. This trend appears to be robust across different dimensions and choices of the form factor $A(z)$. Moreover, for the specific form $A(z)=-az$, a comparison shows that the curvature is typically stronger in five dimensions than in four dimensions.

In addition, the scalar field is everywhere real and finite outside the event horizon. From Eq.~(\ref{gbsoln1D4}), one observes that it approaches zero only at the AdS boundary, while remaining finite throughout the exterior region. This regular behavior signals that the resulting planar hairy black hole geometry is physically well-defined. The associated scalar potential is also smooth across the bulk and tends to the AdS value $V=-6/\ell^{2}$ near the boundary for any choice of the hair parameter $a$. In the limit $a\rightarrow0$, the potential becomes constant, whereas for nonzero $a$ it decreases steadily as $z$ grows. Moreover, in all cases, the potential is again bounded above by its ultraviolet boundary value. All of these conclusions carry over directly to the hairy soliton configuration. In particular, the soliton metric function $g_s(z)$ is obtained from the black hole blackening factor $g_b(z)$ through the replacement $z_h \to z_0$. As a result, the corresponding Kretschmann scalar remains finite throughout the region outside $z_0$, confirming that the hairy soliton geometry is likewise smooth and free of curvature singularities. Similarly, the potential profiles corresponding to different values of $z_h$ and $a$ almost coincide, once again demonstrating that the potential depends only very weakly on these parameters.

\subsection{Thermodynamics of the hairy black hole for $A(z)=-a z$}
In four dimensions, the various terms in the regularized action take a slightly different form compared to the five-dimensional case. Explicitly, now we have 
\begin{eqnarray}
& & S_{ES}^{b} = \frac{1}{16 \pi G_4} \int_{\mathcal{M}} \mathrm{d^4}x~\ \sqrt{-g} V(z) \,, ~~~
S_{GH}^{b} = \frac{1}{8 \pi G_4} \int_{\partial \mathcal{M}} \mathrm{d^3}x \ \sqrt{-\gamma} \ \Theta\,,  \nonumber \\
& & S_{BK}^{b} = -\frac{1}{16 \pi G_4} \int_{\partial \mathcal{M}} \mathrm{d^3}x \ \sqrt{-\gamma} \left(4 - R^{(3)}\right)\,, ~~~
S_{ct}^{b} =
 \frac{2}{16 \pi G_4} \int_{\partial \mathcal{M}} \mathrm{d^3}x \ \sqrt{-\gamma} \left( \sum_{i=1}^{i=3} b_i \phi^{2i} \right) \,.
\label{countertermn1D4}
\end{eqnarray}
The scalar counterterm $S_{ct}^{b}$ is again required to remove additional divergences from the scalar field, and the parameters in the scalar counterterm are determined by requiring complete cancellation of infrared divergences. Using the following boundary expansion
\begin{eqnarray}
& & g(z) = 1-\frac{4 a^3 z^3}{3 \left(e^{2 a z_h} (2 a z_h (a
   z_h-1)+1)-1\right)}+O\left(z^4\right) \,, \nonumber \\
& & \phi(z) = 4 \sqrt{2} \sqrt{a} \sqrt{z}+\frac{1}{3} \sqrt{2} a^{3/2}
   z^{3/2}+O\left(z^{5/2}\right)   \,, \nonumber \\
& & V(z) = -6-20 a z-32 a^2 z^2+O\left(z^3\right) = 2\Lambda + \frac{m^2 \phi^2}{2}    + \cdots
\label{boundaryexpn1D4}
\end{eqnarray}
with $m^2=-5/4$, that satisfies the BF bound, and substituting Eq.~(\ref{gbsoln1D4}) into Eq.~(\ref{countertermn1D4}), we get
\begin{eqnarray}
S_{ES}^{b} & = & \frac{\beta_b L_b V_1}{16 \pi G_4} \left(\frac{e^{-2 a z} \left(4 a^4 z+4 a^3+a z \left(e^{2 a z}+1\right)
   C_b-\left(e^{2 a z}-1\right) C_b\right)}{2 a^3 z^3}  \right)\bigg\rvert_{z=\epsilon}^{z=z_h} \,, \\
S_{GH}^{b} & = &  \frac{\beta_b L_b V_1}{16 \pi G_4} \left( 4 a^3-\frac{6 a}{\epsilon^2}-C_b+\frac{6}{\epsilon^3} \right)\,, \\
S_{BK}^{b} & = & \frac{\beta_b L_b V_1}{16 \pi G_4} \left( 18 a^3-\frac{18 a^2}{\epsilon}+\frac{12 a}{\epsilon^2}+\frac{2 C_b}{3}-\frac{4}{\epsilon^3}  \right) \,,\\
S_{ct}^{b} & = & \frac{\beta_b L_b V_1}{16 \pi G_4} \left(-\frac{62 a^3}{3}+\frac{18 a^2}{\epsilon}-\frac{8 a}{\epsilon^2}  \right) \,,
\label{countertermexpn1D4}
\end{eqnarray}
where $V_1$ is the length of the $x_1$-direction. The renormalized free energy of the hairy black hole is then given by 
\begin{eqnarray}
\mathcal{F}_b & = & - \frac{L_b V_1 C_b}{48 \pi G_4} \,, \nonumber \\
 & = & -\frac{a^3 V_1 L_b}{12 \pi  G_4 \left(e^{2 a z_h} \left(2 a z_h \left(a
   z_h-1\right)+1\right)-1\right)}\,,
\label{freeenergyexpn1D4}
\end{eqnarray}
The
stress-energy tensor now takes the form
\begin{eqnarray}
T_{\mu\nu} =\frac{1}{8 \pi G_4} \left[ \Theta \gamma_{\mu\nu} - \Theta_{\mu\nu}- 2 \gamma_{\mu\nu} + \gamma_{\mu\nu} \left( b_1 \phi^2 + b_2 \phi^4 + b_3 \phi^6 \right) \right]  \,.
\label{stresstensorn1D4}
\end{eqnarray}
Using this $T_{\mu\nu}$, the conserved energy associated with the hairy black hole is computed as
\begin{eqnarray}
 M_b & = &  \frac{L_b V_1 C_b}{24 \pi G_4} \,, \nonumber \\
 & = & \frac{a^3 V_1 L_b}{6 \pi  G_4 \left(e^{2 a z_h} \left(2 a z_h \left(a
   z_h-1\right)+1\right)-1\right)} \,.
\label{massexpn1D4}
\end{eqnarray}
Similarly, the temperature and entropy of the black hole are given by
\begin{eqnarray}
T_b & = & \frac{a^3 z_h^2 e^{2 a z_h}}{\pi  \left(e^{2 a z_h} \left(2 a z_h \left(a
   z_h-1\right)+1\right)-1\right)}\,,  \nonumber \\
S_b  & = & \frac{L_b V_1 e^{-2 a z_h}}{4 G_4 z_h^2}\,.
\label{tempentn1D4}
\end{eqnarray}
Using Eqs.~(\ref{freeenergyexpn1D4}), (\ref{massexpn1D4}), and (\ref{tempentn1D4}), one can readily check that the thermodynamic quantities obey the expected relation
$\mathcal{F}_b = M_b - T_b S_b$. The pressure may also be computed analytically and is found to satisfy the usual identity $\mathcal{F}_b = -P_b$. Moreover, as $a \to 0$, all these results continuously approach those of the standard four-dimensional planar Schwarzschild black hole, providing an additional consistency check of our solutions. Also note that the conserved energy $M_b$ is proportional to the integration constant $C_b$, implying further the primary nature of black hole hair.

\begin{figure}[h!]
\begin{minipage}[b]{0.5\linewidth}
\centering
\includegraphics[width=2.8in,height=2.3in]{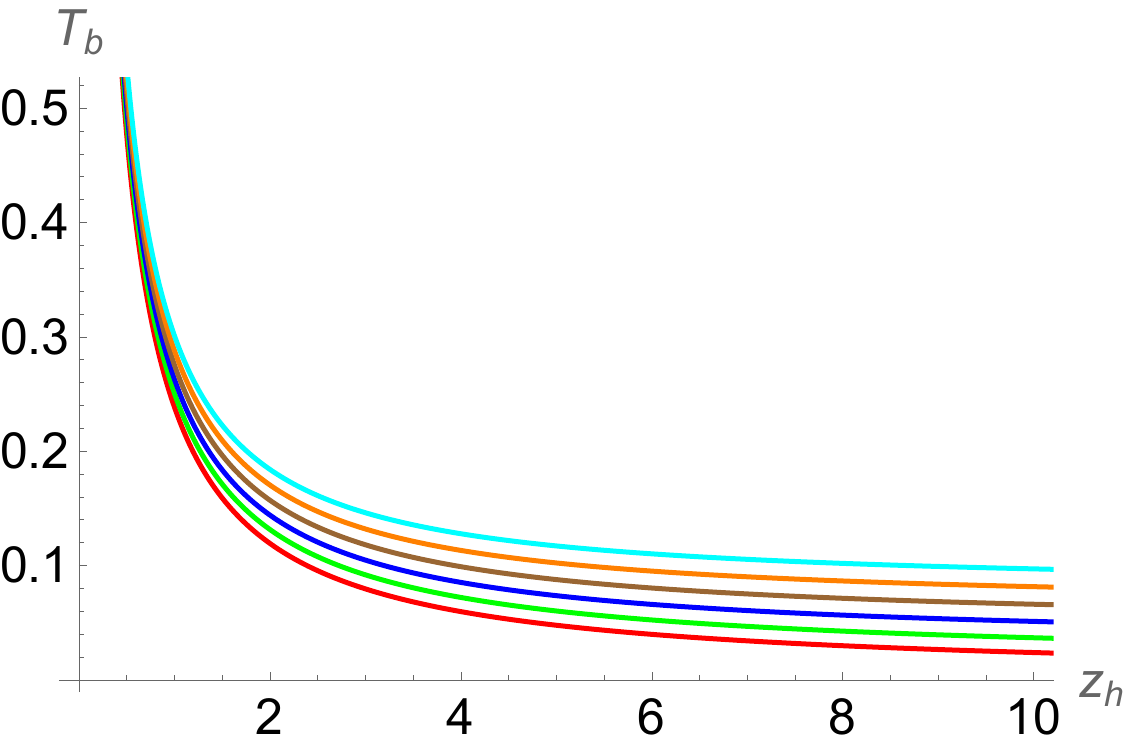}
\caption{ \small Hawking temperature $T_b$ as a function of horizon radius $z_h$ for various values of $a$. Here $G_4=1$ is used. Red, green, blue, brown, orange, and cyan curves correspond to $a=0$, $0.1$, $0.2$, $0.3$, $0.4$, and $0.5$, respectively.}
\label{zhvsTempBHn1D4}
\end{minipage}
\hspace{0.4cm}
\begin{minipage}[b]{0.5\linewidth}
\centering
\includegraphics[width=2.8in,height=2.3in]{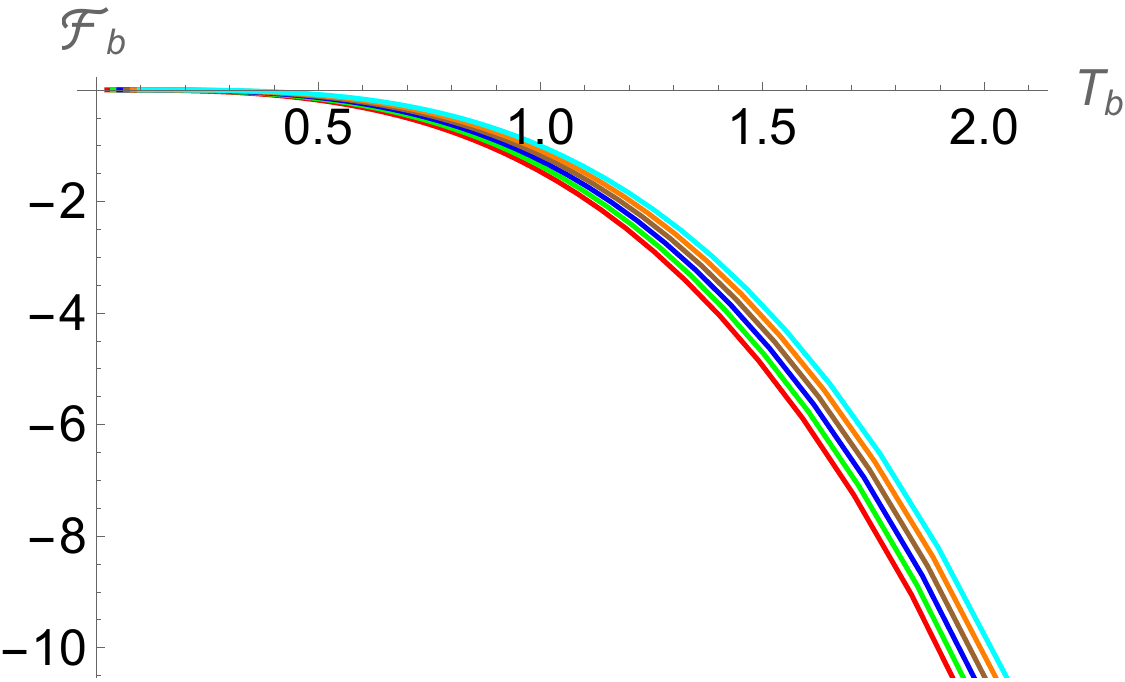}
\caption{\small Free energy $\mathcal{F}_b$ of the hairy black hole as a function of temperature $T_b$ for various values of $a$. Here $G_4=1$ and $V_1=1$ are used. Red, green, blue, brown, orange, and cyan curves correspond to $a=0$, $0.1$, $0.2$, $0.3$, $0.4$, and $0.5$, respectively.}
\label{TempvsFreeBHn1D4}
\end{minipage}
\end{figure}

The thermodynamic phase structure of the above constructed hairy black hole is shown in Figs.~\ref{zhvsTempBHn1D4} and \ref{TempvsFreeBHn1D4}. Like in its five-dimensional counterpart, there exists a single black hole branch. This branch again has a positive specific heat, with its entropy and temperature decreasing with $z_h$, implying that the constructed hairy black hole solutions are locally thermodynamically stable. Moreover, for a fixed horizon radius, the temperature of the black hole increases with $a$. We further analyze its free energy behavior to probe its global stability. The results are shown in Fig.~\ref{TempvsFreeBHn1D4}, where the thermal profile of the hairy black hole free energy for different values of $a$ is shown. We observe that the free energy is always negative. This indicates that the hairy black hole phase always has lower free energy than the thermal-AdS, and is thermodynamically favored at all temperatures.

\subsection{Thermodynamics of the hairy soliton for $A(z)=-a z$}
For the hairy AdS soliton, the near-boundary expansion and the associated scalar mass are identical to those given in Eq.~(\ref{boundaryexpn1D4}) after replacing $z_h$ with $z_0$. The individual contributions to the regularized Euclidean action in Eq.~(\ref{countertermn1D4}) are given by
\begin{eqnarray}
S_{ES}^{s} & = & \frac{\beta_s L_s V_1}{16 \pi G_4} \left(\frac{e^{-2 a z} \left(4 a^4 z+4 a^3+a z \left(e^{2 a z}+1\right)
   C_s-\left(e^{2 a z}-1\right) C_s\right)}{2 a^3 z^3}  \right)\bigg\rvert_{z=\epsilon}^{z=z_0} \,, \\
S_{GH}^{s} & = &  \frac{\beta_s L_s V_1}{16 \pi G_4} \left( 4 a^3-\frac{6 a}{\epsilon^2}-C_s+\frac{6}{\epsilon^3} \right)\,, \\
S_{BK}^{s} & = & \frac{\beta_s L_s V_1}{16 \pi G_4} \left( 18 a^3-\frac{18 a^2}{\epsilon}+\frac{12 a}{\epsilon^2}+\frac{2 C_s}{3}-\frac{4}{\epsilon^3}  \right) \,,\\
S_{ct}^{s} & = & \frac{\beta_s L_s V_1}{16 \pi G_4} \left(-\frac{62 a^3}{3}+\frac{18 a^2}{\epsilon}-\frac{8 a}{\epsilon^2}  \right) \,,
\label{countertermexpsoln1D4}
\end{eqnarray}
with $C_s$ given by
\begin{eqnarray}
C_s & = & \frac{4 a^3}{e^{2 a z_0} \left(2 a z_0 \left(a z_0-1\right)+1\right)-1} \,,
\label{Cssoln1D4}
\end{eqnarray}
The renormalized free energy of the hairy soliton is then obtained as
\begin{eqnarray}
\mathcal{F}_s & = & - \frac{L_s V_1 C_s}{48 \pi G_4} \,, \nonumber \\
 & = & -\frac{a^3 L_s V_1}{12 \pi  G_4 \left(e^{2 a z_0} \left(2 a z_0 \left(a
   z_0-1\right)+1\right)-1\right)}\,.
\label{freeenergyexpsoln1D4}
\end{eqnarray}
And the conserved energy associated with the hairy soliton is
\begin{eqnarray}
 M_s & = & - \frac{L_s V_1 C_s}{48 \pi G_4} \,.
\label{massexpsoln1D4}
\end{eqnarray}
This expression is again in agreement with Eq.~(\ref{freeenergyexpsoln1D4}), i.e., $\mathcal{F}_s=M_s$. We see that, analogous to the five-dimensional case, the energy of the hairy soliton is again negative. We may further contrast the energy of the hairy soliton with that of the corresponding hairy black hole through $\Delta M = M_b - M_s$. The fact that $\Delta M$ is always positive shows that the Horowitz-Myers conjecture continues to hold in the present setup as well.

\subsection{Phase transition between hairy black hole and soliton for $A(z)=-a z$}
\begin{figure}[h!]
\centering
\includegraphics[width=3.6in,height=2.6in]{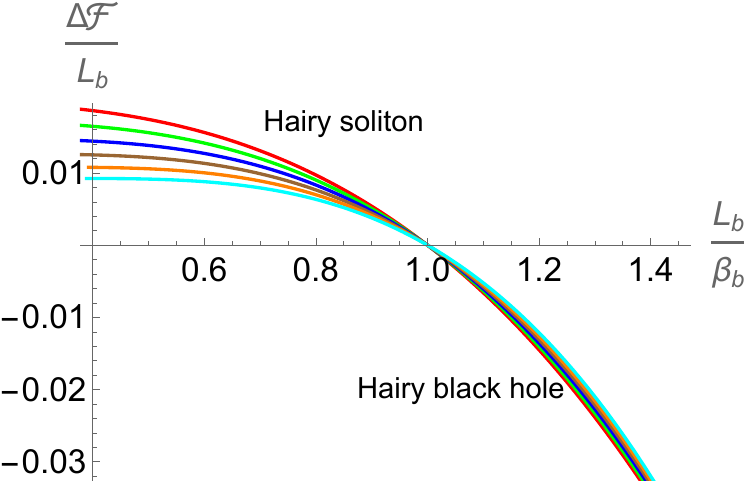}
\caption{ \small Free energy difference $\Delta\mathcal{F}$ as a function of periodicity ratio $L_b/\beta_b$ for various values of $a$. Here $G_4=1$, $V_1=1$, and $z_0=1$ are used. Red, green, blue, brown, orange, and cyan curves correspond to $a=0$, $0.1$, $0.2$, $0.3$, $0.4$, and $0.5$, respectively.}
\label{Lbbybetabvsfreediffvsan1D4}
\end{figure}
\begin{figure}[h!]
\centering
\includegraphics[width=3.8in,height=2.8in]{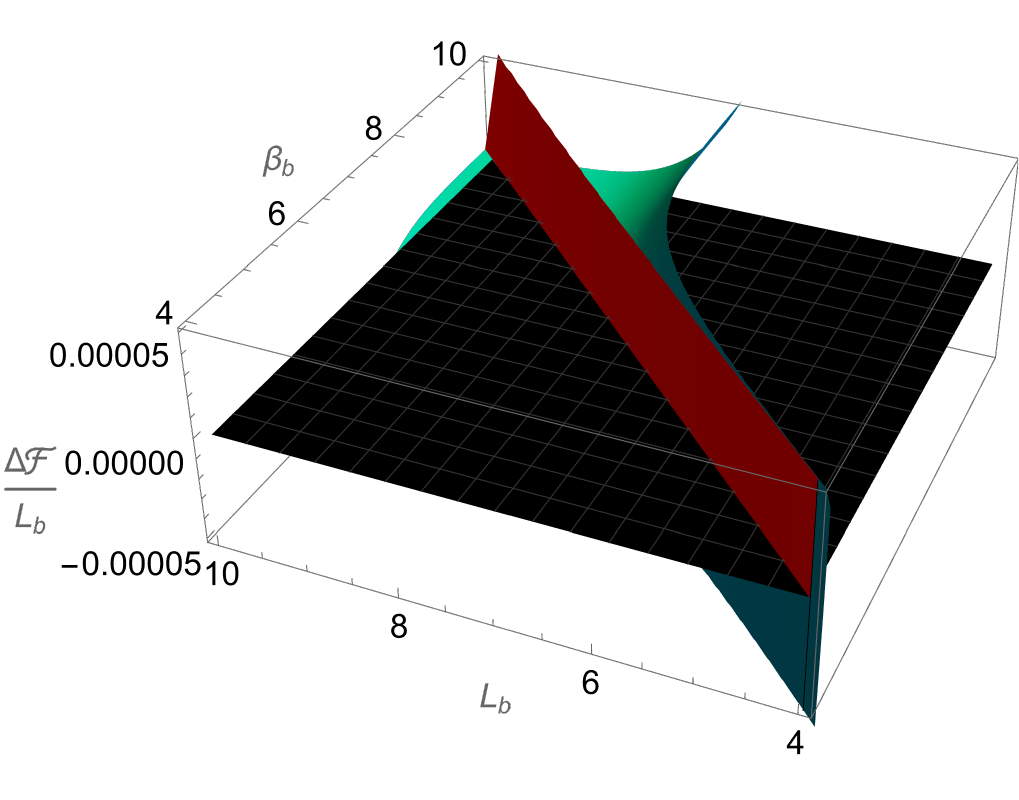}
\caption{ \small Free energy difference $\Delta\mathcal{F}$ as a function of $L_b$ and $\beta_b$ for $a=0.1$ (red surface) and $a=0.5$ (cyan surface). The black plane indicates $\Delta\mathcal{F}=0$ surface.  Here $G_4=1$ and $V_1=1$ are used.}
\label{Lbbybetabvsfreediffvs3Dn1D4}
\end{figure}

%
The free energy difference $\Delta \mathcal{F}$ between the hairy black hole and the hairy soliton is given by
\begin{eqnarray}
\Delta \mathcal{F} &=& \frac{L_b V_1}{48 \pi G_4}\left(C_s - C_b \right)\,,
\end{eqnarray}
and is displayed in Fig.~\ref{Lbbybetabvsfreediffvsan1D4}. 
In obtaining these results, we have again imposed equal temporal and spatial periodicities, 
$\{\beta_b=\beta_s,~L_b=L_s\}$, and for concreteness we set $z_0=1$. We observe that the resulting phase structure closely resembles the five-dimensional case. 
In particular, the dominant phase changes as the periodic ratio $L_b/\beta_b$ is varied from small to large values.  The hairy black hole phase is thermodynamically preferred in the regime $\beta_b < L_b$, 
whereas the hairy soliton phase becomes favored when $L_b < \beta_b$. 
The transition occurs precisely at $\beta_b = L_b$, independent of the value of the hair parameter $a$. Similarly, to complete our discussion, we present in Fig.~\ref{Lbbybetabvsfreediffvs3Dn1D4} a three-dimensional plot of $\Delta\mathcal{F}$. The free energy surface again intersects the $\Delta\mathcal{F}=0$ plane along the straight line $\beta_b = L_b$.

\section{Conclusions}
\label{conclusions}
In this paper, we have analytically constructed a family of black hole and soliton solutions with primary scalar hair in the Einstein-scalar gravity system. We solved the coupled Einstein-scalar equations of motion analytically and obtained an
infinite family of exact scalar hair black hole and soliton solutions in all spacetime dimensions. The obtained gravity solution is expressed in terms of a scale function $A(z)$, and based on its relevance in holographic QCD, we considered its form to be $A(z)=-a z^n$. The parameter $a$ controls the strength of the scalar hair, and in the limit $a\rightarrow 0$, all the hairy black hole and soliton solutions reduce to their nonhairy counterpart. Importantly, the constructed solutions are smooth, have finite curvature, and are devoid of any additional singularities. Our solutions therefore add to the relatively small class of hairy AdS solitons with regular scalar-field profiles.

We then analyzed the thermodynamics phase structure of the constructed hairy solutions. We obtained analytic expressions of the Gibbs free energy, mass, and other thermodynamic observables of hairy black holes and solitons using the holographic renormalization procedure, and found that they
satisfy the standard thermodynamic relations. In all cases, the energy of the hairy soliton turned out to be negative, indicating it to be the ground state of the theory. Interestingly, we further found that the free energy of the hairy black hole and soliton exchange dominates as the ratio $L_b/\beta_b$ varies from low to high values, indicating a first-order phase transition between these two solutions. The transition occurred at $\beta_b=L_b$, with the soliton phase dominating the phase structure when $L_b<\beta_b$, while the black hole phase dominates when $L_b>\beta_b$. This result remains true irrespective of the form of $A(z)$ considered in this work, as well as in different dimensions.

It is also worth emphasizing an important aspect that is particularly relevant for holographic QCD. The reason for considering the soliton backgrounds is not that confinement cannot already be realized within the potential reconstruction framework through an appropriate choice of $A(z)$. Indeed, suitable choices of $A(z)$ can generate confining behavior without the need for compactifying a spatial direction. For instance, one can get a confining phase by considering the deformed thermal-AdS background (\ref{thermalAds}) by choosing $A(z)=-a z^2$ (or other variant of $A(z)$). The motivation for considering the AdS soliton is instead that it provides an alternative regular, horizonless solution of the same Einstein-scalar system with the same asymptotic structure. Since both the deformed thermal-AdS geometry and the AdS soliton can represent confining phases of the dual theory, one should identify which of these backgrounds corresponds to the lowest-energy state of the system. In such a situation, it is natural to regard the geometry with lower energy as the physically preferred background. As we have explicitly demonstrated in our work, the AdS soliton consistently has lower free energy than the corresponding deformed thermal-AdS background. Therefore, although both geometries exhibit confining behavior, the soliton solution emerges as the more favorable background and naturally represents the relevant low-energy phase of the system. For this reason, it is worthwhile to explore its properties in greater detail.

We also believe that the conclusion that the AdS soliton has lower free energy than the corresponding thermal-AdS background is likely to extend to soft-wall AdS/QCD models as well. Although soft-wall models are not derived as fully consistent gravitational solutions, their confining backgrounds are typically based on geometries of the form given in Eq.~(\ref{thermalAds}). Therefore, the same reasoning may apply there too. If this expectation is correct, it raises an interesting question regarding the identification of the appropriate confined phase in soft-wall models, since the AdS soliton may be energetically preferred over the commonly used deformed thermal-AdS background.

There are many directions in which the present work can be extended. It would be interesting to extend this work by adding a $U(1)$ charge and construct examples of hairy charged solitons. We expect that, analogous to the nonhairy situation, the addition of $U(1)$ charge and the interplay of it with the scalar hair would make the thermodynamic phase structure between black hole and soliton even more interesting. One could also probe interesting questions related to the QCD confined phase, such as its chaotic and integrable features, or questions related to transport coefficients, using the hairy soliton phase. Similarly, following \cite{Berenguer:2026ftk}, it would be interesting to construct a one-parameter family of Euclidean geometries interpolating continuously between the soliton and the black hole to study how confinement can be switched off continuously. Another direction would be to see confinement/deconfinement transition and scrambling in soliton/black hole setting \cite{Nongmaithem:2026rop, Yang:2023qxx, Chakraborty:2025crb}. Following \cite{Shaghoulian:2015lcn, BravoGaete:2017dso}, it would also be interesting to find a possible relation between the soliton energy and the black hole entropy through a higher-dimensional Cardy-like formula.

\section*{Acknowledgments}
The work of S.M.~is supported by the core research grant from the Science and Engineering Research Board (now Anusandhan National Research Foundation), a statutory body under the Department of Science and Technology, Government of India, under grant agreement number CRG/2023/007670. 

\section*{Appendix A: Hairy black hole and soliton solutions in $D=6$}

This appendix contains the details of the $D=6$ solutions. The $D=6$ case is interesting as the dual boundary theory of the hairy soliton background lives in $\mathbb{R}^1\times\mathbb{R}^3\times\mathbb{S}^1$, which is more appropriate for the QCD confined phase in four extended dimensions. For simplicity, we mainly concentrate on $A(z)=-a z$ here. Since the geometric and thermodynamic computation of hairy black hole and soliton is completely analogous to $D=4$ and $D=5$ cases, we will be very brief here and mainly present the essential results. The expressions for the blackening function $g_b(z)$ and
the scalar field $\phi(z)$ are given by
\begin{eqnarray}
 g_b(z) & = & 1-C_b \frac{ \left(e^{4 a z} (4 a z (2 a z (4 a z (a z-1)+3)-3)+3)-3\right)
   }{128 a^5},  \nonumber \\
 \phi(z)  & = & 2 \sqrt{2} \left(\sqrt{z} \sqrt{a \left(a+\frac{2}{z}\right)}
   \left(\sqrt{z}-\frac{2 \log \left(\sqrt{a z+2}-\sqrt{a}
   \sqrt{z}\right)}{\sqrt{a} \sqrt{a z+2}}\right)+\log (2)\right)\,,
\label{gbsoln1D6}
\end{eqnarray}
where
\begin{eqnarray}
C_b  & = & \frac{128 a^5}{e^{4 a z_h} \left(4 a z_h \left(2 a z_h \left(4 a z_h \left(a
   z_h-1\right)+3\right)-3\right)+3\right)-3} \,.
   \label{inteconstD6n1}
\label{Cbsoln1D6}
\end{eqnarray}
The analogous expressions for the soliton background can be obtained by replacing $z_h$ with $z_0$. The hairy geometry is again regular everywhere. 

Using the holographic renormalization, we can similarly obtain the expressions of various thermodynamic quantities. For the black hole background, these are given by: 
\begin{eqnarray}
& & T_b = \frac{C_b z_h^4 e^{4 a z_h}}{4 \pi }\,, ~~~
S_b  = \frac{L_b V_3 e^{-4 a z_h}}{4 G_6 z_h^4}\,,~~~
M_b = \frac{L_b V_3 C_b}{20 \pi  G_6}\,,~~~\mathcal{F}_b = - \frac{L_b V_3 C_b}{80 \pi  G_6}\,,~~~P_b=-\mathcal{F}_b\,.
\label{tempentn1D6 }
\end{eqnarray}
These expressions satisfy the relation $\mathcal{F}_b=M_b - T_b S_b$. These thermodynamic expressions further suggest that there is only one black hole branch, which is locally stable at all temperatures. 

\begin{figure}[h!]
\begin{minipage}[b]{0.5\linewidth}
\centering
\includegraphics[width=2.8in,height=2.3in]{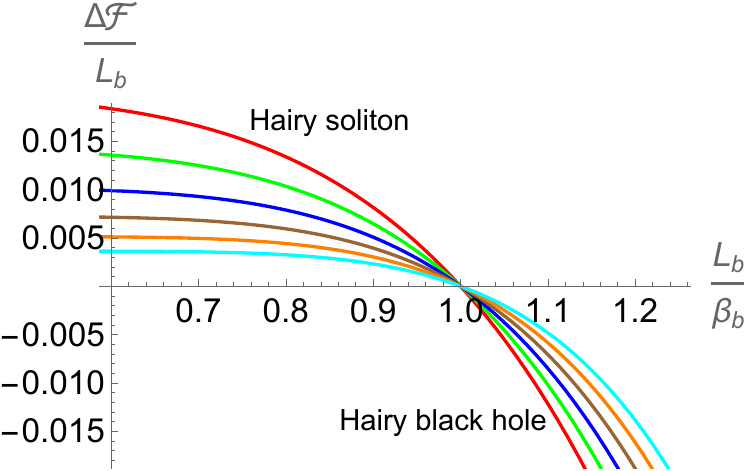}
\caption{ \small Free energy difference $\Delta\mathcal{F}$ as a function of periodicity ratio $L_b/\beta_b$ for various values of $a$. Here $G_6=1$, $V_3=1$, and $z_0=1$ are used. Red, green, blue, brown, orange, and cyan curves correspond to $a=0$, $0.1$, $0.2$, $0.3$, $0.4$, and $0.5$, respectively.}
\label{Lbbybetabvsfreediffvsan1D6}
\end{minipage}
\hspace{0.4cm}
\begin{minipage}[b]{0.5\linewidth}
\centering
\includegraphics[width=2.8in,height=2.3in]{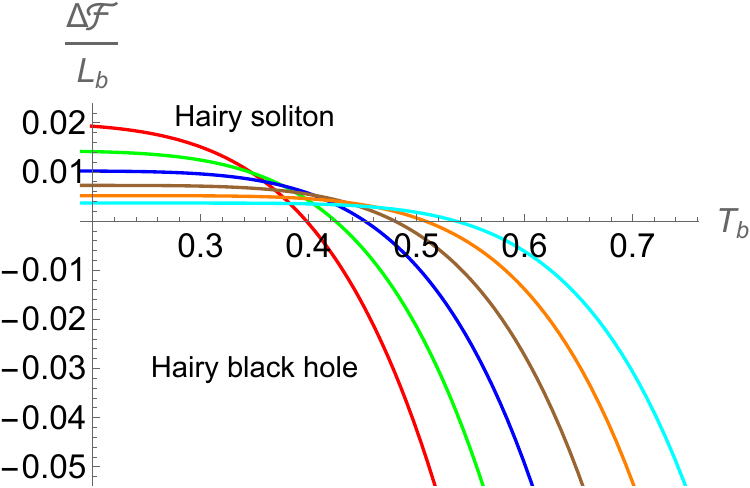}
\caption{\small Free energy difference $\Delta\mathcal{F}$ as a function of black hole temperature $T_b$ for various values of $a$. Here $G_6=1$, $V_3=1$, and $z_0=1$ are used. Red, green, blue, brown, orange, and cyan curves correspond to $a=0$, $0.1$, $0.2$, $0.3$, $0.4$, and $0.5$, respectively.}
\label{Tempvsfreediffvsan1D6}
\end{minipage}
\end{figure}

For the hairy soliton, we similarly have
\begin{eqnarray}
& & M_s = \mathcal{F}_s = - \frac{L_s V_3 C_s}{80 \pi  G_6}\,,
\end{eqnarray}
The energy difference $\Delta M = M_b - M_s$ is again always positive, showing that the Horowitz-Myers conjecture continues to hold.

The free energy difference between the hairy black hole and the soliton is now given by
\begin{eqnarray}
\Delta \mathcal{F} &=& \frac{L_b V_3}{80 \pi G_6}\left(C_s - C_b \right)\,,
\end{eqnarray}
and its behavior is displayed in Figs.~\ref{Lbbybetabvsfreediffvsan1D6} and \ref{Tempvsfreediffvsan1D6}. The resulting phase diagram is quite similar to the $D=5$ case, with the black hole phase dominating the phase diagram at high temperatures, while the soliton phase dominates at low temperatures.

\bibliography{biblio.bib}

\begin{thebibliography}{100}

\bibitem{Hawking:1982dh}
S.~W. Hawking and Don~N. Page.
\newblock {Thermodynamics of Black Holes in anti-De Sitter Space}.
\newblock {\em Commun. Math. Phys.}, 87:577, 1983.

\bibitem{Maldacena:1997re}
Juan~Martin Maldacena.
\newblock {The Large $N$ limit of superconformal field theories and
  supergravity}.
\newblock {\em Adv. Theor. Math. Phys.}, 2:231--252, 1998.

\bibitem{Gubser:1998bc}
S.~S. Gubser, Igor~R. Klebanov, and Alexander~M. Polyakov.
\newblock {Gauge theory correlators from noncritical string theory}.
\newblock {\em Phys. Lett. B}, 428:105--114, 1998.

\bibitem{Witten:1998qj}
Edward Witten.
\newblock {Anti de Sitter space and holography}.
\newblock {\em Adv. Theor. Math. Phys.}, 2:253--291, 1998.

\bibitem{Lemos:1994xp}
J.~P.~S. Lemos.
\newblock {Cylindrical black hole in general relativity}.
\newblock {\em Phys. Lett. B}, 353:46--51, 1995.

\bibitem{Friedman:1993ty}
John~L. Friedman, Kristin Schleich, and Donald~M. Witt.
\newblock {Topological censorship}.
\newblock {\em Phys. Rev. Lett.}, 71:1486--1489, 1993.
\newblock [Erratum: Phys.Rev.Lett. 75, 1872 (1995)].

\bibitem{Cai:1996eg}
Rong-Gen Cai and Yuan-Zhong Zhang.
\newblock {Black plane solutions in four-dimensional space-times}.
\newblock {\em Phys. Rev. D}, 54:4891--4898, 1996.

\bibitem{Vanzo:1997gw}
Luciano Vanzo.
\newblock {Black holes with unusual topology}.
\newblock {\em Phys. Rev. D}, 56:6475--6483, 1997.

\bibitem{Birmingham:1998nr}
Danny Birmingham.
\newblock {Topological black holes in Anti-de Sitter space}.
\newblock {\em Class. Quant. Grav.}, 16:1197--1205, 1999.

\bibitem{Mann:1996gj}
Robert~B. Mann.
\newblock {Pair production of topological anti-de Sitter black holes}.
\newblock {\em Class. Quant. Grav.}, 14:L109--L114, 1997.

\bibitem{Brill:1997mf}
Dieter~R. Brill, Jorma Louko, and Peter Peldan.
\newblock {Thermodynamics of (3+1)-dimensional black holes with toroidal or
  higher genus horizons}.
\newblock {\em Phys. Rev. D}, 56:3600--3610, 1997.

\bibitem{Priyadarshinee:2021rch}
Supragyan Priyadarshinee, Subhash Mahapatra, and Indrani Banerjee.
\newblock {Analytic topological hairy dyonic black holes and thermodynamics}.
\newblock {\em Phys. Rev. D}, 104(8):084023, 2021.

\bibitem{Sheykhi:2007wg}
Ahmad Sheykhi.
\newblock {Thermodynamics of charged topological dilaton black holes}.
\newblock {\em Phys. Rev. D}, 76:124025, 2007.

\bibitem{Witten:1998zw}
Edward Witten.
\newblock {Anti-de Sitter space, thermal phase transition, and confinement in
  gauge theories}.
\newblock {\em Adv. Theor. Math. Phys.}, 2:505--532, 1998.

\bibitem{Chamblin:1999tk}
Andrew Chamblin, Roberto Emparan, Clifford~V. Johnson, and Robert~C. Myers.
\newblock {Charged AdS black holes and catastrophic holography}.
\newblock {\em Phys. Rev. D}, 60:064018, 1999.

\bibitem{Chamblin:1999hg}
Andrew Chamblin, Roberto Emparan, Clifford~V. Johnson, and Robert~C. Myers.
\newblock {Holography, thermodynamics and fluctuations of charged AdS black
  holes}.
\newblock {\em Phys. Rev. D}, 60:104026, 1999.

\bibitem{Cvetic:1999ne}
Mirjam Cvetic and Steven~S. Gubser.
\newblock {Phases of R charged black holes, spinning branes and strongly
  coupled gauge theories}.
\newblock {\em JHEP}, 04:024, 1999.

\bibitem{Kubiznak:2012wp}
David Kubiznak and Robert~B. Mann.
\newblock {P-V criticality of charged AdS black holes}.
\newblock {\em JHEP}, 07:033, 2012.

\bibitem{Caldarelli:1999xj}
Marco~M. Caldarelli, Guido Cognola, and Dietmar Klemm.
\newblock {Thermodynamics of Kerr-Newman-AdS black holes and conformal field
  theories}.
\newblock {\em Class. Quant. Grav.}, 17:399--420, 2000.

\bibitem{Kubiznak:2016qmn}
David Kubiznak, Robert~B. Mann, and Mae Teo.
\newblock {Black hole chemistry: thermodynamics with Lambda}.
\newblock {\em Class. Quant. Grav.}, 34(6):063001, 2017.

\bibitem{Gunasekaran:2012dq}
Sharmila Gunasekaran, Robert~B. Mann, and David Kubiznak.
\newblock {Extended phase space thermodynamics for charged and rotating black
  holes and Born-Infeld vacuum polarization}.
\newblock {\em JHEP}, 11:110, 2012.

\bibitem{Hendi:2012um}
S.~H. Hendi and M.~H. Vahidinia.
\newblock {Extended phase space thermodynamics and P-V criticality of black
  holes with a nonlinear source}.
\newblock {\em Phys. Rev. D}, 88(8):084045, 2013.

\bibitem{Zou:2013owa}
De-Cheng Zou, Shao-Jun Zhang, and Bin Wang.
\newblock {Critical behavior of Born-Infeld AdS black holes in the extended
  phase space thermodynamics}.
\newblock {\em Phys. Rev. D}, 89(4):044002, 2014.

\bibitem{Bohra:2020qom}
Hardik Bohra, David Dudal, Ali Hajilou, and Subhash Mahapatra.
\newblock {Chiral transition in the probe approximation from an
  Einstein-Maxwell-dilaton gravity model}.
\newblock {\em Phys. Rev. D}, 103(8):086021, 2021.

\bibitem{Jena:2024cqs}
Siddhi~Swarupa Jena, Jyotirmoy Barman, Bruno Toniato, David Dudal, and Subhash
  Mahapatra.
\newblock {A dynamical Einstein-Born-Infeld-dilaton model and holographic
  quarkonium melting in a magnetic field}.
\newblock {\em JHEP}, 12:096, 2024.

\bibitem{Jena:2025xcf}
Siddhi~Swarupa Jena, Arpan Bhattacharjee, David Dudal, and Subhash Mahapatra.
\newblock {Probing quarkonium diffusion in a magnetized quark-gluon plasma}.
\newblock {\em Phys. Rev. D}, 112(8):086010, 2025.

\bibitem{Surya:2001vj}
Sumati Surya, Kristin Schleich, and Donald~M. Witt.
\newblock {Phase transitions for flat AdS black holes}.
\newblock {\em Phys. Rev. Lett.}, 86:5231--5234, 2001.

\bibitem{Horowitz:1998ha}
Gary~T. Horowitz and Robert~C. Myers.
\newblock {The AdS / CFT correspondence and a new positive energy conjecture
  for general relativity}.
\newblock {\em Phys. Rev. D}, 59:026005, 1998.

\bibitem{Galloway:2001uv}
G.~J. Galloway, S.~Surya, and E.~Woolgar.
\newblock {A Uniqueness theorem for the AdS soliton}.
\newblock {\em Phys. Rev. Lett.}, 88:101102, 2002.

\bibitem{Galloway:2002ai}
G.~J. Galloway, S.~Surya, and E.~Woolgar.
\newblock {On the geometry and mass of static, asymptotically AdS space-times,
  and the uniqueness of the AdS soliton}.
\newblock {\em Commun. Math. Phys.}, 241:1--25, 2003.

\bibitem{Woolgar:2016axs}
Eric Woolgar.
\newblock {The rigid Horowitz-Myers conjecture}.
\newblock {\em JHEP}, 03:104, 2017.

\bibitem{Nishioka:2006gr}
Tatsuma Nishioka and Tadashi Takayanagi.
\newblock {AdS Bubbles, Entropy and Closed String Tachyons}.
\newblock {\em JHEP}, 01:090, 2007.

\bibitem{Nishioka:2009zj}
Tatsuma Nishioka, Shinsei Ryu, and Tadashi Takayanagi.
\newblock {Holographic Superconductor/Insulator Transition at Zero
  Temperature}.
\newblock {\em JHEP}, 03:131, 2010.

\bibitem{Anabalon:2021tua}
Andres Anabalon and Simon~F. Ross.
\newblock {Supersymmetric solitons and a degeneracy of solutions in AdS/CFT}.
\newblock {\em JHEP}, 07:015, 2021.

\bibitem{Banerjee:2007by}
Nabamita Banerjee and Suvankar Dutta.
\newblock {Phase transition of electrically charged Ricci-flat black holes}.
\newblock {\em JHEP}, 07:047, 2007.

\bibitem{Anabalon:2022ksf}
Andr{\'e}s Anabal{\'o}n, Patrick Concha, Julio Oliva, Constanza Quijada, and
  Evelyn Rodr{\'\i}guez.
\newblock {Phase transitions for charged planar solitons in AdS}.
\newblock {\em Phys. Lett. B}, 835:137521, 2022.

\bibitem{Quijada:2023fkc}
Constanza Quijada, Andr{\'e}s Anabal{\'o}n, Robert~B. Mann, and Julio Oliva.
\newblock {Triple points of gravitational AdS solitons and black holes}.
\newblock {\em Phys. Rev. D}, 110(2):L021902, 2024.

\bibitem{Cai:2007wz}
Rong-Gen Cai, Sang~Pyo Kim, and Bin Wang.
\newblock {Ricci flat black holes and Hawking-Page phase transition in
  Gauss-Bonnet gravity and dilaton gravity}.
\newblock {\em Phys. Rev. D}, 76:024011, 2007.

\bibitem{Anabalon:2019tcy}
Andres Anabalon, Dumitru Astefanesei, David Choque, and Jose~D. Edelstein.
\newblock {Phase transitions of neutral planar hairy AdS black holes}.
\newblock {\em JHEP}, 07:129, 2020.

\bibitem{Gubser:2008px}
Steven~S. Gubser.
\newblock {Breaking an Abelian gauge symmetry near a black hole horizon}.
\newblock {\em Phys. Rev. D}, 78:065034, 2008.

\bibitem{Hartnoll:2008vx}
Sean~A. Hartnoll, Christopher~P. Herzog, and Gary~T. Horowitz.
\newblock {Building a Holographic Superconductor}.
\newblock {\em Phys. Rev. Lett.}, 101:031601, 2008.

\bibitem{Gubser:2008ny}
Steven~S. Gubser and Abhinav Nellore.
\newblock {Mimicking the QCD equation of state with a dual black hole}.
\newblock {\em Phys. Rev. D}, 78:086007, 2008.

\bibitem{DeWolfe:2010he}
Oliver DeWolfe, Steven~S. Gubser, and Christopher Rosen.
\newblock {A holographic critical point}.
\newblock {\em Phys. Rev. D}, 83:086005, 2011.

\bibitem{Gursoy:2008za}
U.~Gursoy, E.~Kiritsis, L.~Mazzanti, and F.~Nitti.
\newblock {Holography and Thermodynamics of 5D Dilaton-gravity}.
\newblock {\em JHEP}, 05:033, 2009.

\bibitem{Jarvinen:2011qe}
Matti Jarvinen and Elias Kiritsis.
\newblock {Holographic Models for QCD in the Veneziano Limit}.
\newblock {\em JHEP}, 03:002, 2012.

\bibitem{Rougemont:2023gfz}
Romulo Rougemont, Joaquin Grefa, Mauricio Hippert, Jorge Noronha, Jacquelyn
  Noronha-Hostler, Israel Portillo, and Claudia Ratti.
\newblock {Hot QCD phase diagram from holographic
  Einstein{\textendash}Maxwell{\textendash}Dilaton models}.
\newblock {\em Prog. Part. Nucl. Phys.}, 135:104093, 2024.

\bibitem{Astefanesei:2023sep}
Dumitru Astefanesei, Paulina Cabrera, Robert~B. Mann, and Ra{\'u}l Rojas.
\newblock {Extended phase space thermodynamics for hairy black holes}.
\newblock {\em Phys. Rev. D}, 108(10):104047, 2023.

\bibitem{Guo:2021ere}
Guangzhou Guo, Peng Wang, Houwen Wu, and Haitang Yang.
\newblock {Thermodynamics and phase structure of an Einstein-Maxwell-scalar
  model in extended phase space}.
\newblock {\em Phys. Rev. D}, 105(6):064069, 2022.

\bibitem{Giribet:2014fla}
Gaston Giribet, Andres Goya, and Julio Oliva.
\newblock {Different phases of hairy black holes in AdS5 space}.
\newblock {\em Phys. Rev. D}, 91(4):045031, 2015.

\bibitem{Hennigar:2015wxa}
Robie~A. Hennigar and Robert~B. Mann.
\newblock {Reentrant Phase Transitions and van der Waals Behaviour for Hairy
  Black Holes}.
\newblock {\em Entropy}, 17(12):8056--8072, 2015.

\bibitem{Ruffini:1971bza}
Remo Ruffini and John~A. Wheeler.
\newblock {Introducing the black hole}.
\newblock {\em Phys. Today}, 24(1):30, 1971.

\bibitem{Bekenstein:1971hc}
Jacob~D. Bekenstein.
\newblock {Nonexistence of baryon number for static black holes}.
\newblock {\em Phys. Rev. D}, 5:1239--1246, 1972.

\bibitem{Israel:1967wq}
Werner Israel.
\newblock {Event horizons in static vacuum space-times}.
\newblock {\em Phys. Rev.}, 164:1776--1779, 1967.

\bibitem{Wald:1971iw}
Robert~M. Wald.
\newblock {Final states of gravitational collapse}.
\newblock {\em Phys. Rev. Lett.}, 26:1653--1655, 1971.

\bibitem{Carter:1971zc}
B.~Carter.
\newblock {Axisymmetric Black Hole Has Only Two Degrees of Freedom}.
\newblock {\em Phys. Rev. Lett.}, 26:331--333, 1971.

\bibitem{Robinson:1975bv}
D.~C. Robinson.
\newblock {Uniqueness of the Kerr black hole}.
\newblock {\em Phys. Rev. Lett.}, 34:905--906, 1975.

\bibitem{Mazur:1982db}
P.~O. Mazur.
\newblock {PROOF OF UNIQUENESS OF THE KERR-NEWMAN BLACK HOLE SOLUTION}.
\newblock {\em J. Phys. A}, 15:3173--3180, 1982.

\bibitem{Teitelboim:1972qx}
C.~Teitelboim.
\newblock {Nonmeasurability of the quantum numbers of a black hole}.
\newblock {\em Phys. Rev. D}, 5:2941--2954, 1972.

\bibitem{Volkov:1990sva}
M.~S. Volkov and D.~V. Galtsov.
\newblock {Black holes in Einstein Yang-Mills theory. (In Russian)}.
\newblock {\em Sov. J. Nucl. Phys.}, 51:747--753, 1990.

\bibitem{Bizon:1990sr}
P.~Bizon.
\newblock {Colored black holes}.
\newblock {\em Phys. Rev. Lett.}, 64:2844--2847, 1990.

\bibitem{Kuenzle:1990is}
H.~P. Kuenzle and A.~K.~M. Masood-ul Alam.
\newblock {Spherically symmetric static SU(2) Einstein Yang-Mills fields}.
\newblock {\em J. Math. Phys.}, 31:928--935, 1990.

\bibitem{Garfinkle:1990qj}
David Garfinkle, Gary~T. Horowitz, and Andrew Strominger.
\newblock {Charged black holes in string theory}.
\newblock {\em Phys. Rev. D}, 43:3140, 1991.
\newblock [Erratum: Phys.Rev.D 45, 3888 (1992)].

\bibitem{Greene:1992fw}
Brian~R. Greene, Samir~D. Mathur, and Christopher~M. O'Neill.
\newblock {Eluding the no hair conjecture: Black holes in spontaneously broken
  gauge theories}.
\newblock {\em Phys. Rev. D}, 47:2242--2259, 1993.

\bibitem{Torii:1993vm}
Takashi Torii and Kei-ichi Maeda.
\newblock {Black holes with nonAbelian hair and their thermodynamical
  properties}.
\newblock {\em Phys. Rev. D}, 48:1643--1651, 1993.

\bibitem{Herdeiro:2014goa}
Carlos A.~R. Herdeiro and Eugen Radu.
\newblock {Kerr black holes with scalar hair}.
\newblock {\em Phys. Rev. Lett.}, 112:221101, 2014.

\bibitem{Berti:2013gfa}
Emanuele Berti, Vitor Cardoso, Leonardo Gualtieri, Michael Horbatsch, and
  Ulrich Sperhake.
\newblock {Numerical simulations of single and binary black holes in
  scalar-tensor theories: circumventing the no-hair theorem}.
\newblock {\em Phys. Rev. D}, 87(12):124020, 2013.

\bibitem{Ovalle:2020kpd}
J.~Ovalle, R.~Casadio, E.~Contreras, and A.~Sotomayor.
\newblock {Hairy black holes by gravitational decoupling}.
\newblock {\em Phys. Dark Univ.}, 31:100744, 2021.

\bibitem{Mahapatra:2022xea}
Subhash Mahapatra and Indrani Banerjee.
\newblock {Rotating hairy black holes and thermodynamics from gravitational
  decoupling}.
\newblock {\em Phys. Dark Univ.}, 39:101172, 2023.

\bibitem{Navarro:2026lrf}
Henrique Navarro and Roldao da~Rocha.
\newblock {Coherent quantum hairy black holes from gravitational decoupling}.
\newblock 2 2026.

\bibitem{Guimaraes:2025jsh}
V.~F. Guimar{\~a}es, R.~T. Cavalcanti, and R.~da~Rocha.
\newblock {Hair imprints of the gravitational decoupling and hairy black hole
  spectroscopy}.
\newblock {\em Class. Quant. Grav.}, 42(17):175011, 2025.

\bibitem{Meert:2021khi}
Pedro Meert and Roldao da~Rocha.
\newblock {Gravitational decoupling, hairy black holes and conformal
  anomalies}.
\newblock {\em Eur. Phys. J. C}, 82(2):175, 2022.

\bibitem{daRocha:2026kko}
Roldao da~Rocha.
\newblock {Hairy black hole solutions in nonlocal quadratic gravity}.
\newblock 1 2026.

\bibitem{Herdeiro:2015waa}
Carlos A.~R. Herdeiro and Eugen Radu.
\newblock {Asymptotically flat black holes with scalar hair: a review}.
\newblock {\em Int. J. Mod. Phys. D}, 24(09):1542014, 2015.

\bibitem{Torii:1998ir}
Takashi Torii, Kengo Maeda, and Makoto Narita.
\newblock {No scalar hair conjecture in asymptotic de Sitter space-time}.
\newblock {\em Phys. Rev. D}, 59:064027, 1999.

\bibitem{Torii:2001pg}
Takashi Torii, Kengo Maeda, and Makoto Narita.
\newblock {Scalar hair on the black hole in asymptotically anti-de Sitter
  space-time}.
\newblock {\em Phys. Rev. D}, 64:044007, 2001.

\bibitem{Martinez:2004nb}
Cristian Martinez, Ricardo Troncoso, and Jorge Zanelli.
\newblock {Exact black hole solution with a minimally coupled scalar field}.
\newblock {\em Phys. Rev. D}, 70:084035, 2004.

\bibitem{Winstanley:2002jt}
Elizabeth Winstanley.
\newblock {On the existence of conformally coupled scalar field hair for black
  holes in (anti-)de Sitter space}.
\newblock {\em Found. Phys.}, 33:111--143, 2003.

\bibitem{Kolyvaris:2010yyf}
Theodoros Kolyvaris, George Koutsoumbas, Eleftherios Papantonopoulos, and
  George Siopsis.
\newblock {A New Class of Exact Hairy Black Hole Solutions}.
\newblock {\em Gen. Rel. Grav.}, 43:163--180, 2011.

\bibitem{Dias:2011tj}
Oscar J.~C. Dias, Pau Figueras, Shiraz Minwalla, Prahar Mitra, Ricardo
  Monteiro, and Jorge~E. Santos.
\newblock {Hairy black holes and solitons in global $AdS_5$}.
\newblock {\em JHEP}, 08:117, 2012.

\bibitem{Bhattacharyya:2010yg}
Sayantani Bhattacharyya, Shiraz Minwalla, and Kyriakos Papadodimas.
\newblock {Small Hairy Black Holes in $AdS_5 x S^5$}.
\newblock {\em JHEP}, 11:035, 2011.

\bibitem{Kleihaus:2013tba}
Burkhard Kleihaus, Jutta Kunz, Eugen Radu, and Bintoro Subagyo.
\newblock {Axially symmetric static scalar solitons and black holes with scalar
  hair}.
\newblock {\em Phys. Lett. B}, 725:489--494, 2013.

\bibitem{Kolyvaris:2011fk}
Theodoros Kolyvaris, George Koutsoumbas, Eleftherios Papantonopoulos, and
  George Siopsis.
\newblock {Scalar Hair from a Derivative Coupling of a Scalar Field to the
  Einstein Tensor}.
\newblock {\em Class. Quant. Grav.}, 29:205011, 2012.

\bibitem{Anabalon:2013qua}
Andres Anabalon, Dumitru Astefanesei, and Robert Mann.
\newblock {Exact asymptotically flat charged hairy black holes with a dilaton
  potential}.
\newblock {\em JHEP}, 10:184, 2013.

\bibitem{Mahapatra:2020wym}
Subhash Mahapatra, Supragyan Priyadarshinee, Gosala~Narasimha Reddy, and
  Bhaskar Shukla.
\newblock {Exact topological charged hairy black holes in AdS Space in
  $D$-dimensions}.
\newblock {\em Phys. Rev. D}, 102(2):024042, 2020.

\bibitem{Priyadarshinee:2023cmi}
Supragyan Priyadarshinee and Subhash Mahapatra.
\newblock {Analytic three-dimensional primary hair charged black holes and
  thermodynamics}.
\newblock {\em Phys. Rev. D}, 108(4):044017, 2023.

\bibitem{Daripa:2024ksg}
Ayan Daripa and Subhash Mahapatra.
\newblock {Analytic three-dimensional primary hair charged black holes with
  Coulomb-like electrodynamics and their thermodynamics}.
\newblock {\em Phys. Rev. D}, 109(12):124039, 2024.

\bibitem{Dudal:2017max}
David Dudal and Subhash Mahapatra.
\newblock {Thermal entropy of a quark-antiquark pair above and below
  deconfinement from a dynamical holographic QCD model}.
\newblock {\em Phys. Rev. D}, 96(12):126010, 2017.

\bibitem{Bohra:2019ebj}
Hardik Bohra, David Dudal, Ali Hajilou, and Subhash Mahapatra.
\newblock {Anisotropic string tensions and inversely magnetic catalyzed
  deconfinement from a dynamical AdS/QCD model}.
\newblock {\em Phys. Lett. B}, 801:135184, 2020.

\bibitem{Mahapatra:2018gig}
Subhash Mahapatra and Pratim Roy.
\newblock {On the time dependence of holographic complexity in a dynamical
  Einstein-dilaton model}.
\newblock {\em JHEP}, 11:138, 2018.

\bibitem{He:2013qq}
Song He, Shang-Yu Wu, Yi~Yang, and Pei-Hung Yuan.
\newblock {Phase Structure in a Dynamical Soft-Wall Holographic QCD Model}.
\newblock {\em JHEP}, 04:093, 2013.

\bibitem{Arefeva:2018hyo}
Irina Aref'eva and Kristina Rannu.
\newblock {Holographic Anisotropic Background with Confinement-Deconfinement
  Phase Transition}.
\newblock {\em JHEP}, 05:206, 2018.

\bibitem{Arefeva:2018cli}
Irina Aref'eva, Kristina Rannu, and Pavel Slepov.
\newblock {Orientation Dependence of Confinement-Deconfinement Phase Transition
  in Anisotropic Media}.
\newblock {\em Phys. Lett. B}, 792:470--475, 2019.

\bibitem{Dudal:2021jav}
D.~Dudal, A.~Hajilou, and S.~Mahapatra.
\newblock {A quenched 2-flavour
  Einstein{\textendash}Maxwell{\textendash}Dilaton gauge-gravity model}.
\newblock {\em Eur. Phys. J. A}, 57(4):142, 2021.

\bibitem{Alanen:2009xs}
J.~Alanen, K.~Kajantie, and V.~Suur-Uski.
\newblock {A gauge/gravity duality model for gauge theory thermodynamics}.
\newblock {\em Phys. Rev. D}, 80:126008, 2009.

\bibitem{Cai:2012xh}
Rong-Gen Cai, Song He, and Danning Li.
\newblock {A hQCD model and its phase diagram in Einstein-Maxwell-Dilaton
  system}.
\newblock {\em JHEP}, 03:033, 2012.

\bibitem{Toniato:2025gts}
Bruno Toniato, David Dudal, Subhash Mahapatra, Roldao da~Rocha, and
  Siddhi~Swarupa Jena.
\newblock {Holographic QCD model for heavy and exotic mesons at finite density:
  A self-consistent dynamical approach}.
\newblock {\em Phys. Rev. D}, 111(12):126021, 2025.

\bibitem{Karch:2006pv}
Andreas Karch, Emanuel Katz, Dam~T. Son, and Mikhail~A. Stephanov.
\newblock {Linear confinement and AdS/QCD}.
\newblock {\em Phys. Rev. D}, 74:015005, 2006.

\bibitem{Gursoy:2007cb}
U.~Gursoy and E.~Kiritsis.
\newblock {Exploring improved holographic theories for QCD: Part I}.
\newblock {\em JHEP}, 02:032, 2008.

\bibitem{Dudal:2018ztm}
David Dudal and Subhash Mahapatra.
\newblock {Interplay between the holographic QCD phase diagram and entanglement
  entropy}.
\newblock {\em JHEP}, 07:120, 2018.

\bibitem{Mahapatra:2019uql}
Subhash Mahapatra.
\newblock {Interplay between the holographic QCD phase diagram and mutual
  {\textbackslash}{\&} $n$-partite information}.
\newblock {\em JHEP}, 04:137, 2019.

\bibitem{Jena:2022nzw}
Siddhi~Swarupa Jena, Bhaskar Shukla, David Dudal, and Subhash Mahapatra.
\newblock {Entropic force and real-time dynamics of holographic quarkonium in a
  magnetic field}.
\newblock {\em Phys. Rev. D}, 105(8):086011, 2022.

\bibitem{Herzog:2006ra}
Christopher~P. Herzog.
\newblock {A Holographic Prediction of the Deconfinement Temperature}.
\newblock {\em Phys. Rev. Lett.}, 98:091601, 2007.

\bibitem{Gubser:2000nd}
Steven~S. Gubser.
\newblock {Curvature singularities: The Good, the bad, and the naked}.
\newblock {\em Adv. Theor. Math. Phys.}, 4:679--745, 2000.

\bibitem{Coleman:1991ku}
Sidney~R. Coleman, John Preskill, and Frank Wilczek.
\newblock {Quantum hair on black holes}.
\newblock {\em Nucl. Phys. B}, 378:175--246, 1992.

\bibitem{Gonzalez:2013aca}
P.~A. Gonz{\'a}lez, Eleftherios Papantonopoulos, Joel Saavedra, and Yerko
  V{\'a}squez.
\newblock {Four-Dimensional Asymptotically AdS Black Holes with Scalar Hair}.
\newblock {\em JHEP}, 12:021, 2013.

\bibitem{Anabalon:2012sn}
Andres Anabalon, Fabrizio Canfora, Alex Giacomini, and Julio Oliva.
\newblock {Black Holes with Primary Hair in gauged N=8 Supergravity}.
\newblock {\em JHEP}, 06:010, 2012.

\bibitem{Charmousis:2014zaa}
Christos Charmousis, Theodoros Kolyvaris, Eleftherios Papantonopoulos, and
  Minas Tsoukalas.
\newblock {Black Holes in Bi-scalar Extensions of Horndeski Theories}.
\newblock {\em JHEP}, 07:085, 2014.

\bibitem{Kitagawa:2026tcl}
Masaki Kitagawa, Naoki Tsukamoto, and Ryotaro Kase.
\newblock {Dyonic hairy black holes in $U(1)$ gauge-invariant
  scalar-vector-tensor theories: Cubic and quartic interactions}.
\newblock 3 2026.

\bibitem{Anabalon:2016izw}
Andres Anabalon, Dumitru Astefanesei, and David Choque.
\newblock {Hairy AdS Solitons}.
\newblock {\em Phys. Lett. B}, 762:80--85, 2016.

\bibitem{Breitenlohner:1982jf}
Peter Breitenlohner and Daniel~Z. Freedman.
\newblock {Stability in Gauged Extended Supergravity}.
\newblock {\em Annals Phys.}, 144:249, 1982.

\bibitem{DeWolfe:2011ts}
Oliver DeWolfe, Steven~S. Gubser, and Christopher Rosen.
\newblock {Dynamic critical phenomena at a holographic critical point}.
\newblock {\em Phys. Rev. D}, 84:126014, 2011.

\bibitem{tHooft:1974pnl}
Gerard 't~Hooft.
\newblock {A Two-Dimensional Model for Mesons}.
\newblock {\em Nucl. Phys. B}, 75:461--470, 1974.

\bibitem{Polyakov:1976fu}
Alexander~M. Polyakov.
\newblock {Quark Confinement and Topology of Gauge Groups}.
\newblock {\em Nucl. Phys. B}, 120:429--458, 1977.

\bibitem{Gross:1992tu}
David~J. Gross.
\newblock {Two-dimensional QCD as a string theory}.
\newblock {\em Nucl. Phys. B}, 400:161--180, 1993.

\bibitem{Balasubramanian:1999re}
Vijay Balasubramanian and Per Kraus.
\newblock {A Stress tensor for Anti-de Sitter gravity}.
\newblock {\em Commun. Math. Phys.}, 208:413--428, 1999.

\bibitem{Liu:2013gja}
Hai-Shan Liu and H.~L{\"u}.
\newblock {Scalar Charges in Asymptotic AdS Geometries}.
\newblock {\em Phys. Lett. B}, 730:267--270, 2014.

\bibitem{Lu:2014maa}
H.~Lu, C.~N. Pope, and Qiang Wen.
\newblock {Thermodynamics of AdS Black Holes in Einstein-Scalar Gravity}.
\newblock {\em JHEP}, 03:165, 2015.

\bibitem{Li:2020spf}
Li~Li.
\newblock {On Thermodynamics of AdS Black Holes with Scalar Hair}.
\newblock {\em Phys. Lett. B}, 815:136123, 2021.

\bibitem{Berenguer:2026ftk}
Marti Berenguer, Johanna Erdmenger, Nick Evans, Wanxiang Fan, and Florian
  Vasel.
\newblock {Confinement and chiral symmetry breaking in holography: a smooth
  switch-off}.
\newblock 1 2026.

\bibitem{Nongmaithem:2026rop}
Jasper Nongmaithem, Deepak~Kumar Kar, Ayush Kumar, Shreeya~Rajesh Parab, David
  Dudal, and Subhash Mahapatra.
\newblock {Holographic Krylov complexity in confining backgrounds: particle,
  string, and frames}.
\newblock 10 2026.

\bibitem{Yang:2023qxx}
Jiayue Yang and Andrew~R. Frey.
\newblock {Complexity, scaling, and a phase transition}.
\newblock {\em JHEP}, 09:029, 2023.

\bibitem{Chakraborty:2025crb}
Adrita Chakraborty and Balbeer Singh.
\newblock {Study of chaos and scrambling in hairy AdS Soliton}.
\newblock 9 2025.

\bibitem{Shaghoulian:2015lcn}
Edgar Shaghoulian.
\newblock {Black hole microstates in AdS}.
\newblock {\em Phys. Rev. D}, 94(10):104044, 2016.

\bibitem{BravoGaete:2017dso}
Moises Bravo~Gaete, Luis Guajardo, and Mokhtar Hassaine.
\newblock {A Cardy-like formula for rotating black holes with planar horizon}.
\newblock {\em JHEP}, 04:092, 2017.

\end{thebibliography}
\bibliographystyle{unsrt}

\end{document}